\documentclass[twocolumn]{article}
\usepackage{indentfirst,amsmath,latexsym,amssymb}
\newcommand{\bs}{\boldsymbol}

\newcommand{\be}{\begin{equation}}
\newcommand{\ee}{\end{equation}}
\newcommand{\ra}{\rightarrow}

\newcommand{\qed}{\hfill $\bullet$}

\newtheorem{prop}{Proposition}
\newtheorem{thm}[prop]{Theorem}
\newtheorem{lem}[prop]{Lemma}
\newtheorem{cor}[prop]{Corollary}

\newenvironment{prf}{\trivlist \item[\hskip \labelsep{\bf Proof}]}{\qed \endtrivlist}

\newenvironment{alg}{\trivlist \item[\hskip \labelsep{\bf Algorithm}]}{\qed \endtrivlist}
\newenvironment{defn}[1]{\trivlist \item[\hskip \labelsep{\bf Definition #1}]}{\qed \endtrivlist}

\begin{document}
%\pagestyle{empty}
%\small

%DEFNS FROM NEW10G3

\newcommand{\je}[1]{j={#1}, \ldots \nu}
\newcommand{\jem}[1]{j={#1}, \ldots \nu-1}

\newcommand{\bc}{\mathbf{c}}
\newcommand{\bldf}{\mathbf{f}}

\newcommand{\cu}[1]{\mathbf{U}^{#1}}
\newcommand{\bu}[1]{\mathbf{u}^{#1}}
\newcommand{\bx}{{\mathbf{x}}}
\newcommand{\by}{{\mathbf{y}}}
\newcommand{\bz}{\mathbf{z}}
\newcommand{\bh}{\mathbf{h}}
\newcommand{\bg}{\mathbf{g}}
\newcommand{\bw}{\mathbf{w}}
\newcommand{\bp}{\mathbf{p}}
\newcommand{\bq}{\mathbf{q}}
\newcommand{\co}{\mathbf{O}}
\newcommand{\bo}{\mathbf{o}}
\newcommand{\bldp}{\mathbf{p}}

\newcommand{\sx}{{\mathbb{S}}_X}   %new
\newcommand{\sxp}{{\mathbb{S}}_{X'}}   %new
\newcommand{\imomg}{{\rm im}\,\omega}   %new

\newcommand{\cz}{\mathbf{Z}}

\newcommand{\uu}[1]{u_{#1}, \ldots u_\nu}
\newcommand{\cuu}[1]{U_{#1} \times \ldots U_\nu}
\newcommand{\pu}[1]{\p{#1} u_{#1}}
\newcommand{\cphi}[2][k]{\Phi^{#2}_{#1}}
\newcommand{\cthe}[2][k]{\Theta^{#2}_{#1}}

\newcommand{\thek}[2][k]{\theta_{#1,#2}}  %can't use \the; predefined.

\newcommand{\tldcthe}[2][k]{{\tilde{\Theta}}^{#2}_{#1}}
\newcommand{\tldthe}[2][k]{{\tilde{\theta}}_{{#1},{#2}}}
\newcommand{\tldu}[1]{{\tilde{U}}_{#1}}  %new

\newcommand{\dotcthe}[2][k]{{\dot{\Theta}}^{#2}_{#1}}
\newcommand{\dotthe}[2][k]{{\dot{\theta}}_{{#1},{#2}}}

\newcommand{\grphd}{{\mathcal{D}}}
\newcommand{\grphf}{{\mathcal{F}}}
\newcommand{\grpht}{{\mathcal{T}}}
\newcommand{\grphr}{{\mathcal{R}}}
\newcommand{\calvd}{\mathcal{V}(\grphd)}
\newcommand{\caled}{\mathcal{E}(\grphd)}
\newcommand{\calvt}{\mathcal{V}(\grpht)}
\newcommand{\calet}{\mathcal{E}(\grpht)}
\newcommand{\calvs}{\mathcal{V}(\grphs)}
\newcommand{\cales}{\mathcal{E}(\grphs)}
\newcommand{\caler}{\mathcal{E}(\grphr)}
\newcommand{\calvr}{\mathcal{V}(\grphr)}
\newcommand{\calef}{\mathcal{E}(\grphf)}
\newcommand{\calvf}{\mathcal{V}(\grphf)}
\newcommand{\ffa}{(\grphf,f_A)}  %changed
\newcommand{\dda}{(\grphd,d_A)}  %changed
\newcommand{\gtta}{(\grpht,t_A)}
\newcommand{\autd}{{\rm Aut}(\grphd)}
\newcommand{\autdda}{{\rm Aut}(\grphd,d_A)}
\newcommand{\ct}{{\mathbf{T}}}
\newcommand{\bt}{{\mathbf{t}}}
\newcommand{\teeta}{(\ct,\bt_A)}  %changed
\newcommand{\ce}{{\mathbf{E}}}
\newcommand{\cet}{{\mathbf{E}}_\grpht}
\newcommand{\ces}{{\mathbf{E}}_\grphs}
\newcommand{\cesp}{{\mathbf{E}}_{\grphs'}}
\newcommand{\cer}{{\mathbf{E}}_\grphr}

% PARAMETERS
\newcommand{\qprm}{q}

\newcommand{\htc}{homogeneous trellis code}
\newcommand{\hts}{homogeneous trellis shift}
\newcommand{\hlt}{homogeneous Latin trellis}
\newcommand{\rlt}{regular Latin trellis}
\newcommand{\shlt}{sharply homogeneous Latin trellis}
\newcommand{\rhlt}{regular homogeneous Latin trellis}
\newcommand{\hc}{homogeneous code}
\newcommand{\hs}{homogeneous shift}
\newcommand{\htr}{homogeneous trellis}
\newcommand{\sht}{sharply homogeneous trellis}
\newcommand{\rht}{regular homogeneous trellis}
\newcommand{\st}{sharply transitive}
\newcommand{\ifof}{if and only if}
\newcommand{\mc}{monotonicity condition}
\newcommand{\tg}{translation group}
\newcommand{\tn}{translation net}
\newcommand{\tstn}{translation $(t,3)$ net}
\newcommand{\rseg}{regular symmetry edge group}
\newcommand{\sh}{sharply homogeneous}
\newcommand{\lt}{Latin trellis}
\newcommand{\ls}{Latin square}
\newcommand{\mols}{mutually orthogonal Latin squares}
\newcommand{\lsr}{Latin square representation}
\newcommand{\sqs}{square subgraph}
\newcommand{\ts}{trellis section}
\newcommand{\sg}{symmetry group}
\newcommand{\ag}{automorphism group}
\newcommand{\ch}{coupling homomorphism}

\newcommand{\cy}{{\mathbf{Y}}}
\newcommand{\cf}{{\mathbf{F}}}
\newcommand{\calo}{{\mathcal{O}}}

\newcommand{\bone}{{\mathbf{1}}}
\newcommand{\bzero}{{\mathbf{0}}}

\newcommand{\bv}[1]{\mathbf{v}^{#1}}
\newcommand{\bvhat}[1]{{\hat{\mathbf{v}}}^{#1}}
\newcommand{\cv}[1]{\mathbf{V}^{#1}}
\newcommand{\sm}{Schreier matrix}
\newcommand{\dsm}{dual Schreier matrix}
\newcommand{\xj}{{\{X_j\}}}
\newcommand{\yk}{{\{Y_k\}}}
\newcommand{\mhatj}{{{\hat{M}}_j}}
\newcommand{\xhatj}{{{\hat{X}}_j}}
\newcommand{\bjpset}{{\{B_j^+\}}}
\newcommand{\bjmset}{{\{B_j^-\}}}
\newcommand{\bkmset}{{\{B_k^-\}}}
\newcommand{\bjp}{{B_j^+}}
\newcommand{\bkm}{{B_k^-}}
\newcommand{\bjm}{{B_j^-}}
\newcommand{\grphtjp}{{\grpht_j^+}}
\newcommand{\grphtkm}{{\grpht_k^-}}
\newcommand{\sctl}{strongly controllable}
\newcommand{\scgs}{strongly controllable group shift}
\newcommand{\scgt}{strongly controllable group trellis}
\newcommand{\scgc}{strongly controllable group code}
\newcommand{\lcgt}{$\ell$-controllable group trellis}
\newcommand{\lctl}{$l$-controllable}
\newcommand{\ellctl}{$\ell$-controllable}
\newcommand{\ellctlsg}{$\ell$-controllable shift group}
\newcommand{\ellctlss}{$\ell$-controllable shift structure}
\newcommand{\sctlsg}{strongly controllable shift group}
\newcommand{\gpss}{group with a shift structure}
\newcommand{\sst}{shift structure}
\newcommand{\lcmgt}{$\ell$-controllable minimal group trellis}
\newcommand{\prc}{proper representation chain}
\newcommand{\pt}{permutation tower}
\newcommand{\lgc}{Latin group code}
\newcommand{\lclgc}{$\ell$-controllable Latin group code}
\newcommand{\sclgc}{strongly controllable Latin group code}
\newcommand{\lsg}{Latin shift group}
\newcommand{\htvphi}[1]{{\hat{\varphi}}^{(#1)}}
\newcommand{\calt}{{\mathcal{T}}}
\newcommand{\hhat}{{\hat{h}}}
\newcommand{\bhhat}{{\hat{\mathbf{h}}}}
\newcommand{\hchk}{{\check{h}}}
\newcommand{\bhchk}{{\check{\mathbf{h}}}}
\newcommand{\ghat}{{\hat{g}}}
\newcommand{\phihat}{{\hat{\phi}}}
\newcommand{\bsig}{{\bs{\sigma}}}
\newcommand{\blam}{{\bs{\lambda}}}
\newcommand{\bdel}{{\pmb{\Delta}}}
\newcommand{\bdelhat}{{\hat{\pmb{\Delta}}}}
\newcommand{\chihat}{{\hat{\chi}}}
\newcommand{\what}{{\hat{w}}}
\newcommand{\bwhat}{{\hat{\mathbf{w}}}}
\newcommand{\zhat}{{\hat{z}}}
\newcommand{\bzhat}{{\hat{\mathbf{z}}}}
\newcommand{\cphihat}{{\hat{\Phi}}}
\newcommand{\calx}{{\mathcal{X}}}
\newcommand{\cals}{{\mathcal{S}}}
\newcommand{\tldr}{{\tilde{r}}}
\newcommand{\calg}{{\mathcal{G}}}
\newcommand{\tldcalg}{{\tilde{\mathcal{G}}}}
\newcommand{\calv}{{\mathcal{V}}}
\newcommand{\cale}{{\mathcal{E}}}
\newcommand{\call}{{\mathcal{L}}}
\newcommand{\tta}{{\tt{a}}}
\newcommand{\ttz}{{\tt{z}}}
\newcommand{\ttga}{{G_{\tta}}}
\newcommand{\ttka}{{K_{\tta}}}
\newcommand{\ttcla}{{L_{\tta}}}
\newcommand{\ttla}{{l_{\tta}}}
\newcommand{\ttkz}{{K_{{\tt{z}}}}}
\newcommand{\cxhat}{{\hat{X}}}
\newcommand{\hatdelta}{{\hat{\Delta}}}
\newcommand{\dotcx}{{\dot{X}}}
\newcommand{\dotcq}{{\dot{Q}}}
\newcommand{\hatcx}{{\hat{X}}}
\newcommand{\rmdef}{\stackrel{\rm def}{=}}

\title{On Strongly Controllable Group Codes and Mixing Group Shifts: \\
Solvable Groups, Translation Nets, and Algorithms}
\author{Kenneth M. Mackenthun Jr.}
\date{October 5, 2008}
\maketitle

\begin{abstract}
The branch group of a \scgc\ is a shift group.  We show
that a shift group can be characterized in a very simple way.  In
addition it is shown that if a \scgc\
is labeled with \ls s, a \sclgc, then the shift group is solvable.  Moreover
the mathematical structure of a \ls\ (as a translation net) and
the shift group of a \sclgc\ are closely related.  Thus a \sclgc\ can be
viewed as a natural extension of a \ls\ to a sequence space.
Lastly we construct shift groups.  We show that it is sufficient to
construct a simpler group, the state group of a shift group.  We
give an algorithm to find the state group, and from this it is easy
to construct a \sclgc.
\end{abstract}

\section{Introduction}

Kitchens introduced the fundamental idea of a group shift
and showed that a group shift is a shift of finite type \cite{KIT}.
A group shift is essentially a time invariant group code.
Forney and Trott showed that a group code has a well defined
state space and can be represented on a trellis, and a \scgc\
can be realized with a shift register \cite{FT}.  In a following
article, among other results, Loeliger and
Mittelholzer gave an abstract characterization of the group which can
appear as the branch group of a \scgc,
which they call a {\it group with a shift structure} \cite{LM}.

In this paper, we give a simple characterization of a \gpss,
or {\it shift group}.  We show that a shift group $G$ involves
a normal chain $\xj$ and a tower of isomorphisms using groups
in the normal chain.  In addition, there are two important
normal subgroups $X_0$ and $Y_0$ of $G$ which have normal chains
which also characterize the shift group.  These results are shown
in Section \ref{sec2}.

In Section \ref{sec3}, we use the theory of \tn s to show that if
a group code is \sctl\ and is labeled with \ls s, the shift group is solvable.
We show that \ls s which can appear in a \lgc\ are isotopic
to those constructed by the automorphism method of Mann \cite{MN1}.
It is shown that if a group code is \sctl\ and if $X_0\cap Y_0=\bone$,
$X_0\simeq Y_0$, and $X_0$ is elementary abelian, then a complete
set of \mols\ can be used to label the group code (throughout the paper,
we use $\bone$ for the identity of a group).
We show that the structure of a shift group is closely related
to the structure of a \ls\ as a \tn.

In Section \ref{sec4}, we show that a shift group with $X_0\cap Y_0=\bone$
can be represented as a subdirect product group.  Then we give
necessary and sufficient conditions for a subdirect product group
to be a shift group.  These conditions show that to find a shift group
it is sufficient to construct the state group of a shift group.  We give a
characterization of the state group.

Lastly in Section \ref{sec5}, we give an algorithm to find
the state group of a shift group; this can be used to find a \lgc.

%HERE
\section{Shift groups}
\label{sec2}

Let $\calg$ be any graph with vertices $\calv$
(also called states) and edges $\cale$;
in shorthand we write $\calg=(\calv,\cale)$.
We say a graph $\calg$ is {\it $l$-controllable} if for any ordered pair of
states $(s,s')$ in $\calg$, there is a path of length $l$ from $s$ to $s'$
in $\calg$.  A graph that is \lctl\ for some integer $l$ is said to be
{\it strongly controllable}.  The least integer $l$ for which a strongly
controllable graph $\calg$ is \lctl\ is denoted as $\ell$, and we say $\calg$
is \ellctl.  In this paper, we only study the case $l=\ell$.

The preceding definition uses the idea of controllability in
systems theory and the theory of convolutional codes.  There is a
similar notion in the theory of symbolic dynamics, drawn from ergodic
theory.  A graph $\calg$ is {\it primitive} if there is a positive
integer $M$ such that for any ordered pair of states $(s,s')$ in $\calg$
and any $m\ge M$, there is a path of length $m$ from $s$ to $s'$ in
$\calg$ \cite{SMT}.  If a graph has an edge into each state, then an
$\ell$-controllable graph is primitive with $M=\ell$.

In this paper, we
consider a particular graph constructed using a group $B$, where
the edges $\cale$ form group $B$, and the vertices $\calv$ form a
quotient group in $B$.  We denote this graph as $\calg_B$.  We now
discuss this construction in more detail.

Let $B$ be a finite group which contains normal subgroups $B^+$ and
$B^-$ such that $B/B^-$ is isomorphic to $B/B^+$ via an isomorphism
$\psi:  B/B^-\ra B/B^+$.  Let $\pi^+$ be the (natural) map which
sends each element of $B$ to the coset of $B^+$ that it belongs to;
likewise for $\pi^-:  B\ra B/B^-$.  Let $\calg_B=(\calv,\cale)$ be the
graph with vertices $\calv=B/B^+$ and edges $\cale=B$, such that
each edge $e\in\cale$ has initial state ${\tt i}(e)=\pi^+(e)$
and terminal state ${\tt t}(e)=\psi\circ\pi^-(e)$.
(This discussion is taken from Problem 2.2.16 of \cite{LMr}, which is
based on \cite{FT,LM}.)
It is known that the edge shift of graph $\calg_B$
is a group shift, and moreover, any group shift which is also an edge
shift can be modeled in this way \cite{LMr,FT}.

We want to determine when graph $\calg_B$ is \ellctl.
As in \cite{LM}, consider all paths
$e_0,e_1,\ldots e_j,\ldots$ in $\calg_B$ which begin
in the identity state, i.e., ${\tt i}(e_0)=B^+$.  Let $\bjp$, $j\ge 0$,
be the set of all edges $e_j$ on such paths.  Similarly,
consider all paths $\ldots e_{-j},\ldots e_{-1},e_0$ in $\calg_B$ which end
in the identity state, i.e., ${\tt t}(e_0)=B^-$.  Let $\bjm$, $j\ge 0$,
be the set of all edges $e_{-j}$ on such paths.
Note that $B_0^+=B^+$ and $B_0^-=B^-$.  Also note that $B_j^+\lhd B$
and $B_j^-\lhd B$ \cite{LM}.
The next result follows directly from work in \cite{LM}.

\begin{prop}
\label{prop1}
The graph $\calg_B$ is $\ell$-controllable \ifof\
$B_\ell^+=B$, or equivalently, \ifof\ $B_\ell^-=B$.
\end{prop}

We denote the normal series $B_{-1}^+, B_0^+, B_1^+, \ldots B_\ell^+$
by the notation $\bjpset$, where $B_{-1}^+$ is the identity $\bone$ of $B$, and
the normal series $B_{-1}^-, B_0^-, B_1^-, \ldots B_\ell^-$
by the notation $\bjmset$, where $B_{-1}^-$ is the identity $\bone$ of $B$.
Loeliger and Mittelholzer give a definition of a \gpss\ which uses
$\bjpset$, $\bjmset$, and intersection terms \cite{LM}.  Here we
study a simpler definition which uses just $\bjpset$ and $B_0^-$
\cite{MAC,MAC1}.  Consider a group $G$ with a normal series
$$ \bone=X_{-1}\subset X_0\subset X_1 \subset \cdots \subset X_\ell=G,  $$
where $X_{-1}$ is the identity $\bone$ of $G$.
We denote the normal series $X_{-1},X_0,\ldots X_\ell$ by $\xj$.

\begin{defn}{1}
%\label{ss}
We say a group $G$ has a {\it shift structure} $(\xj,Y_0,\varphi)$
if there is a normal chain $\xj$ with $X_\ell=G$ and each $X_j\lhd G$,
a normal subgroup $Y_0$, and an isomorphism $\varphi$ from $G/Y_0$
onto $G/X_0$ such that
\be
\label{small1}
\varphi(X_j Y_0/Y_0)=X_{j+1}/X_0
\ee
for $-1\le j<\ell$.

We say $G$ is a {\it shift group} if it has a \sst\ $(\xj,Y_0,\varphi)$.
\end{defn}

{\it Remark:}  Note that $G/Y_0\simeq G/X_0$ implies $|Y_0|=|X_0|$ \cite{LM}.
Furthermore, using (\ref{small1}) for $j=\ell-1$, we have
$\varphi(X_{\ell-1} Y_0/Y_0)=X_\ell/X_0$.  This means $X_{\ell-1} Y_0=G$.
Lastly, note that (\ref{small1}) holds trivially for $j=-1$.

\begin{thm}
If the graph $\calg_B$ is $\ell$-controllable, then $B$ has a \sst\
$(\bjpset,B_0^-,\psi)$.
\end{thm}

\begin{prf}
If the graph $\calg_B$ is $\ell$-controllable, there is a sequence
$\bjpset$ with $B_\ell^+=B$.  It is easy to see that each $\bjp\lhd B$ \cite{LM}.
We know that the terminal states of $\bjp$ are the initial states of
$B_{j+1}^+$.  But
the terminal states of $\bjp$ are $\psi(\bjp B_0^-/B_0^-)$, and the
initial states of $B_{j+1}^+$ are $B_{j+1}^+/B_0^+$.  Thus we must have
$$
\psi(\frac{\bjp B_0^-}{B_0^-})=\frac{B_{j+1}^+}{B_0^+}
$$
for all $j$, $-1\le j<\ell$.  This proves (\ref{small1}) of
Definition 1.
\end{prf}

Let $G$ be a \gpss\ $(\xj,Y_0,\varphi)$.  We define $\calg_G$ to be
a graph analogous to $\calg_B$, that is, $\calg_G$ is the graph $(\calv,\cale)$
with vertices $\calv=G/X_0$ and edges $\cale=G$, such that
each edge $e\in\cale$ has initial state ${\tt i}(e)=\pi_X(e)$
and terminal state ${\tt t}(e)=\varphi\circ\pi_Y(e)$, where $\varphi$ is
an isomorphism $\varphi:  G/Y_0\ra G/X_0$, and $\pi_X$, $\pi_Y$ are
the natural maps $\pi_X:  G\ra G/X_0$, $\pi_Y:  G\ra G/Y_0$.

\begin{thm}
Let $(\xj,Y_0,\varphi)$ be a \sst\ for some group $G$.  Then the
graph $\calg_G$ is $\ell$-controllable.
\end{thm}

\begin{prf}
We show that $\xj$ gives a sequence of edges
which form well defined paths.  We must show that the terminal states
of $X_j$ are the initial states of $X_{j+1}$.  But the terminal states
of $X_j$ are $\varphi(X_j Y_0/Y_0)$, and the initial states of $X_{j+1}$
are $X_{j+1}/X_0$.  Since
$$
\varphi(X_j Y_0/Y_0)=X_{j+1}/X_0
$$
by assumption, $\xj$ gives a well defined sequence of edges.
But we know that $X_\ell=G$; thus $\calg_G$ is $\ell$-controllable
by Proposition \ref{prop1}.
\end{prf}

The proofs of the above two theorems are patterned after corresponding
proofs in \cite{LM}.  These two theorems give the following important
corollary.

\begin{cor}
The graph $\calg_B$ is $\ell$-controllable \ifof\ $B$ is a shift group
with \sst\ $(\bjpset,B_0^-,\psi)$.  An analogous result holds for graph
$\calg_G$.
\end{cor}

We pause here to give two useful technical lemmas.
The following lemma is an easy extension of the first isomorphism theorem.

\begin{lem}
\label{lem16}
Let $H$ and $H'$ be groups and consider any homomorphism $f$ from $H$ onto $H'$.
If $H_1'$ is any normal subgroup of $H'$, then $f^{-1}(H_1')$ is a normal
subgroup of $H$ and $H/f^{-1}(H_1')\simeq H'/H_1'$.
\end{lem}

\begin{lem}
\label{wkhorse}
Let groups $Q'$, $Q$, $R'$, and $R$ satisfy $Q\subset R$, $Q'\subset Q$,
$R'\subset R$, $Q'\subset R'$, $R'\cap Q=Q'$, and $R=QR'$.
Assume that $Q\lhd R$, $R'\lhd R$.  There are three results:
\be
\label{wkhse1}
\frac{Q}{Q'}\simeq\frac{R}{R'}
\ee
with assignment $qQ'\mapsto qR'$ for $q\in Q$,
\be
\label{wkhse2}
\frac{R}{Q}\simeq\frac{R'}{Q'},
\ee
and
\be
\label{wkhse3}
\frac{R}{Q'}\simeq\frac{Q}{Q'}\times\frac{R'}{Q'}.
\ee
\end{lem}

\begin{prf}
It is clear $R'R=R'Q$.  Then
\be
\label{eqwkhse}
\frac{R'R}{R'}=\frac{R'Q}{R'}.
\ee
By the second isomorphism theorem, there is an isomorphism
$$
\nu:  \frac{R}{R'\cap R}\ra\frac{R'R}{R'},
$$
and an isomorphism
$$
\nu':  \frac{Q}{R'\cap Q}\ra\frac{R'Q}{R'}.
$$
Then using (\ref{eqwkhse}) there is an isomorphism
$$
\nu^{-1}\circ\nu':  \frac{Q}{R'\cap Q}\ra\frac{R}{R'\cap R},
$$
or just
$$
\nu^{-1}\circ\nu':  \frac{Q}{Q'}\ra\frac{R}{R'}.
$$
Thus there is an isomorphism
$$
\frac{Q}{Q'}\simeq\frac{R}{R'},
$$
with assignment $qQ'\mapsto qR'$ for $q\in Q$.  This proves (\ref{wkhse1}).

We now show (\ref{wkhse2}).  Since $R=QR'$, each coset of $Q$ in $R$
must contain a representative in $R'$.  But the
representatives of $R'$ in $Q$ are $Q'$.  Thus each coset of $Q$ in $R$
contains one and only one coset of $Q'$.  Then it is clear that
we can define a 1-1 correspondence between cosets of $Q'$ in $R'$ and
cosets of $Q$ in $R$, and this gives the isomorphism in (\ref{wkhse2}).

We know that $Q'\lhd R$.  From the preceding paragraph,
each coset of $Q$ in $R$ contains one and only one coset of $Q'$ in $R'$.
Then
$$
\frac{R}{Q'}=\left(\frac{Q}{Q'}\right)\left(\frac{R'}{Q'}\right).
$$
Since each coset of $Q$ in $R$ contains one and only one coset of $Q'$, this
means $(Q/Q')\cap (R'/Q')=\bone$.  Also we have $Q/Q'\lhd R/Q'$ and $R'/Q'\lhd R/Q'$.
Then (\ref{wkhse3}) follows.
\end{prf}

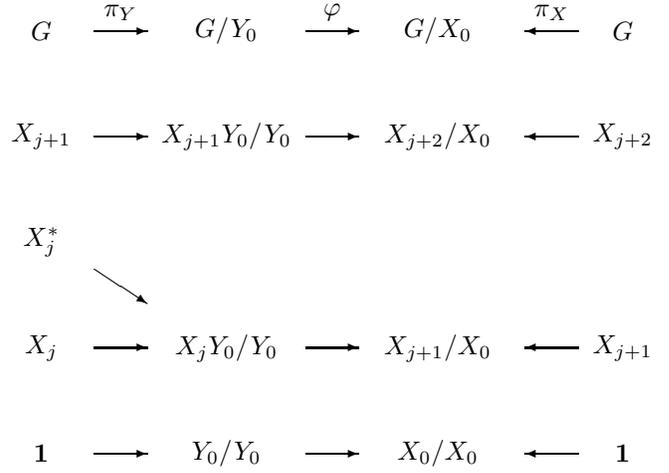
\begin{figure}[htbp]
\centering
\vspace{3ex}

\begin{picture}(220,160)

\put(0,0){\makebox(0,0){$\bone$}}
\put(0,40){\makebox(0,0){$X_j$}}
\put(0,80){\makebox(0,0){$X_j^*$}}
\put(0,120){\makebox(0,0){$X_{j+1}$}}
\put(0,160){\makebox(0,0){$G$}}

\put(70,0){\makebox(0,0){$Y_0/Y_0$}}
\put(70,40){\makebox(0,0){$X_j Y_0/Y_0$}}
\put(70,120){\makebox(0,0){$X_{j+1} Y_0/Y_0$}}
\put(70,160){\makebox(0,0){$G/Y_0$}}
%\put(70,80){\makebox(0,0){$\frac{X_j Y_0}{Y_0}$}}

\put(150,0){\makebox(0,0){$X_0/X_0$}}
\put(150,40){\makebox(0,0){$X_{j+1}/X_0$}}
\put(150,120){\makebox(0,0){$X_{j+2}/X_0$}}
\put(150,160){\makebox(0,0){$G/X_0$}}

\put(220,0){\makebox(0,0){$\bone$}}
\put(220,40){\makebox(0,0){$X_{j+1}$}}
\put(220,120){\makebox(0,0){$X_{j+2}$}}
\put(220,160){\makebox(0,0){$G$}}

\put(20,0){\vector(1,0){20}}
%\put(30,4){\makebox(0,0)[b]{$f_{j+1}$}}
\put(20,40){\vector(1,0){20}}
%\put(30,84){\makebox(0,0)[b]{$f_{j+1}$}}
\put(20,120){\vector(1,0){20}}
%\put(30,124){\makebox(0,0)[b]{$f_{j+1}$}}
\put(20,70){\vector(3,-2){20}}
%\put(30,63){\makebox(0,0)[b]{$f_{j+1}$}}
\put(20,160){\vector(1,0){20}}
\put(30,164){\makebox(0,0)[b]{$\pi_Y$}}

\put(100,160){\vector(1,0){20}}
\put(110,164){\makebox(0,0)[b]{$\varphi$}}
\put(100,120){\vector(1,0){20}}
%\put(110,124){\makebox(0,0)[b]{$\varphi$}}
\put(100,40){\vector(1,0){20}}
%\put(110,84){\makebox(0,0)[b]{$\varphi$}}
\put(100,0){\vector(1,0){20}}
%\put(110,4){\makebox(0,0)[b]{$\varphi$}}

\put(203,0){\vector(-1,0){20}}
\put(203,40){\vector(-1,0){20}}
\put(203,120){\vector(-1,0){20}}
\put(203,160){\vector(-1,0){20}}
\put(193,164){\makebox(0,0)[b]{$\pi_X$}}

\end{picture}

\caption{Relationship of groups in $G$.}
\label{fig1}

\end{figure}

We now discuss Figure \ref{fig1}, which shows the relationship of
some important groups in $G$.  Note that groups in the same column
are subgroups of the group at the top.

Examine the left side of Figure \ref{fig1}.  Fix $j$, $-1\le j<\ell$.  The natural
map $\pi_Y:  G\ra G/Y_0$ is defined by the assignment $g\mapsto gY_0$.
Let $\pi_Y|X_{j+1}$ be the restriction of $\pi_Y$
to $X_{j+1}$.  Then $\pi_Y|X_{j+1}$ is an onto homomorphism
$\pi_Y|X_{j+1}:  X_{j+1}\ra X_{j+1} Y_0/Y_0$.
Now $X_j Y_0/Y_0$ is a normal subgroup of $X_{j+1} Y_0/Y_0$.  Then
from Lemma \ref{lem16}, $(\pi_Y|X_{j+1})^{-1}(X_j Y_0/Y_0)$ is a normal
subgroup of $X_{j+1}$ (we call it $X_j^*$), and
\be
\label{cs1}
\frac{X_{j+1}}{X_j^*}\simeq\frac{X_{j+1} Y_0/Y_0}{X_j Y_0/Y_0}.
\ee

\begin{lem}
\label{lem17}
For $-1\le j<\ell$, we have $X_j^*\lhd G$, $X_j^*=X_j (X_{j+1}\cap Y_0)$,
\be
\label{fst}
\frac{X_j^*}{X_j}\simeq\frac{X_{j+1}\cap Y_0}{X_j\cap Y_0},
\ee
and
\be
\label{scd}
\frac{X_j^*}{X_{j+1}\cap Y_0}\simeq\frac{X_j}{X_j\cap Y_0}.
\ee
\end{lem}

\begin{prf}
Note that $X_j^*$ is just
$X_{j+1}\cap \pi_Y^{-1}(X_j Y_0/Y_0)=X_{j+1}\cap X_j Y_0$.
Thus $X_j^*\lhd G$.  Moreover, by the Dedekind Law (cf. Problem
2.49 of \cite{ROT}), $X_{j+1}\cap X_j Y_0=X_j (X_{j+1}\cap Y_0)$,
giving $X_j^*=X_j (X_{j+1}\cap Y_0)$.
Thus $X_j^*$ is just the cosets of $X_j$ with representatives in
$X_{j+1}\cap Y_0$, or as well the cosets of $X_{j+1}\cap Y_0$
with representatives in $X_j$.  Applying Lemma \ref{wkhorse} shows
(\ref{fst}) and (\ref{scd}).
\end{prf}

\begin{prop}
We have $X_{\ell-1}^*=X_{\ell-1} (X_\ell\cap Y_0)=X_{\ell-1}Y_0=X_\ell$.
Also $X_{-1}^*=X_{-1} (X_0\cap Y_0)=X_0\cap Y_0$.
\end{prop}

Fix $j$, $-1\le j<\ell-1$.  In the center of Figure \ref{fig1},
because of the isomorphism $\varphi$, we have
\be
\label{cs2}
\frac{X_{j+1} Y_0/Y_0}{X_j Y_0/Y_0}\simeq\frac{X_{j+2}/X_0}{X_{j+1}/X_0}.
\ee
On the right side of Figure \ref{fig1}, we can apply the
correspondence theorem or third isomorphism theorem \cite{ROT}.
For example, we have
\be
\label{cs3}
\frac{X_{j+2}/X_0}{X_{j+1}/X_0}\simeq\frac{X_{j+2}}{X_{j+1}}.
\ee
Using (\ref{cs1}), (\ref{cs2}), and (\ref{cs3}), we conclude
\be
\label{cs4}
\frac{X_{j+1}}{X_j^*}\simeq\frac{X_{j+2}}{X_{j+1}}.
\ee
Thus there is an isomorphism from $X_{j+1}/X_j^*$ to $X_{j+2}/X_{j+1}$.
Further we see there is a homomorphism from $X_{j+1}/X_j$ to $X_{j+2}/X_{j+1}$.

\begin{thm}
\label{thm2}
If a group $G$ has a shift structure $(\xj,Y_0,\varphi)$, then
the chief series $\xj$ has a refinement given by
\be
\label{cut1}
\cdots\subset X_j\subset X_j^*\subset X_{j+1}\subset X_{j+1}^*\subset\cdots,
\ee
where each $X_j^*\lhd G$, such that
\be
\label{cut2}
\frac{X_{j+1}}{X_j^*}\simeq\frac{X_{j+2}}{X_{j+1}}
\ee
for $-1\le j<\ell-1$, where
\be
\label{cut3}
X_j^*=X_j (X_{j+1}\cap Y_0)
\ee
for $-1\le j<\ell$.  Note that $X_\ell/X_{\ell-1}^*\simeq\bone$.
\end{thm}

\begin{prf}
This has been shown by (\ref{cs4}) and Lemma \ref{lem17}.
\end{prf}

{\it Remark:}  Note that $X_j^*=X_j$ \ifof\ $X_{j+1}\cap Y_0=X_j\cap Y_0$,
and in this case $|X_{j+1}|/|X_j|=|X_{j+2}|/|X_{j+1}|$.  We have
$X_{-1}^*=X_{-1}(X_0\cap Y_0)=X_0\cap Y_0$.  Since $X_{\ell-1}Y_0=G$,
we have $X_{\ell-1}^*=X_{\ell-1}(X_\ell\cap Y_0)=G$.
By definition of $\ell$, we have $X_{\ell-1}<<X_\ell=G$,
and therefore $X_{\ell-1}<<X_{\ell-1}^*$.  (For groups $A$ and $B$, we define
$A>>B$ and $B<<A$ if $B$ is a strictly proper subgroup of $A$.)

Using the following lemma, we can refine the normal chain in (\ref{cut1})
and collapse the tower of isomorphisms in (\ref{cut2}) into $X_0$.

\begin{lem}
\label{lem8}
Let $G$ be a \gpss\ $(\xj,Y_0,\varphi)$.  Fix $j$, $-1\le j<\ell-1$.
If there is a normal chain
\be
\label{capa}
X_{j+1}=Q_{j+1}^0\lhd Q_{j+1}^1\lhd Q_{j+1}^2\lhd\cdots\lhd Q_{j+1}^{p-1}\lhd Q_{j+1}^p=X_{j+2},
\ee
then there is a normal chain
\begin{multline}
\label{capb}
X_j\lhd Q_j^a\lhd\cdots\lhd Q_j^b\lhd Q_j^0\lhd Q_j^1\lhd Q_j^2\lhd\cdots \\
\lhd Q_j^{p-1}\lhd Q_j^p=X_{j+1},
\end{multline}
where $Q_j^0=X_j^*$ and the normal chain
\be
\label{specnc}
X_j\lhd Q_j^a\lhd\cdots\lhd Q_j^b\lhd Q_j^0
\ee
is an arbitrary refinement of the trivial normal chain $X_j\lhd Q_j^0$.
We have $X_j=Q_j^0$ \ifof\ $X_j=X_j^*$; in this case any refinement in
(\ref{specnc}) is trivial.  Although there is no restriction on the choice of the
normal chain in (\ref{specnc}), there are dependent relations among the
$Q_j^n$ and $Q_{j+1}^n$, $0\le n\le p$.  We have
\be
\label{eqx4}
\frac{Q_j^n}{Q_j^m}\simeq\frac{Q_{j+1}^n}{Q_{j+1}^m}
\ee
for $m,n$ satisfying $0\le m\le n\le p$.
Moreover $Q_j^n\lhd G$ if $Q_{j+1}^n\lhd G$, for $n$ satisfying $0\le n\le p$.
In addition, $Q_j^n$ and $Q_{j+1}^n$ are related by the isomorphism
$\varphi$,
\be
\label{eqx5}
\varphi(Q_j^n Y_0/Y_0)=Q_{j+1}^n/X_0,
\ee
for $n$ satisfying $0\le n\le p$.  For the normal chain in
(\ref{specnc}), we have
\begin{multline}
\label{eqx6}
\varphi(X_j Y_0/Y_0)=\varphi(Q_j^a Y_0/Y_0)=\cdots=\varphi(Q_j^b Y_0/Y_0)=  \\
\varphi(Q_j^0 Y_0/Y_0)=X_{j+1}/X_0.
\end{multline}

Conversely, if there is a normal chain as in (\ref{capb}) with
$Q_j^0=X_j^*$, then there is a normal chain as in (\ref{capa}),
and $Q_{j+1}^n\lhd G$ if $Q_j^n\lhd G$, for $n$ satisfying $0\le n\le p$,
and properties (\ref{eqx4})-(\ref{eqx6}) hold.
\end{lem}

\begin{prf}
Fix $j$, $-1\le j<\ell-1$.
We first show that if (\ref{capa}) holds, then (\ref{capb}) holds.
As in (\ref{capa}), let
$$
Q_{j+1}^0\lhd Q_{j+1}^1\lhd Q_{j+1}^2\lhd\cdots\lhd Q_{j+1}^p
$$
be a normal chain with each $Q_{j+1}^n\lhd G$.
We know $X_0\lhd G$ and $X_0\subset Q_{j+1}^0$.
Then from the correspondence theorem, there is a normal chain
$$
\frac{Q_{j+1}^0}{X_0}\lhd \frac{Q_{j+1}^1}{X_0}\lhd \frac{Q_{j+1}^2}{X_0}\lhd
\cdots\lhd \frac{Q_{j+1}^p}{X_0}
$$
where
\be
\label{eqx1}
\frac{Q_{j+1}^n/X_0}{Q_{j+1}^m/X_0}\simeq\frac{Q_{j+1}^n}{Q_{j+1}^m},
\ee
for $m\ge 0,n\ge 0$ satisfying $0\le m\le n\le p$, and each $Q_{j+1}^n/X_0\lhd G/X_0$.

Since $\varphi:  G/Y_0\ra G/X_0$ is an isomorphism, for each
$n$, $0\le n\le p$, there is a subgroup $\dotcq_j^n/Y_0$ such that
$\varphi(\dotcq_j^n/Y_0)=Q_{j+1}^n/X_0$.  Thus the isomorphism $\varphi$ gives
a normal chain
\be
\label{eqxx}
\frac{\dotcq_j^0}{Y_0}\lhd \frac{\dotcq_j^1}{Y_0}\lhd \frac{\dotcq_j^2}{Y_0}\lhd
\cdots\lhd \frac{\dotcq_j^p}{Y_0},
\ee
where each $\dotcq_j^n/Y_0\lhd G/Y_0$, and
\be
\label{eqx2}
\frac{\dotcq_j^n/Y_0}{\dotcq_j^m/Y_0}\simeq\frac{Q_{j+1}^n/X_0}{Q_{j+1}^m/X_0}.
\ee
Since $G$ is a shift group, we have $\dotcq_j^0/Y_0=X_jY_0/Y_0$ and
$\dotcq_j^p/Y_0=X_{j+1}Y_0/Y_0$.

As before, consider the natural map $\pi_Y:  G\ra G/Y_0$ defined
by $g\mapsto gY_0$, and its restriction $\pi_Y|X_{j+1}$.
Define $Q_j^n\rmdef (\pi_Y|X_{j+1})^{-1}(\dotcq_j^n/Y_0)$.
Then $Q_j^0=X_j^*$ and $Q_j^p=X_{j+1}$.  Then using (\ref{eqxx}) and the
correspondence theorem, we have a normal chain
\be
\label{eqx2a}
Q_j^0\lhd Q_j^1\lhd Q_j^2\lhd\cdots\lhd Q_j^p,
\ee
where
\be
\label{eqx3}
\frac{Q_j^n}{Q_j^m}\simeq\frac{\dotcq_j^n/Y_0}{\dotcq_j^m/Y_0}.
\ee
Since $Q_j^0=X_j^*$, we have $X_j\subset Q_j^0$, and combining this
with (\ref{eqx2a}) gives (\ref{capb}).
Note that $Q_j^n=X_{j+1}\cap\pi_Y^{-1}(\dotcq_j^n/Y_0)$,
and thus each $Q_j^n\lhd G$.  Collecting (\ref{eqx1}), (\ref{eqx2}),
and (\ref{eqx3}) gives (\ref{eqx4}).  Finally we have that
$\varphi(\dotcq_j^n/Y_0)=Q_{j+1}^n/X_0$.  But
$(\pi_Y|X_{j+1})(Q_j^n)=\dotcq_j^n/Y_0$.  This means
$Q_j^n Y_0/Y_0=\dotcq_j^n/Y_0$, giving (\ref{eqx5}).

Note that (\ref{eqx6}) holds since $X_jY_0=X_j^*Y_0$.

Now assume (\ref{capb}) holds.  We can show that (\ref{capa}) holds
by essentially reversing the above steps.
\end{prf}

We see there are two cases to consider in Lemma \ref{lem8} depending
on whether $X_j^*=X_j$ or $X_j^*>>X_j$.
Formally, we introduce a parameter $\varepsilon_j$ for $-1\le j<\ell$.
We set $\varepsilon_j=1$ if $X_j^*>>X_j$, and $\varepsilon_j=0$ if $X_j^*=X_j$.

In the next theorem, we use Lemma \ref{lem8} to find a refinement of
(\ref{cut1}).  It is convenient to write the refinement using slightly
different notation than in Lemma \ref{lem8}.  Thus in place of (\ref{capa}),
we write the portion of the refinement between $X_{j+1}$ and $X_{j+2}$ as
\begin{multline}
\label{reft2}
X_{j+1}=X_{j+1}^{(i_{j+1})}\lhd X_{j+1}^{(i_{j+1}+1)}
\lhd X_{j+1}^{(i_{j+1}+2)}\lhd\cdots  \\
\lhd X_{j+1}^{(\ell'-1)}\lhd X_{j+1}^{(\ell')}=X_{j+2},
\end{multline}
where $i_{j+1}$ and $\ell'$ are positive integers.
Using (\ref{reft2}) in Lemma \ref{lem8}, we obtain the portion of the
refinement between $X_j$ and $X_{j+1}$ as
\begin{multline}
\label{reft1}
X_j\lhd X_j^{(i_{j+1})}\lhd X_j^{(i_{j+1}+1)}\lhd X_j^{(i_{j+1}+2)}\lhd\cdots  \\
\lhd X_j^{(\ell'-1)}\lhd X_j^{(\ell')}=X_{j+1},
\end{multline}
where $X_j^{(i_{j+1})}=X_j^*$.
We only use Lemma \ref{lem8} for a trivial refinement in (\ref{specnc}),
that is, when $X_j=Q_j^a=\cdots=Q_j^b$.  In (\ref{reft1}),
we have $X_j^{(i_{j+1})}=X_j^*$ if $\varepsilon_j=1$, and
$X_j^{(i_{j+1})}=X_j^*=X_j$ if $\varepsilon_j=0$.

In general for each $j$, $-1\le j\le\ell-1$, we define a refinement
in which the superscript
$m$ of $X_j^{(m)}$ runs from integer $i_j$ to integer $\ell'$.
For $0\le j\le\ell$, we define
$X_{j-1}^{(\ell')}\rmdef X_j\rmdef X_j^{(i_j)}$;
then $X_{\ell-1}^{(\ell')}=X_{\ell}=X_{\ell}^{(i_{\ell})}$.
We also define $X_{-1}\rmdef X_{-1}^{(i_{-1})}$.
In this notation,
the portion of the refinement between $X_j$ and $X_{j+1}$ is
\begin{multline}
\label{reft3}
X_j=X_j^{(i_j)}\lhd X_j^{(i_j+1)}\lhd X_j^{(i_j+2)}\lhd\cdots  \\
\lhd X_j^{(\ell'-1)}\lhd X_j^{(\ell')}=X_{j+1}.
\end{multline}
Comparing (\ref{reft1}) and (\ref{reft3}) shows that we must have
$X_j=X_j^{(i_j)}=X_j^{(i_{j+1})}=X_j^*$ if $\varepsilon_j=0$ and
$X_j^{(i_j+1)}=X_j^{(i_{j+1})}=X_j^*$ if $\varepsilon_j=1$.
This means $i_j+\varepsilon_j=i_{j+1}$.  If we use the above
procedure and apply Lemma \ref{lem8} recursively starting with
the normal chain
$$
X_{\ell-1}=X_{\ell-1}^{(i_{\ell-1})}\lhd X_{\ell-1}^{(\ell')}=X_{\ell},
$$
we obtain
\be
\label{thm14a1plus}
i_j=\ell'-\sum_{j\le i<\ell}\varepsilon_i
\ee
for $-1\le j<\ell$.  Define
$$
\ell'\rmdef \sum_{-1\le i<\ell}\varepsilon_i.
$$
Then from (\ref{thm14a1plus}) we see $i_{-1}=0$.  If $j=\ell$,
we define $i_j=i_\ell\rmdef\ell'$ trivially.
Thus as $j$ runs from $-1$ to $\ell$, $i_j$ takes all values in
the range $[0,\ell']$.

\begin{thm}
\label{thm14a}
Let a group $G$ have a shift structure $(\xj,Y_0,\varphi)$.
There is a refinement of $\xj$, and of the normal chain in (\ref{cut1}),
given by
\begin{multline}
\label{eqbb}
X_{-1}=X_{-1}^{(i_{-1})}\lhd\cdots\lhd X_{-1}^{(\ell')}=X_0=X_0^{(i_0)}\lhd\cdots  \\
\lhd X_{j-1}^{(\ell')}=X_j=X_j^{(i_j)}\lhd X_j^{(i_j+1)}\lhd X_j^{(i_j+2)}\lhd\cdots  \\
\lhd X_j^{(\ell'-1)}\lhd X_j^{(\ell')}=X_{j+1}=X_{j+1}^{(i_{j+1})}\lhd\cdots  \\
\lhd X_{\ell-1}^{(i_{\ell-1})}\lhd X_{\ell-1}^{(i_{\ell-1}+1)}=X_{\ell-1}^{(\ell')}=X_{\ell}=X_{\ell}^{(i_{\ell})},
\end{multline}
where each $X_j^{(i_j+n)}\lhd G$ and $X_j^{(i_j+1)}=X_j^*$ if $\varepsilon_j=1$.
Moreover
\be
\label{thm14a1}
\frac{X_{-1}^{(i_j+n)}}{X_{-1}^{(i_j+m)}}\simeq\frac{X_j^{(i_j+n)}}{X_j^{(i_j+m)}}
\ee
for $-1\le j<\ell$ and $m,n$ satisfying $i_j\le i_j+m\le i_j+n\le\ell'$.
In addition, the isomorphism $\varphi$ satisfies
\be
\label{thm14a2}
\varphi(X_j^{(i_j+\varepsilon_j+n)} Y_0/Y_0)=X_{j+1}^{(i_{j+1}+n)}/X_0
\ee
for $-1\le j<\ell$ and $n$ satisfying $i_j+\varepsilon_j\le i_j+\varepsilon_j+n\le\ell'$.
We have $\varphi(X_j^{(i_j)} Y_0/Y_0)=X_{j+1}^{(i_{j+1})}/X_0$
if $\varepsilon_j=1$ or $\varepsilon_j=0$, for $-1\le j<\ell$.
\end{thm}

\begin{prf}
Starting from the normal chain
$X_{\ell-1}=X_{\ell-1}^{(i_{\ell-1})}\lhd X_{\ell-1}^{(\ell')}=X_{\ell}$,
where $X_{\ell-1}\lhd G$ and $X_\ell\lhd G$, we can use Lemma \ref{lem8}
to go `backwards' and for each $j$, $-1\le j<\ell-1$, obtain a
normal chain from $X_j$ to $X_{j+1}$ as in (\ref{eqbb}),
where  each $X_j^{(i_j+n)}\lhd G$ for $n$ satisfying
$i_j\le i_j+n\le\ell'$, and $X_j^{(i_j+1)}=X_j^*$ if $\varepsilon_j=1$.

Since $i_{j+1}=i_j+\varepsilon_j$, we can restate (\ref{eqx5}) of Lemma \ref{lem8} as in
(\ref{thm14a2}), for $n$ satisfying $i_j+\varepsilon_j\le i_j+\varepsilon_j+n\le\ell'$.

It only remains to show (\ref{thm14a1}).  We can do this by induction.
We assume (\ref{thm14a1}) holds for $q+1$, that is, we assume
\be
\label{zeq40}
\frac{X_{q+1}^{(i_j+n)}}{X_{q+1}^{(i_j+m)}}\simeq\frac{X_j^{(i_j+n)}}{X_j^{(i_j+m)}}
\ee
for $q+1\le j<\ell$ and $m,n$ satisfying $i_j\le i_j+m\le i_j+n\le\ell'$.
Note that the left hand side of (\ref{zeq40}) is well defined since
$i_{q+1}\le i_j$ for $q+1\le j$.
Then we show (\ref{thm14a1}) holds for $q$, that is, we show
\be
\label{zeq41}
\frac{X_q^{(i_j+n)}}{X_q^{(i_j+m)}}\simeq\frac{X_j^{(i_j+n)}}{X_j^{(i_j+m)}}
\ee
for $q\le j<\ell$ and $m,n$ satisfying $i_j\le i_j+m\le i_j+n\le\ell'$.

Assume that $j$ satisfies $q+1\le j<\ell$ and $m,n$ satisfy
$i_j\le i_j+m\le i_j+n\le\ell'$.  Assume that (\ref{zeq40}) holds.
We can write the portion of the normal chain in (\ref{eqbb})
between $X_q$ and $X_{q+1}$ as
\begin{multline}
\label{zeq42minus}
X_q=
X_q^{(i_q)}\lhd X_q^{(i_q+1)}\lhd X_q^{(i_q+2)}\lhd\cdots  \\
\lhd X_q^{(\ell'-1)}\lhd X_q^{(\ell')}=X_{q+1},
\end{multline}
and between $X_{q+1}$ and $X_{q+2}$ as
\begin{multline}
\label{zeq42}
X_{q+1}=
X_{q+1}^{(i_{q+1})}\lhd X_{q+1}^{(i_{q+1}+1)}\lhd X_{q+1}^{(i_{q+1}+2)}\lhd\cdots  \\
\lhd X_{q+1}^{(\ell'-1)}\lhd X_{q+1}^{(\ell')}=X_{q+2}.
\end{multline}
Then using Lemma \ref{lem8} with (\ref{zeq42}) in place of (\ref{capa}) and
(\ref{zeq42minus}) in place of (\ref{capb}), we have from (\ref{eqx4})
\be
\label{zeq43}
\frac{X_q^{(i_j+n)}}{X_q^{(i_j+m)}}\simeq\frac{X_{q+1}^{(i_j+n)}}{X_{q+1}^{(i_j+m)}}.
\ee
Note that all terms in (\ref{zeq43}) are well defined since
$i_q\le i_{q+1}\le i_j$ for $q+1\le j$.
Combining (\ref{zeq43}) with (\ref{zeq40}) gives
\be
\label{zeq44}
\frac{X_q^{(i_j+n)}}{X_q^{(i_j+m)}}\simeq\frac{X_j^{(i_j+n)}}{X_j^{(i_j+m)}}.
\ee
We know that (\ref{zeq44}) holds
for $q+1\le j<\ell$ and $m,n$ satisfying $i_j\le i_j+m\le i_j+n\le\ell'$.
But (\ref{zeq44}) also holds trivially for $j=q$.  Then (\ref{zeq44}) holds
for $q\le j<\ell$ and $m,n$ satisfying $i_j\le i_j+m\le i_j+n\le\ell'$,
giving (\ref{zeq41}).

We start the induction by proving (\ref{zeq41}) for $q=\ell-2$.
But from Theorem \ref{thm2} or Lemma \ref{lem8}, we know there are
normal chains $X_{\ell-1}\lhd X_\ell$ and
$X_{\ell-2}\lhd X_{\ell-2}^*\lhd X_{\ell-1}$ with
$$
\frac{X_{\ell-1}}{X_{\ell-2}^*}\simeq\frac{X_\ell}{X_{\ell-1}}.
$$
Rewriting this as
$$
\frac{X_{\ell-2}^{(\ell')}}{X_{\ell-2}^{(i_{\ell-1})}}\simeq
\frac{X_{\ell-1}^{(\ell')}}{X_{\ell-1}^{(i_{\ell-1})}}
$$
gives (\ref{zeq41}) for $q=\ell-2$.
\end{prf}

We illustrate Theorem \ref{thm14a} in Figure \ref{fig1a}.  In the example
shown, we have $\varepsilon_{j+1}=0$ and $\varepsilon_j=1$.  Then if we let
$i_{j+2}=i+1$, we have $i_{j+1}=i+1$ and $i_j=i$.  For this
example then, we have $X_{j+1}^{(i+1)}=X_{j+1}^*=X_{j+1}$ and $X_j^{(i+1)}=X_j^*>>X_j$.
Note that the quotient groups formed by entries at the intersection of
each column of the same two rows are isomorphic.  For example,
$$
\frac{X_{-1}^{(\ell'-1)}}{X_{-1}^{(i+2)}}\simeq\frac{X_j^{(\ell'-1)}}{X_j^{(i+2)}}
\simeq\frac{X_{j+1}^{(\ell'-1)}}{X_{j+1}^{(i+2)}}
\simeq\frac{X_{j+2}^{(\ell'-1)}}{X_{j+2}^{(i+2)}}.
$$
Figure \ref{fig1a} is reminiscent of the shift register structure
used to realize \scgc s \cite{FT,LM}.

\begin{figure}[htbp]
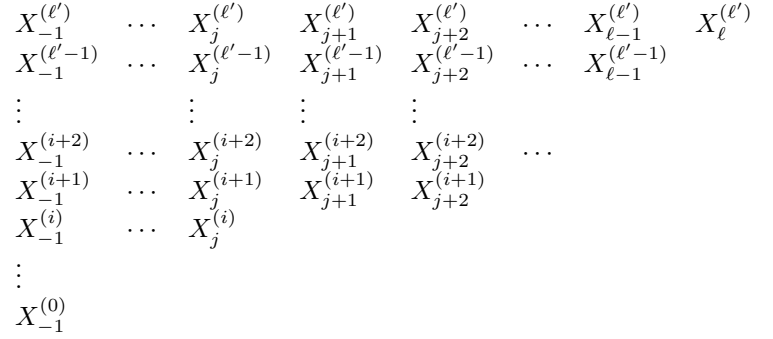

\centering
\vspace{3ex}

$
\begin{array}{llllllll}
X_{-1}^{(\ell')}    &  \cdots         &  X_j^{(\ell')}    & X_{j+1}^{(\ell')}  & X_{j+2}^{(\ell')}  & \cdots & X_{\ell-1}^{(\ell')}   & X_\ell^{(\ell')}  \\
X_{-1}^{(\ell'-1)}  &  \cdots         &  X_j^{(\ell'-1)}  & X_{j+1}^{(\ell'-1)}& X_{j+2}^{(\ell'-1)}& \cdots & X_{\ell-1}^{(\ell'-1)} &         \\
\vdots              &                 &  \vdots           & \vdots             & \vdots             &        &                        &         \\
X_{-1}^{(i+2)}      &  \cdots         &  X_j^{(i+2)}      & X_{j+1}^{(i+2)}    & X_{j+2}^{(i+2)}    & \cdots &                        &         \\
X_{-1}^{(i+1)}      &  \cdots         &  X_j^{(i+1)}      & X_{j+1}^{(i+1)}    & X_{j+2}^{(i+1)}    &        &                        &         \\
X_{-1}^{(i)}        &  \cdots         &  X_j^{(i)}        &                    &                    &        &                        &         \\
\vdots              &                 &                   &                    &                    &        &                        &         \\
X_{-1}^{(0)}        &                 &                   &                    &                    &        &                        &
\end{array}
$

\caption{Illustration of Theorem \ref{thm14a}.}
\label{fig1a}

\end{figure}

We are particularly interested in the portion of the normal chain from
$X_{-1}$ to $X_0$:
\begin{multline}
\label{nrml1}
X_{-1}=X_{-1}^{(i_{-1})}\lhd X_{-1}^{(i_{-1}+1)}\lhd\cdots\lhd X_{-1}^{(i_{-1}+n)}\lhd\cdots  \\
\lhd X_{-1}^{(\ell'-1)}\lhd X_{-1}^{(\ell')}=X_0.
\end{multline}
In (\ref{nrml1}),
the superscript $m$ of $X_{-1}^{(m)}$ takes all values in the interval
$[i_{-1},\ell']$ or $[0,\ell']$.  Using (\ref{thm14a1plus}), for $j$
satisfying $-1\le j\le\ell$, we know $i_j$ takes all values in the interval
$[0,\ell']$.  Then for $-1\le j\le\ell$, the term $X_{-1}^{(i_j)}$
appears in (\ref{nrml1}), and we can make the definition
$$
\Delta_j\rmdef X_{-1}^{(i_j)}.
$$
Then
%\begin{multline}
%\label{r0}
%X_{-1}=\Delta_{-1}\lhd \Delta_0\lhd\cdots\lhd\Delta_j\lhd  \\
%\cdots\lhd\Delta_{\ell-1}\lhd \Delta_\ell=X_0
%\end{multline}
\be
\label{r0}
X_{-1}=\Delta_{-1}\lhd \Delta_0\lhd\cdots\lhd\Delta_j\lhd
\cdots\lhd\Delta_{\ell-1}\lhd \Delta_\ell=X_0
\ee
is a refinement of (\ref{nrml1}) which at most just repeats terms in
(\ref{nrml1}).  Since each $X_{-1}^{(i_{-1}+n)}\lhd G$, we know that each
$\Delta_j\lhd G$.

Given the \sst\ $(\xj,Y_0,\varphi)$ of a shift group $G$,
the normal chain in (\ref{eqbb}) is uniquely determined, and so the normal
chains (\ref{nrml1}) and (\ref{r0}) are uniquely determined.  We
say the normal chain in (\ref{r0}) is a {\it signature chain} of shift group $G$.
Also, given the \sst\ of a shift group $G$,
we can form the intersection group $X_j\cap Y_0$ for each $j$, and
this gives the normal chain
\begin{multline}
\label{ncy0}
\bone=(X_{-1}\cap Y_0)\subset (X_0\cap Y_0)\subset (X_1\cap Y_0)\subset \cdots \\
(X_{\ell-1}\cap Y_0)\subset (X_\ell\cap Y_0)=Y_0,
\end{multline}
where each $X_j\cap Y_0 \lhd G$.
We say the normal chain in (\ref{ncy0}) is a {\it cosignature chain} of
shift group $G$.  The cosignature chain is also uniquely determined
by the \sst\ of a shift group.

We now give some properties of the signature and cosignature chain.

\begin{thm}
\label{thm3}
Let group $G$ have a shift structure $(\xj,Y_0,\varphi)$.
Fix $j$, $-1\le j<\ell$.
The signature chain has the property that
\be
\label{r1}
\frac{X_{j+1}}{X_j}\simeq\frac{X_0}{\Delta_j},
\ee
\be
\label{r2}
\frac{X_{j+1}}{X_j^*}\simeq\frac{X_0}{\Delta_{j+1}},
\ee
and
\be
\label{r3}
\frac{X_j^*}{X_j}\simeq\frac{\Delta_{j+1}}{\Delta_j}.
\ee
The cosignature chain has the property that
\be
\label{r1a}
\frac{X_j Y_0}{X_j}\simeq\frac{Y_0}{X_j\cap Y_0},
\ee
\be
\label{r2a}
\frac{X_{j+1} Y_0}{X_{j+1}}\simeq\frac{Y_0}{X_{j+1}\cap Y_0},
\ee
\be
\label{r3a}
\frac{X_j^*}{X_j}\simeq\frac{X_{j+1}\cap Y_0}{X_j\cap Y_0},
\ee
and
\be
\label{r4a}
\frac{X_j Y_0}{X_j^*}\simeq\frac{X_{j+1} Y_0}{X_{j+1}}\simeq\frac{Y_0}{X_{j+1}\cap Y_0},
\ee
where (\ref{r1a})-(\ref{r3a}) are analogous to (\ref{r1})-(\ref{r3}).
Lastly, we have
\be
\label{r4b}
\Delta_0=X_{-1}^*=X_0\cap Y_0,
\ee
\be
\label{r4}
\frac{\Delta_{j+1}}{\Delta_j}\simeq\frac{X_{j+1}\cap Y_0}{X_j\cap Y_0},
\ee
and
\be
\label{r5}
|\Delta_{j+1}|=|X_{j+1}\cap Y_0|.
\ee
\end{thm}

\begin{prf}
Results (\ref{r1})-(\ref{r3}) follow from (\ref{thm14a1}) of Theorem
\ref{thm14a} using the definition of $\Delta_j$.

Results (\ref{r1a}) and (\ref{r2a}) follow from the second isomorphism
theorem, and result (\ref{r3a}) follows from (\ref{fst}) of
Lemma \ref{lem17}.  But we know
$$
\frac{X_j Y_0}{X_j^*}=\frac{X_j^* Y_0}{X_j^*}\simeq\frac{Y_0}{X_j^*\cap Y_0}
=\frac{Y_0}{X_{j+1}\cap Y_0}\simeq\frac{X_{j+1} Y_0}{X_{j+1}},
$$
giving (\ref{r4a}).

We now show (\ref{r4b}).  We have $X_{-1}^{(i_{-1}+1)}=X_{-1}^*$
if $\varepsilon_{-1}=1$, and
$X_{-1}^{(i_{-1})}=X_{-1}^*=X_{-1}$ if $\varepsilon_{-1}=0$.
Also $i_0$ and $i_{-1}$ are related by $i_0=i_{-1}+\varepsilon_{-1}$.
Thus $X_{-1}^{(i_0)}=X_{-1}^*$ if $\varepsilon_{-1}=1$ or $\varepsilon_{-1}=0$.
But $\Delta_0=X_{-1}^{(i_0)}$ and $X_{-1}^*=X_0\cap Y_0$; then
(\ref{r4b}) follows.  We have (\ref{r4}) holds using
(\ref{r3}) and (\ref{r3a}).  Now use induction with
(\ref{r4b}) and (\ref{r4}) to obtain (\ref{r5}).
\end{prf}

{\it Remark:}  Note from (\ref{r4}) that if
$\Delta_{-1}=\cdots=\Delta_j=\bone$ and $\Delta_{j+1}\ne\bone$,
then $X_0\cap Y_0=\cdots=X_j\cap Y_0=\bone$ and
$\Delta_{j+1}\simeq X_{j+1}\cap Y_0$.
Since $X_{\ell-1}<< X_{\ell-1}^*$, we always have $|\Delta_{\ell-1}|<|\Delta_\ell|$.

\begin{cor}
\label{cor11a}
Assume $G$ is a shift group with shift structure $(\xj,Y_0,\varphi)$.
The factor groups of the signature chain $\{\Delta_j\}$
are isomorphic to the factor groups of the cosignature chain $\{X_j\cap Y_0\}$
in 1-1 order, i.e., as in (\ref{r4}).  The signature chain
$\{\Delta_j\}$ is a composition series of $X_0$
\ifof\ the cosignature chain $\{X_j\cap Y_0\}$ is a composition series of $Y_0$.
The signature chain $\{\Delta_j\}$ is a solvable series of $X_0$
(meaning factor groups are abelian)
\ifof\ the cosignature chain $\{X_j\cap Y_0\}$ is a solvable series of $Y_0$.
\end{cor}

\begin{prf}
We prove the second statement:  a normal series is a composition
series \ifof\ its factor groups are either simple or trivial
(cf. Problem 5.7 of \cite{ROT}).
\end{prf}

Loeliger and Mittelholzer give an example of a shift group in which
$X_0\simeq\cz_2\times\cz_2$ and $Y_0\simeq\cz_4$, with
$\Delta_0=X_0\cap Y_0=\cz_2$ (cf. Example 3.2 of
\cite{LM}).  Even though $X_0$ and $Y_0$ are not isomorphic, it
can be verified that the results in Corollary \ref{cor11a} hold.

We have the following easy corollary of Theorem \ref{thm3}.

\begin{cor}
\label{cor11}
If $G$ is a \gpss\ $(\xj,Y_0,\varphi)$, the factor groups $X_{j+1}/X_j$ in the
normal chain $\xj$ are abelian if $X_0$ is abelian.  In this case then,
$\xj$ is a solvable series and $G$ is solvable.
\end{cor}

We show the relevance of this corollary in the next section.

%%%%%%%%%%%%%%%%%%%%%
We now generalize Theorem \ref{thm14a} and Corollary \ref{cor11}.
In the next theorem, we find a refinement of (\ref{eqbb})
using Lemma \ref{lem8}.
As before, it is convenient to write the refinement using slightly
different notation than in Lemma \ref{lem8}.  Thus in place of (\ref{capa}),
we write the portion of the refinement between $X_{j+1}$ and $X_{j+2}$ as
\begin{multline}
\label{xreft2}
X_{j+1}=\hatcx_{j+1}^{(r_{j+1})}\lhd \hatcx_{j+1}^{(r_{j+1}+1)}
\lhd \hatcx_{j+1}^{(r_{j+1}+2)}\lhd\cdots  \\
\lhd \hatcx_{j+1}^{(\kappa-1)}\lhd \hatcx_{j+1}^{(\kappa)}=X_{j+2},
\end{multline}
where $r_{j+1}$ and $\kappa$ are positive integers.
Using (\ref{xreft2}) in Lemma \ref{lem8}, we obtain the portion of the
refinement between $X_j$ and $X_{j+1}$ as
\begin{multline}
\label{xreft1}
X_j\lhd Q_j^a\lhd\cdots\lhd Q_j^b\lhd\hatcx_j^{(r_{j+1})}
\lhd\hatcx_j^{(r_{j+1}+1)}\lhd \hatcx_j^{(r_{j+1}+2)}\lhd\cdots  \\
\lhd \hatcx_j^{(\kappa-1)}\lhd \hatcx_j^{(\kappa)}=X_{j+1},
\end{multline}
where $\hatcx_j^{(r_{j+1})}=X_j^*$.
In this case, we use Lemma \ref{lem8} for a nontrivial refinement in (\ref{specnc});
in fact we select
$$
X_j\lhd Q_j^a\lhd\cdots\lhd Q_j^b\lhd\hatcx_j^{(r_{j+1})}
$$
to be a composition chain of $X_j\lhd\hatcx_j^{(r_{j+1})}$.  In (\ref{xreft1}),
we have $\hatcx_j^{(r_{j+1})}=X_j^*$ if $\varepsilon_j=1$, and
$\hatcx_j^{(r_{j+1})}=X_j^*=X_j$ if $\varepsilon_j=0$.

In general for each $j$, $-1\le j\le\ell-1$, we define a refinement
in which the superscript
$m$ of $\hatcx_j^{(m)}$ runs from integer $r_j$ to integer $\kappa$.
For $0\le j\le\ell$, we define
$\hatcx_{j-1}^{(\kappa)}\rmdef X_j\rmdef \hatcx_j^{(r_j)}$;
then $\hatcx_{\ell-1}^{(\kappa)}=X_{\ell}=\hatcx_{\ell}^{(r_{\ell})}$.
We also define $X_{-1}\rmdef \hatcx_{-1}^{(r_{-1})}$.
In this notation,
the portion of the refinement between $X_j$ and $X_{j+1}$ is
\begin{multline}
\label{xreft3}
X_j=\hatcx_j^{(r_j)}\lhd \hatcx_j^{(r_j+1)}\lhd \hatcx_j^{(r_j+2)}\lhd\cdots  \\
\lhd \hatcx_j^{(r_j+\delta_j-1)}\lhd \hatcx_j^{(r_j+\delta_j)}\lhd\cdots  \\
\lhd \hatcx_j^{(\kappa-1)}\lhd \hatcx_j^{(\kappa)}=X_{j+1}.
\end{multline}
Comparing (\ref{xreft1}) and (\ref{xreft3}) shows that we must have
$X_j=\hatcx_j^{(r_j)}=\hatcx_j^{(r_{j+1})}=X_j^*$ if $\varepsilon_j=0$.
If $\varepsilon_j=1$, there is an integer parameter $\delta_j>0$ such that
$\hatcx_j^{(r_j+\delta_j)}=\hatcx_j^{(r_{j+1})}=X_j^*$.
This means $r_j+\delta_j=r_{j+1}$ if $\varepsilon_j=1$.  If $\varepsilon_j=0$,
so that $r_j=r_{j+1}$, we set $\delta_j\rmdef 0$.  If we use the above
procedure and apply Lemma \ref{lem8} recursively starting with
the normal chain
\begin{multline*}
X_{\ell-1}=\hatcx_{\ell-1}^{(r_{\ell-1})}\lhd \hatcx_{\ell-1}^{(r_{\ell-1}+1)}
\lhd\cdots\lhd \hatcx_{\ell-1}^{(r_{\ell-1}+\delta_{\ell-1}-1)}  \\
\lhd \hatcx_{\ell-1}^{(r_{\ell-1}+\delta_{\ell-1})}=\hatcx_{\ell-1}^{(\kappa)}=X_{\ell},
\end{multline*}
a composition chain of $X_{\ell-1}\lhd X_\ell$, we obtain
\be
\label{xreft4}
r_j=\kappa-\sum_{j\le i<\ell}\delta_i
\ee
for $-1\le j<\ell$.  Define
$$
\kappa\rmdef \sum_{-1\le i<\ell}\delta_i.
$$
Then from (\ref{xreft4}) we see $r_{-1}=0$.  If $j=\ell$,
we define $r_j=r_\ell\rmdef\kappa$ trivially.
Thus as $j$ runs from $-1$ to $\ell$, $r_j$ takes values in
the range $[0,\kappa]$.

%%%%%%%%%%%%%%%%%%%%%%%%%%%%%%%%%%%%%%

%%%%%%%%%%%%%%%%%%%%%%%%%%%

\begin{thm}
\label{thm15}
Let a group $G$ have a shift structure $(\xj,Y_0,\varphi)$.
There is a refinement of $\xj$, and of the normal chain $\{X_j^{(i_j+n')}\}$
in (\ref{eqbb}), given by
\begin{multline}
\label{eqbc}
X_{-1}=\hatcx_{-1}^{(r_{-1})}\lhd\cdots\lhd \hatcx_{-1}^{(\kappa)}=X_0=\hatcx_0^{(r_0)}\lhd\cdots  \\
\lhd \hatcx_{j-1}^{(\kappa)}=X_j=\hatcx_j^{(r_j)}\lhd \hatcx_j^{(r_j+1)}\lhd \hatcx_j^{(r_j+2)}\lhd\cdots  \\
\lhd \hatcx_j^{(\kappa-1)}\lhd \hatcx_j^{(\kappa)}=X_{j+1}=\hatcx_{j+1}^{(r_{j+1})}\lhd\cdots  \\
\lhd \hatcx_{\ell-1}^{(r_{\ell-1})}\lhd \hatcx_{\ell-1}^{(r_{\ell-1}+1)}=\hatcx_{\ell-1}^{(\kappa)}=X_{\ell}=\hatcx_{\ell}^{(r_{\ell})},
\end{multline}
where $\hatcx_j^{(r_j+\delta_j)}=X_j^*$ if $\varepsilon_j=1$.  The normal chain
(\ref{eqbc}) is a composition series of $G$.  Moreover
$$
\frac{\hatcx_{-1}^{(r_j+n+1)}}{\hatcx_{-1}^{(r_j+n)}}\simeq\frac{\hatcx_j^{(r_j+n+1)}}{\hatcx_j^{(r_j+n)}}
$$
for $-1\le j<\ell$ and $n$ satisfying $r_j\le r_j+n<\kappa$.
In addition, the isomorphism $\varphi$ satisfies
\be
%label{thm14a2}
\varphi(\hatcx_j^{(r_j+\delta_j+n)} Y_0/Y_0)=\hatcx_{j+1}^{(r_{j+1}+n)}/X_0
\ee
for $-1\le j<\ell$ and $n$ satisfying $r_j+\delta_j\le r_j+\delta_j+n\le\kappa$.
We have
$$
\varphi(\hatcx_j^{(r_j+n)} Y_0/Y_0)=\hatcx_{j+1}^{(r_{j+1})}/X_0
$$
for $-1\le j<\ell$ and $n=0,\ldots,\delta_j-1$.  The term
$X_j^{(i_j+n')}$ in (\ref{eqbb}), $n'=0,\ldots,\ell'-i_j$, is the term
$\hatcx_j^{(r_j+n)}$ in the refinement (\ref{eqbc}),
where $n=\sum_{j\le i<j+n'} \delta_i$.
\end{thm}

\begin{prf}
The proof is similar to the proof of Theorem \ref{thm14a}.
\end{prf}

\begin{cor}
\label{cor15}
Let a group $G$ have a \sst\ $(\xj,Y_0,\varphi)$.  Then $G$ is solvable
\ifof\ $X_0$ is solvable.
\end{cor}

\begin{prf}
If $G$ is solvable, then every subgroup is solvable, so $X_0$ is
solvable.  For the converse result, note that
we can construct a figure like Figure \ref{fig1a}.
Going backwards, first find a normal chain from $X_{\ell-1}$ to $X_\ell$
for which factor groups are simple.  By Lemma \ref{lem8}, there is a chain
from $X_{\ell-2}^*$ to $X_{\ell-1}$ with the same factor groups.  Now find
a chain from $X_{\ell-2}$ to $X_{\ell-2}^*$ for which factor groups are
simple.  This gives a chain from $X_{\ell-2}$ to $X_{\ell-1}$ with simple
factor groups.  Continue in this way to $X_0$.  Then there is a chain
from $X_{-1}^*$ to $X_0$ for which factor groups are simple.  Now find
a chain from $X_{-1}$ to $X_{-1}^*$ for which factor groups are
simple.  This gives a chain for $X_0$ in which all factor groups
are simple, i.e., this is a composition chain of $X_0$.  But if $X_0$
is solvable, then this composition chain must have primary cyclic factor
groups.  Going in reverse, this implies that factor groups of chain from
$X_j$ to $X_{j+1}$ are primary cyclic, for $0\le j<\ell$.  This implies
$G$ is solvable.
\end{prf}

Since $G=X_{\ell-1}Y_0$ has normal subgroup $X_0Y_0$, we can regard
$G$ as like a wreath product with base group $X_0Y_0$.

We illustrate some of the results in this section in Figure \ref{fig2}.
The group $G=X_{\ell-1}Y_0$ is composed of cosets of $X_0Y_0$, and also
cosets of $X_0$ and cosets of $Y_0$.  For $j=0,\ldots\ell-1$, the normal
subgroup $X_jY_0$ is composed of cosets of $X_0Y_0$, and also
cosets of $X_0$ and $Y_0$.  In Figure \ref{fig2}, we draw $G$ and $X_jY_0$
as a group of cosets of $Y_0$, with $Y_0$ laid along the vertical axis.
We have $|X_jY_0|=|X_j||Y_0|/|X_j\cap Y_0|$.  Thus in Figure \ref{fig2},
$X_jY_0$ has a `height' of $|Y_0|$ and a `width' of $|X_j|/|X_j\cap Y_0|$.
Note that we have
\begin{align*}
|X_j| &=|X_0|\prod_{k=1}^j \frac{|X_k|}{|X_{k-1}|} \\
      &=\frac{|X_0|^{j+1}}{|\Delta_{j-1}|\cdots |\Delta_0|}.
\end{align*}
Thus the signature chain or cosignature chain determines $|X_j|$ and
$|X_jY_0|$.  Using Figure \ref{fig2}, it is easy to visualize
many of the results
in Theorems \ref{thm2} and \ref{thm3}.  The following
is clear from the structure of $G$ and $\calg_G$ (see also \cite{FT,LM}).

\begin{prop}
\label{square}
A coset of $X_0$ and a coset of $Y_0$ are disjoint unless they are in the
the same coset of $X_0Y_0$, in which case they have $|X_0\cap Y_0|$
elements in common.
\end{prop}

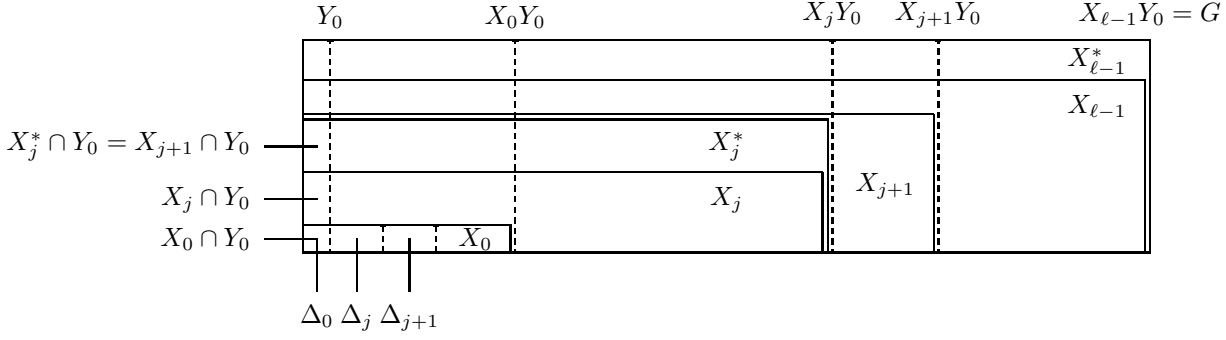
\begin{figure*}[htbp]    %include this eventually
%\begin{figure*}[p]   %drop this eventually
\centering
\vspace{3ex}

\begin{picture}(320,105)(-20,-25)

%boxes
\put(0,0){\framebox(78,10)}
\put(0,0){\framebox(196,30)}
\put(0,0){\framebox(198,50)}
\put(0,0){\framebox(238,52)}
\put(0,0){\framebox(318,65)}
\put(0,0){\framebox(320,80)}

%lines for X_jY_0
\put(10,0){\dashbox{2.0}(0,80)}
\put(80,0){\dashbox{2.0}(0,80)}
\put(200,0){\dashbox{2.0}(0,80)}
\put(240,0){\dashbox{2.0}(0,80)}

%dashed lines for X_0
\put(30,0){\dashbox{2.0}(0,10)}
\put(50,0){\dashbox{2.0}(0,10)}

%labels for X_jY_0
\put(10,85){\makebox(0,0)[b]{$Y_0$}}
\put(80,85){\makebox(0,0)[b]{$X_0 Y_0$}}
\put(200,85){\makebox(0,0)[b]{$X_j Y_0$}}
\put(240,85){\makebox(0,0)[b]{$X_{j+1} Y_0$}}
\put(320,85){\makebox(0,0)[b]{$X_{\ell-1} Y_0=G$}}

%lines and captions on left
\put(-15,5){\line(1,0){20}}
\put(-20,5){\makebox(0,0)[r]{$X_0\cap Y_0$}}
\put(-15,20){\line(1,0){20}}
\put(-20,20){\makebox(0,0)[r]{$X_j\cap Y_0$}}
\put(-15,40){\line(1,0){20}}
\put(-20,40){\makebox(0,0)[r]{$X_j^*\cap Y_0=X_{j+1}\cap Y_0$}}

%labels inside boxes
\put(65,5){\makebox(0,0){$X_0$}}
\put(160,20){\makebox(0,0){$X_j$}}
\put(160,40){\makebox(0,0){$X_j^*$}}
\put(220,25){\makebox(0,0){$X_{j+1}$}}
\put(300,55){\makebox(0,0){$X_{\ell-1}$}}
\put(300,72){\makebox(0,0){$X_{\ell-1}^*$}}

%lines and captions on bottom
\put(5,-15){\line(0,1){20}}
\put(5,-20){\makebox(0,0)[t]{$\Delta_0$}}
\put(20,-15){\line(0,1){20}}
\put(20,-20){\makebox(0,0)[t]{$\Delta_j$}}
\put(40,-15){\line(0,1){20}}
\put(40,-20){\makebox(0,0)[t]{$\Delta_{j+1}$}}

\end{picture}

\caption{Diagram of shift group $G$ with shift structure $(\xj,Y_0,\varphi)$.}
\label{fig2}

\end{figure*}

%FIGURE 2 IS HERE

\section{Group codes}
\label{sec3}

Trott and Sarvis speculated there might be a connection between a \htc,
Latin square, and translation net \cite{TS,JPS}.  In this section,
we show such a connection for a group code, the most important
example of a \htc.

Let $\calg$ be any graph.  We define a {\it labeled graph}
$(\calg,\call)$ as a graph $\calg$ and a mapping $\call:  \cale\ra A$
where $A$ is an {\it alphabet}.  Let $B$ be a group,
and let $\calg_B$ be a graph constructed
as in Section \ref{sec2} using $\cale=B$ and $\calv=B/B^+$,
where $B^+\lhd B$.  We define a {\it group code} as a labeled graph
$(\calg_B,\omega)$ where $\omega$ is a homomorphism $\omega:  \cale\ra A$
and alphabet $A$ is a group; this is essentially the definition used in \cite{LM}.
We say the group code is \ellctl\ if graph $\calg_B$
is \ellctl.  In particular, here we consider a group $G$ with a \sst\
$(\xj,Y_0,\varphi)$.  Then graph $\calg_G$, formed using $\cale=G$
and $\calv=G/X_0$, is \ellctl\ and group code $(\calg_G,\omega)$ is \ellctl.
We only consider the case $|X_0\cap Y_0|=1$
where there are no multiple edges.  If $|X_0\cap Y_0|>1$, the
discussion below can be applied to quotient group $G/X_0\cap Y_0$.

Since $X_0\cap Y_0=\bone$, and $X_0\lhd G$, $Y_0\lhd G$, we have
$X_0Y_0\simeq X_0\times Y_0$.  From Definition 1, we
know that $|X_0|=|Y_0|$.  Thus it is natural to think of
$X_0Y_0$ as a {\it square} whose rows are $\{gX_0|g\in X_0Y_0\}$
and columns are $\{gY_0|g\in X_0Y_0\}$.  The elements of row
$gX_0$ are edges that split from state $gX_0$ in $G/X_0$.
The elements of column $gY_0$ are edges that merge to state
$\varphi(gY_0)$ in $G/X_0$.
In $G/X_0Y_0$, we can think of coset $hX_0Y_0$ as a square.
The rows of the square are $\{gX_0|g\in hX_0Y_0\}$; elements in
$gX_0$ split from state $gX_0$.  The columns of the square are
$\{gY_0|g\in hX_0Y_0\}$; elements in $gY_0$ merge to state
$\varphi(gY_0)$.  Proposition \ref{square} shows that
a row and column do not intersect unless they are from the
same square, in which case they intersect once.  If we regard
$\calg_G$ as a trellis section, such squares are often called
subtrellises \cite{TS}.

Suppose we can form a group code in which all squares can be
labeled so they are {\it Latin squares}.  In this case, the edges that
split from any state all have different labels, and the edges
that merge to any state all have different labels.  This type
of labeling is useful in practical trellis codes
\cite{UNG,FY2,MT,RSH,TS}.
We call such a group code a {\it Latin group code}.  Again it is
to be understood that the term \lgc\ means the \ls s are formed
using squares defined as above.  Also it is understood the shift
group $G$ of a \lgc\ has $|X_0\cap Y_0|=1$.

In a \lgc, since $\omega$ is a homomorphism $\omega:  G\ra A$, we
must have the assignment $\omega:  X_0Y_0\mapsto A_0$, where
$A_0\lhd A$ and $|A_0|=|X_0|$.  Given a coset $gA_0$ of $A_0$, all
squares in $\omega^{-1}(gA_0)$ have the same labels, and we call
this collection of Latin squares a {\it Latin clique}.
Assume there are $q$ Latin cliques in $\calg$,
called $C_0,\ldots C_{q-1}$; then $|A|/|A_0|=q$.  We define
$\omega^{-1}(A_0)\rmdef G_0$; this gives
$G_0\lhd G$.  We assume that $G_0$ is the stabilizer of
squares in Latin clique $C_0$.  Let $\tta$ be the identity of $A$.
Let $\ttga$ be the kernel of $\omega$, or $\ttga=\omega^{-1}(\tta)$.
Then $\ttga\lhd G$, and $\omega$ is essentially the natural map with
kernel $\ttga$, or essentially $\omega':  G\ra G/\ttga$
with $G/\ttga\simeq A$.  Without loss of generality, we can assume
that label $\tta$ is used in Latin clique $C_0$.  Then $\ttga\subset G_0$.

Let $\Box_0$ be the \ls\ of square $X_0Y_0$; assume $\Box_0\subset C_0$.
We can think of $\Box_0$ as a finite geometry $F$ with three
parallel classes of lines.  The first class are the rows
$\{gX_0|g\in X_0Y_0\}$ of $X_0Y_0$; the second class are the columns
$\{gY_0|g\in X_0Y_0\}$ of $X_0Y_0$.  The third parallel class consists of
lines formed by entries in $\Box_0$ with the same label.  For example,
line $\ttla$ consists of all entries in $\Box_0$ with label $\tta$.
Without loss of generality, we can assume that line $\ttla$ includes
the identity entry, i.e., the label of $\bone$ is $\tta$.
Note that lines from the same class do not intersect, and using
Proposition \ref{square}, lines from different classes intersect
exactly once.  Thus $\Box_0$ is a $(t,s)$ net for $s=3$, where $t=|X_0|$.

We define an action of $G$ on itself by the product $gG$ for each $g\in G$.
In this sense, $X_0Y_0$ acts transitively, in fact regularly,
on the entries in square $X_0Y_0$.  In fact, because
there is a homomorphism $\omega:  G\ra A$, $X_0Y_0$ must also be a \tg\
of \ls\ $\Box_0$, and so $\Box_0$ must be a \tstn.

From the theory of Latin squares \cite{DJ}, a finite geometry $F$
that is a $(t,3)$ net is a \tstn\
\ifof\ $F$ has a \tg\ $Q$ which has a
{\it partial congruence partition} (PCP):  three subgroups
$W_1$, $W_2$, $W_3$ such that $W_i\cap W_j=\bone$ and $W_iW_j=Q$
for $1\le i,j\le 3$, $i\ne j$.  In this case, $W_i$ acts regularly
on lines in the $i^{\rm th}$ parallel class of the $(t,3)$ net,
$1\le i\le 3$.  In general $F$ may have more than
one \tg, and a given \tg\ $Q$ may have more than one PCP \cite{APS}.

We already know that $\Box_0$ has \tg\ $X_0Y_0$.  But any PCP
in $X_0Y_0$ must have $W_1=X_0$ and $W_2=Y_0$ because the only
subgroup of $G$ which acts regularly on a row of $\Box_0$ is $X_0$,
and the only subgroup which acts regularly on a column of $\Box_0$
is $Y_0$.  Thus $\Box_0$ can be a \tstn\
\ifof\ there is some subgroup $W_3\subset X_0Y_0$ which forms a
PCP with $W_1=X_0$ and $W_2=Y_0$.  But $W_3$ must necessarily be
the stabilizer of line $\ttla$, or
$\ttga\cap X_0Y_0\rmdef \ttka$.  Thus $\Box_0$ is
a \tstn\ \ifof\ $X_0\cap \ttka=Y_0\cap \ttka=\bone$ and
$X_0\ttka=Y_0\ttka=X_0Y_0$.

We now digress briefly to discuss the work of Sprague \cite{APS},
Mann \cite{MN1}, and Bailey and Jungnickel \cite{BJ} (see also \cite{DJ}).

\begin{thm}[Sprague]
\label{sp1}
Let $\mathbf{W}=\{W_1,\ldots W_s\}$ be a $(t,s)$ PCP in $Q$.  Then the
following assertions hold:

(1) If $W_1$ is a normal subgroup of $Q$, then $W_2 \simeq \cdots\simeq W_s$.

(2) If $W_1$ and $W_2$ are normal subgroups of $Q$, then one has
$Q\simeq W_1\times W_2$ and $W_1 \simeq W_2 \simeq \cdots\simeq W_s$.

(3) If $\mathbf{W}$ has 3 normal components, then $Q$ is abelian.
\end{thm}

Given a group $H$ and an automorphism $\theta$
of $H$, we can construct a \ls\ {\it based on $H$}.  The point set is
$H\times H$; the rows are $\{(h,1)| h\in H\}$;
the columns are $\{(1,h)| h\in H\}$; and the letters are
the sets $\{(h_1,h_2)| h_1 (\theta(h_2))=k\}$ for elements $k$
of $H$.  We call this the \ls\ based on $H$ constructed by
the {\it automorphism method} of Mann \cite{MN1}.  Define
a set $\Sigma$ of automorphisms of $H$ to be {\it fixed point free}
if $\theta \sigma^{-1}$ is fixed point free for every distinct
pair of elements $\theta, \sigma$ of $\Sigma$.

\begin{thm}[Bailey and Jungnickel]
\label{sp2}
Let $H$ be a group of order $t$, and let $\Sigma$ be a fixed point free
set of $s'$ automorphisms of $H$.
Put $Q=H\times H$.  For $\theta$ in $\Sigma$, put
$W_\theta=\{(h,\theta(h))| h\in H\}$, and put $W_0=\bone\times H$ and
$W_\infty=H\times\bone$.  Then $\{W_0,W_\infty\}\cup\{W_\theta| \theta\in\Sigma\}$
is a $(t,s'+2)$ PCP for $Q$ with normal components $W_0$ and $W_\infty$.
Conversely, every $(t,s'+2)$ PCP with two normal components may be represented
in this way.
\end{thm}

This theorem shows that a fixed point free set of $s'$ automorphisms of $H$
gives rise to a set of $s'$ \mols\ based on $H$.  When
$H$ is elementary abelian, this method gives complete sets of
\mols\ based on $H$, that is, $s'=t-1$ \cite{BJ}.

We now use these results in our discussion.
We know something more about the \tg\ of $X_0Y_0$.  We have
$X_0\lhd X_0Y_0$ and $Y_0\lhd X_0Y_0$.  Then from Theorem \ref{sp1},
we must have $X_0\simeq Y_0\simeq\ttka$.  In fact from Theorem \ref{sp2},
$\ttka$ can be explicitly determined as
\be
\label{rte66}
\ttka=\{x(\mu\circ\theta(x)) | x\in X_0,{\rm \ }\mu\circ\theta:  X_0\ra Y_0\},
\ee
where $\theta$ is an automorphism of $X_0$ and $\mu$ is an isomorphism
from $X_0$ to $Y_0$, $X_0\stackrel{\mu}{\simeq}Y_0$.
Thus each distinct composition $\mu\circ\theta:  X_0\ra Y_0$ gives
a different $\ttka$.  Thus $\Box_0$ can be a \tstn\ \ifof\ there
is an isomorphism $X_0\simeq Y_0$.

Further, since $\ttka=\ttga\cap X_0Y_0$, then $\ttka\lhd G$ and
$\ttka\lhd X_0Y_0$.  Then we know from Theorem \ref{sp1} that $X_0Y_0$ must
be abelian, and since $X_0Y_0\simeq X_0\times Y_0$, both $X_0$ and
$Y_0$ must be abelian.  Note that the possible isomorphisms
$X_0\simeq Y_0$ are well known when $X_0$ is abelian \cite{ROT}.

\begin{thm}
The shift group $G$ of an \lclgc\ $(\calg_G,\omega)$ has $X_0\cap Y_0=\bone$,
$X_0\simeq Y_0$, and $X_0,Y_0$ abelian.
\end{thm}

\begin{cor}
The shift group $G$ of an \lclgc\ $(\calg_G,\omega)$
is a solvable group and $\xj$ is a solvable series.
\end{cor}

\begin{prf}
Use Corollary \ref{cor11}.
\end{prf}

\begin{thm}
\label{thm14}
The Latin squares $\Box_0$ which can appear in an
\lclgc\ $(\calg_G,\omega)$ are exactly those based on $X_0$
constructed by the automorphism method of Mann, where $X_0$ is abelian.
\end{thm}

\begin{prf}
The construction in (\ref{rte66}) gives Latin squares constructed
by the automorphism method of Mann \cite{BJ}.
\end{prf}

The Sarvis conjecture is that each fully connected subtrellis of a homogeneous
Latin trellis corresponds to a principal isotope of a group Latin square
\cite{JPS}; this is equivalent to the conjecture that $\Box_0$ is the
principal isotope of a group Latin square \cite{TS}.
Theorem \ref{thm14} shows the Sarvis conjecture is true
for \ellctl\ Latin group codes because every \ls\ $\Box_0$ constructed
by the automorphism method is isotopic to a group table (it is a
rearrangement of the columns of a group table).
Using the above approach, we can show the Sarvis conjecture is true
for an \ellctl\ \hlt\ as well.

For a group code used to convey binary information, a
{\it bit-oriented group code}, $|X_0|$ must be
some power of 2 because the input information stream is binary.

\begin{thm}
In an \ellctl\ bit-oriented \lgc,
$X_0$ is an abelian $p$-group and $G$ is a $p$-group, $p=2$.
\end{thm}

\begin{prf}
From Theorem \ref{thm3}, we have
$$
\frac{|X_{j+1}|}{|X_j|}=\frac{|X_0|}{|\Delta_j|}.
$$
But $|X_0|$ is a power of 2 and so any subgroup $\Delta_j$ of $X_0$ must
have order a power of 2.  Thus $|X_0|/|\Delta_j|$ is a power of 2, and so
$$
|G|=|X_0|\prod_{j=1}^{\ell-1}\frac{|X_{j+1}|}{|X_j|}
$$
must be a power of 2.
\end{prf}

Trott and Sarvis have observed that $\Box_0$ of all published
\htc s is the group table of $\cz^2\times\cz^2\times\cdots\times\cz^2$
\cite{TS}.  The theorem above indicates that practical (bit-oriented)
\lgc s might be constructed for which this is not true, but
that indeed $\Box_0$ is based on an abelian 2-group.

We say shift group $G$ is a {\it Latin shift group} if it has
a shift structure $(\xj,Y_0,\varphi)$ with $X_0\cap Y_0=\bone$,
$X_0\simeq Y_0$, and $X_0,Y_0$ abelian.

The previous results show some similarities of the mathematical structure
of a \ls\ and \lsg.  We now show a more direct analogy.
Recall that we have shown the following relations for \ls\ $\Box_0$.

\begin{prop}
\label{propx}
The $(t,3)$ net $\Box_0$ has \tg\ $K_0=X_0Y_0$ which is a $(t,3)$ PCP
with the following properties:

(1) $X_0$, $Y_0$, and $\ttka$ are disjoint.

(2) $K_0=X_0Y_0=X_0\ttka=Y_0\ttka$.

(3) $X_0\lhd K_0$, $Y_0\lhd K_0$, $\ttka\lhd K_0$.

(4) $K_0\simeq X_0\times Y_0$, $K_0\simeq X_0\times\ttka$, $K_0\simeq Y_0\times\ttka$.

(5) $X_0\simeq Y_0\simeq \ttka$.
\end{prop}

We now show that similar properties hold for Latin clique $C_0$.
A {\it partial net} is a generalization of a net in which lines from
different classes need not intersect \cite{BS}.  Latin clique $C_0$ is a partial
net with three parallel classes of lines.  The first (second) parallel
class of lines are the rows (columns) of \ls s that comprise $C_0$.  Thus
lines in the first parallel class are the
rows $\{gX_0|g\in G_0\}$, and lines in the second parallel class are the
columns $\{gY_0|g\in G_0\}$.  Note that a row and column
do not intersect unless they are from the same square, in which
case they intersect once.  The third parallel class consists of lines
formed by entries in all squares having the same label.  For example,
line $\ttcla$ consists of all entries with label $\tta$;
of course $\ttla\subset\ttcla$.
Note that a line in the third parallel class intersects each row
and each column exactly once.  Since $\ttga$ is the stabilizer of
$\ttcla$, this means $\ttga\cap X_0=\ttga\cap Y_0=\bone$.  Note
that each row and column has $|X_0|=|Y_0|$ points, and each line
in the third parallel class has $|G_0|/|X_0|$ points.  Then
$|G_0|=|\ttga||X_0|$, giving $G_0\simeq X_0\times\ttga$.
This gives the following result.

\begin{prop}
\label{propy}
The partial net $C_0$ has \tg\ $G_0$
with the following properties:

(1) $X_0$, $Y_0$, and $\ttga$ are disjoint.

(2) $G_0=X_0\ttga=Y_0\ttga$.

(3) $X_0\lhd G_0$, $Y_0\lhd G_0$, $\ttga\lhd G_0$.

(4) $G_0\simeq X_0\times\ttga$, $G_0\simeq Y_0\times\ttga$.

Note we also have $G_0/(X_0Y_0)\simeq\ttga/\ttka$.
\end{prop}

Comparing Proposition \ref{propx} and Proposition \ref{propy}, we see
that (1)-(4) of Proposition \ref{propy} correspond to (1)-(4) of
Proposition \ref{propx}.  Thus we see the mathematical structure of
Latin clique $C_0$ is analogous to the mathematical structure of
\ls\ $\Box_0$.  Also note that from (4) of Proposition \ref{propy},
we can obtain $G_0/X_0\simeq\ttga$ and $G_0/Y_0\simeq\ttga$,
or just $G_0/X_0\simeq G_0/Y_0$, which is
the isomorphism constructed by Sarvis and Trott in their algorithm \cite{ST}.

The shift group $G$ is itself the \tg\ of a partial
net with three parallel classes of lines.  The first (second) parallel
class of lines are the rows (columns) of \ls s that comprise $\calg_G$.  Thus
lines in the first parallel class are the
rows $\{gX_0|g\in G\}$, and lines in the second parallel class are the
columns $\{gY_0|g\in G\}$.  The third parallel class consists of lines
in each square formed by entries having the same label; line $\ttla$
is an example.  Note that lines in different classes intersect
exactly once if they are from the same square, and otherwise do not
intersect.  This means that any collection of lines in the third
parallel class, with exactly one line from each square, forms a
right transveral of $G/X_0$ and $G/Y_0$.

\begin{thm}
%\label{thm20}
In a \lsg\ $G$, there is a set $G_v$ of $G$
which is a right transveral of $G/X_0$ and $G/Y_0$,
where $G_v\supset\ttga\supset\ttka$.
\end{thm}

\begin{prop}
\label{propz}
The graph $\calg_G$ of an \lclgc\ $(\calg_G,\omega)$
has \tg\ $G$ with the following properties:

(1) $X_0$, $Y_0$, and $G_v$ are disjoint.

(2) $G=X_0 G_v=Y_0 G_v$.

(3) $X_0\lhd G$, $Y_0\lhd G$, $\ttka\subset\ttga\subset G_v\subset G$.

(4) $G_v$ is a right transversal of $G/X_0$ and $G/Y_0$.

Note we also have $\ttka\lhd G$ and $\ttga\lhd G$.
\end{prop}

We see that (1)-(4) of Proposition \ref{propz} correspond to (1)-(4) of
Proposition \ref{propy}.  Taken together,
Propositions \ref{propx}, \ref{propy}, and \ref{propz} show that
the \lgc\ has a mathematical structure similar to the \ls.
In this sense, we can say that the \lgc\ is a natural generalization
of a \ls\ to a sequence space.

As previously mentioned, when $X_0$ is elementary abelian,
a complete set of $|X_0|-1$ \mols\ based on $X_0$
can be constructed.  In this case then, we can construct a
{\it mutually orthogonal Latin group code}
in which \ls\ $\Box_0$ is replaced by $|X_0|-1$ \mols, a translation plane.

\section{The subdirect product group and state group}
\label{sec4}

\newcommand{\tldcg}{{\tilde{G}}}
\newcommand{\tldg}{{\tilde{g}}}
\newcommand{\tldch}{{\tilde{H}}}
\newcommand{\tldcx}{{\tilde{X}}}
\newcommand{\tldcs}{{\tilde{S}}}
\newcommand{\tldcy}{{\tilde{Y}}}
\newcommand{\tldcu}{{\tilde{U}}}
\newcommand{\tldcv}{{\tilde{V}}}
\newcommand{\tldxj}{{\{\tldcx_j\}}}
\newcommand{\uj}{{\{U_j\}}}
\newcommand{\tlduj}{{\{\tldcu_j\}}}
\newcommand{\tldvphi}{{\tilde{\varphi}}}
\newcommand{\tldphi}{{\tilde{\phi}}}
\newcommand{\tldclambda}{{\tilde{\Lambda}}}
\newcommand{\tldcgamma}{{\tilde{\Gamma}}}
\newcommand{\tldcdelta}{{\tilde{\Delta}}}
\newcommand{\hatclambda}{{\hat{\Lambda}}}
\newcommand{\hatcy}{{\hat{Y}}}
\newcommand{\hatcg}{{\hat{G}}}
\newcommand{\checkcg}{{\check{G}}}
%\hatphi,\hatcv,\hatch,\hatcgam,\hatbeta
\newcommand{\hatphi}{{\hat{\phi}}}
\newcommand{\hatcv}{{\hat{V}}}
\newcommand{\hatch}{{\hat{H}}}
\newcommand{\hatcgamma}{{\hat{\Gamma}}}
\newcommand{\hatcu}{{\hat{U}}}
\newcommand{\hatuj}{{\{\hatcu_j\}}}
\newcommand{\hatbeta}{{\hat{\beta}}}
\newcommand{\dotphi}{{\dot{\phi}}}
\newcommand{\dotcv}{{\dot{V}}}
\newcommand{\dotch}{{\dot{H}}}
\newcommand{\ddotch}{{\ddot{H}}}
\newcommand{\dotcgamma}{{\dot{\Gamma}}}
\newcommand{\dotcu}{{\dot{U}}}
\newcommand{\dotuj}{{\{\dotcu_j\}}}
\newcommand{\dotbeta}{{\dot{\beta}}}
\newcommand{\doteta}{{\dot{\eta}}}
\newcommand{\dotalpha}{{\dot{\alpha}}}

In this section, we assume group $G$ has a shift structure
$(\xj,Y_0,\varphi)$.  Then $G$ has a normal chain $\xj$ with
$X_\ell=G$ and each $X_j\lhd G$,
a normal subgroup $Y_0$, and an isomorphism $\varphi$ from $G/Y_0$
onto $G/X_0$ such that
\be
\label{email1}
\varphi(X_j Y_0/Y_0)=X_{j+1}/X_0
\ee
for $-1\le j<\ell$.  Define
$$
G_X\rmdef G/Y_0
$$
and
$$
G_Y\rmdef G/X_0.
$$
Defined in this manner, $G_X$ increments along the horizontal axis
in Figure \ref{fig2}, and $G_Y$ increments along the vertical axis.
Groups $G_X$ and $G_Y$ are called {\it state groups}
of shift group $G$.  Define
$$
G_X^j\rmdef \frac{X_jY_0}{Y_0}
$$
for $-1\le j\le\ell$, and
$$
G_Y^j\rmdef X_j/X_0,
$$
for $0\le j\le\ell$.  We see that $G_X^{-1}=Y_0/Y_0=\bone$,
$G_Y^0=X_0/X_0=\bone$, $G_X^{\ell-1}=G_X^\ell=G_X$, and $G_Y^\ell=G_Y$.
Note that $G_X^j\lhd G_X$ for $-1\le j\le\ell$, and
$G_Y^j\lhd G_Y$ for $0\le j\le\ell$.  With these definitions,
we can rewrite (\ref{email1}) as
\be
\label{eq49}
\varphi(G_X^j)=G_Y^{j+1}
\ee
for $-1\le j<\ell$; we can rewrite the isomorphism $\varphi:  G/Y_0\ra G/X_0$
as $\varphi:  G_X\ra G_Y$ or $\varphi(G_X)=G_Y$.

We can think of graph $\calg_G$ as essentially a bipartite graph $\calg_\ell$
with input states $G_Y$ and output states $G_X$.
An element $g\in X_\ell$ splits from input state $gX_0$ and merges to
output state $gY_0$.  In addition, there is an isomorphism
$\varphi:  G_X\ra G_Y$ from output states to input states.
In graph $\calg_\ell$ all the output states
are connected to input states via the isomorphism $\varphi(G_X)=G_Y$.

In the same manner, we can associate a bipartite graph $\calg_j$
with $X_j$, for $0\le j<\ell$.  An element $g\in X_j$ splits from input state $gX_0$
and merges to output state $gY_0$.  Then it is clear that the input
states of $\calg_j$ are cosets in $G_Y^j=X_j/X_0$ and the output
states are cosets in $G_X^j=X_jY_0/Y_0$.  There
are $|X_0|$ edges which split from each input state, and $|X_j\cap Y_0|$
edges which merge to each output state.  Since $|X_j\cap Y_0|<|X_0|$
for $j<\ell$, there are more output states than input states.
Some of the output states are connected to input states
via the isomorphism $\varphi(G_X^{j-1})=G_Y^j$, but some of the output states
are not connected to input states.  In this sense the graph $\calg^j$ is
not ``controllable'' for $0\le j<\ell$.  The graph $\calg_{-1}$ is the
trivial bipartite graph with one edge from input state $X_0$
to output state $Y_0$.

The input states of $\calg_{j+1}$ are $G_Y^{j+1}$, and the output states
are $G_X^{j+1}$.  Then it is clear by construction that
$\calg_j$ is a subgraph of $\calg_{j+1}$, for $-1\le j<\ell$
(the input states of $\calg_\ell$ are $G_Y^\ell=G_Y$ and the output
states are $G_X^\ell=G_X$).  Thus we have exhibited
a sequence of graphs $\calg_j$ that converges to $\calg_\ell$,
where $\calg_j$ is a subgraph of $\calg_{j+1}$ and $\calg_\ell$
is essentially $\calg_G$.  This observation forms
the basis of the algorithm in Section \ref{sec5}.

Note that $X_j\cap Y_0$ plays the same role in $X_j$ as $Y_0$
plays in $G$.  By the second isomorphism theorem, we have
$$
\frac{X_jY_0}{Y_0}\simeq\frac{X_j}{X_j\cap Y_0}
$$
and there is a 1-1 correspondence between cosets of $Y_0$ in
$X_jY_0/Y_0$ and cosets of $X_j\cap Y_0$ in $X_j/X_j\cap Y_0$
(this isomorphism and correspondence can be clearly seen using Figure \ref{fig2}).
Thus we have
\be
\label{email3}
\frac{X_j(X_j\cap Y_0)}{X_j\cap Y_0}=\frac{X_j}{X_j\cap Y_0}
\simeq\frac{X_jY_0}{Y_0}=G_X^j.
\ee
Using (\ref{email3}) and $G_Y^j=X_j/X_0$, we can define a graph
isomorphic to $\calg_j$ which only uses elements in $X_j$.

We further restrict the shift groups $G$
that we consider to those with $X_0\cap Y_0=\bone$.
We say such a shift group is {\it reduced}.  The following
proposition shows that there is essentially no loss in generality in
doing so.

\begin{prop}
Any shift group $G$ with $|X_0\cap Y_0|>1$ is an extension of
$X_0\cap Y_0$ by a shift group $\dot{G}$, where $\dot{G}$ has
$\dot{X_0}\cap \dot{Y_0}=\bone$.
\end{prop}

Each element $g\in G$ is in one
and only one coset of $Y_0$ and one and only one coset of $X_0$.
Let $\gamma:  G\ra G_X\times G_Y$ represent this correspondence
using the assignment $g\mapsto (g_x,g_y)$; note that $\gamma$
is well defined.  The map $\gamma$ is a homomorphism from $G$ into
$G_X\times G_Y$:  if $\gamma(g)=(g_x,g_y)$ and $\gamma(g')=(g_x',g_y')$,
then $gg'$ must be in coset $g_xg_x'$ of $Y_0$ and coset $g_yg_y'$ of $X_0$,
or $\gamma(gg')=(g_xg_x',g_yg_y')$.  Let $\tldcg=\gamma(G)$.  Then
$\tldcg$ is a subgroup of $G_X\times G_Y$.  Since $|X_0\cap Y_0|=1$,
from Proposition \ref{square} a coset of $X_0$ and a coset of $Y_0$
intersect in at most one element of $G$.  Thus the map $\gamma:  G\ra\tldcg$
is a bijection, and in fact $\gamma$ is an isomorphism.
Let $G\stackrel{\gamma}{\simeq} \tldcg$
denote the isomorphism given by the correspondence $\gamma$.

Let $\gamma_x:  G\ra G_X$ be the projection of $\gamma$ onto its
first coordinate, i.e., $\gamma_x:  g\mapsto g_x$.
Similarly, let $\gamma_y:  G\ra G_Y$ be the projection of $\gamma$ onto its
second coordinate, i.e., $\gamma_y:  g\mapsto g_y$.  We know that
$\tldcg$ is a subgroup of the direct product $G_X\times G_Y$.
Moreover, since $\gamma_x:  G\ra G_X$ is onto, and $\gamma_y:  G\ra G_Y$
is onto, we have that $\tldcg$ is a subdirect product
of $G_X$ and $G_Y$.  (As in \cite{MH}, we say $H$ is a {\it subdirect product}
of $H_X$ and $H_Y$ if it is a subgroup of $H_X\times H_Y$ and the
first and second coordinate of $H$ take all values in $H_X$ and
$H_Y$, respectively; we also say $H$ is a subdirect product of
$H_X\times H_Y$.)

Define $\tldcx_j\subset\tldcg$ by $\tldcx_j\rmdef \gamma(X_j)$, for $-1\le j\le\ell$.
Consider the subgroup $X_j$ of $G$ for $0\le j\le\ell$.
We now determine the image $\gamma_x(X_j)$.  But
$\gamma_x(X_j)$ must be the cosets of $Y_0$ in $G_X=G/Y_0$ that intersect
$X_j$; these must be the elements in subgroup $X_jY_0/Y_0$.
Thus we must have $\gamma_x(X_j)=X_jY_0/Y_0$ and $\gamma_x(X_j)$
is onto $X_jY_0/Y_0$.  The image $\gamma_y(X_j)$ is just the cosets
of $X_0$ in $G_Y=G/X_0$ that intersect $X_j$.  Thus $\gamma_y(X_j)=X_j/X_0$
and $\gamma_y(X_j)$ is onto $X_j/X_0$.  Thus we have shown
$\tldcx_j$ is a subdirect product of $X_jY_0/Y_0$ and $X_j/X_0$.

It is easy to see that $\tldcx_{-1}$ is a subdirect product
of $Y_0/Y_0$ and $X_0/X_0$, and in fact $\tldcx_{-1}=\bone\times\bone$.

\begin{prop}
$\tldcg$ is a subdirect product of $G_X\times G_Y$.
$\tldcx_j$ is a subdirect product of $G_X^j\times G_Y^j$, for $0\le j\le\ell$.
$\tldcx_{-1}$ is a subdirect product of $G_X^{-1}\times G_Y^0$, and
$\tldcx_{-1}=\bone\times\bone$.
\end{prop}

As with $X_j$, we can associate a graph $\tldcalg_j$
with $\tldcx_j$.  In $\tldcalg_j$, if $\tldg=(g_x,g_y)\in\tldcx_j$,
then $\tldg$ is an edge from input state $g_y$ to output state $g_x$.
Since $\tldcx_j$ is a subdirect product of $G_X^j\times G_Y^j$, the input
states of $\tldcalg_j$ are $G_Y^j$ and the output states are $G_X^j$.
Let $\tldg=\gamma(g)$.  In graph $\calg_j$, $g$ is an edge from input
state $gX_0$ to output state $gY_0$.  But we must have
$\gamma_y(g)=gX_0=g_y$ and $\gamma_x(g)=gY_0=g_x$.  Thus
$\tldg$ is an edge in $\tldcalg_j$ with input state $g_y$ and
output state $g_x$ \ifof\ $g=\gamma^{-1}(\tldg)$ is an edge in $\calg_j$ with
input state $g_y$ and output state $g_x$.  Thus $\tldcalg_j$ is isomorphic to
$\calg_j$.  For $\tldcalg_j$, there is an isomorphism
$\varphi:  G_X^{j-1}\ra G_Y^j$ from some of the output states to input
states, the same as for $\calg_j$.  As for
$X_j$, it can be shown that $\tldcalg_j$ is a subgraph of $\tldcalg_{j+1}$.
Thus we have found
a sequence of graphs $\tldcalg_j$ that converges to $\tldcalg_\ell$
where $\tldcalg_j$ is a subgraph of $\tldcalg_{j+1}$ and $\tldcalg_\ell$
is essentially $\calg_G$.

We now examine the image of $X_0$ under $\gamma$.
We know $\tldcx_0=\gamma(X_0)$.  We have
$\gamma_x(X_0)=X_0Y_0/Y_0\simeq X_0$ (since $X_0Y_0$ has $X_0\lhd X_0Y_0$,
$Y_0\lhd X_0Y_0$, and $X_0\cap Y_0=\bone$, define the homomorphism
$\kappa:  xy\mapsto x$; then the kernel is $Y_0$ and the first isomorphism
theorem gives the result) and $\gamma_y(X_0)=X_0/X_0=\bone$.
Define $X_0'\rmdef X_0Y_0/Y_0$.  Then $\tldcx_0$ is a subdirect
product of $X_0'\times\bone$.  But in this case we have
$\tldcx_0=X_0'\times\bone$.

Now examine the image of $Y_0$ under $\gamma$.
Define $\tldcy_0\rmdef \gamma(Y_0)$.  We have $\gamma_x(Y_0)=Y_0/Y_0=\bone$
and $\gamma_y(Y_0)=X_0Y_0/X_0\simeq Y_0$.
Define $Y_0''\rmdef X_0Y_0/X_0$.  Then $\tldcy_0$ is a subdirect
product of $\bone\times Y_0''$, and in this case
$\tldcy_0=\bone\times Y_0''$.

These results give
\begin{align}
\gamma(X_0Y_0) &=\tldcx_0\tldcy_0 \\
               &=(X_0'\times\bone)(\bone\times Y_0'') \\
               &=X_0'\times Y_0''.
\end{align}

Note that we will use a prime for subgroups of the $G_X$ coordinate
and a double prime for subgroups of the $G_Y$ coordinate.

\begin{thm}
\label{thm37}
$\tldcg$ is a subdirect product of $G_X\times G_Y$.
$\tldcg$ contains a normal subgroup $\tldcx_0\tldcy_0=X_0'\times Y_0''$
such that
$$
X_0'\times Y_0''\simeq X_0\times Y_0,
$$
where $\tldcx_0=X_0'\times\bone$ and $\tldcy_0=\bone\times Y_0''$.
Then $G_X=G/Y_0$ contains a group $X_0'\simeq X_0$ and $X_0'\lhd G_X$.
Further $G_Y=G/X_0$ contains a group $Y_0''\simeq Y_0$ and $Y_0''\lhd G_Y$.
$\tldcx_0$ are all the elements of $\tldcg$ with second coordinate equal $\bone$.
$\tldcy_0$ are all the elements of $\tldcg$ with first coordinate equal $\bone$.
\end{thm}

\begin{prf}
Since $G_Y=G/X_0$, the only elements of $G$ for which $G_Y=\bone$ are
subgroup $X_0$.  Thus the only elements of $\tldcg$ with second coordinate
$\bone$ are $\tldcx_0$.  Similarly, since $G_X=G/Y_0$, the only elements
of $G$ for which $G_X=\bone$ are $Y_0$.
\end{prf}

\begin{thm}
\label{thm31m}
$\tldcg$ is a subdirect product of groups $G_X$ and $G_Y$
\ifof\ there is an isomorphism
\be
\label{eq69}
\frac{G_X}{X_0'}\simeq K\simeq\frac{G_Y}{Y_0''}
\ee
such that $(g_x,g_y)$, where $g_x\in G_X$ and $g_y\in G_Y$,
is an element of $\tldcg$ \ifof\ $g_x$ and $g_y$ have the
same image $k\in K$ in the homomorphisms $G_X\ra K$,
$G_Y\ra K$.
\end{thm}

\begin{prf}
The only elements of $\tldcg$ that have the identity $\bone$
in the second coordinate are $\tldcx_0=X_0'\times\bone$.
The only elements of $\tldcg$ that have the identity $\bone$
in the first coordinate are $\tldcy_0=\bone\times Y_0''$.  Then the
theorem is just an application of the subdirect product theorem
in Hall's text \cite{MH}.
\end{prf}

In general, when the condition in Theorem \ref{thm31m} holds, we say
$\tldcg$ is a subdirect product of $G_X\times G_Y$ {\it implied by}
the isomorphism (\ref{eq69}).

Fix $j$, $0\le j\le\ell$.  Define $\Lambda_j\rmdef X_j\cap Y_0$.
We now examine the image of $\Lambda_j$ under $\gamma$.
Define $\tldclambda_j\rmdef\gamma(\Lambda_j)$.
The image $\gamma_x(X_j\cap Y_0)$ is the cosets of $Y_0$ in $G_X=G/Y_0$ that intersect
$X_j\cap Y_0$.  Then $\gamma_x(X_j\cap Y_0)=Y_0/Y_0=\bone$.
And $\gamma_y(X_j\cap Y_0)$ is the cosets
of $X_0$ in $G_Y=G/X_0$ that intersect $X_j\cap Y_0$.
Then
$$
\gamma_y(X_j\cap Y_0)=\frac{X_j}{X_0}\cap\frac{X_0Y_0}{Y_0}.
$$
Define
$$
\Lambda_j''\rmdef\frac{X_j}{X_0}\cap\frac{X_0Y_0}{Y_0}.
$$
Thus $\tldclambda_j$ is a subdirect
product of $\bone\times \Lambda_j''$, and so in fact
$\tldclambda_j=\bone\times \Lambda_j''$.
Note that $\Lambda_j''=G_Y^j\cap Y_0''$ and $\Lambda_j''\lhd G_Y$.
Also $\Lambda_0''=X_0/X_0=\bone$ and $\tldclambda_j\lhd\tldcg$.

\begin{thm}
\label{thm38}
Fix $j$, $0\le j\le\ell$.
$\tldcx_j$ is a subdirect product of $G_X^j\times G_Y^j$.
$\tldcx_j$ contains a normal subgroup
$\tldcx_0\tldclambda_j=X_0'\times \Lambda_j''$ such that
$$
X_0'\times \Lambda_j''\simeq X_0\times\Lambda_j,
$$
where $\tldcx_0=X_0'\times\bone$ and $\tldclambda_j=\bone\times \Lambda_j''$.
Then $G_X^j=X_jY_0/Y_0$ contains a group $X_0'\simeq X_0$ and $X_0'\lhd G_X$.
Further $G_Y^j=X_j/X_0$ contains a group $\Lambda_j''$ such that
$\Lambda_j''\simeq\Lambda_j$, $\Lambda_j''=G_Y^j\cap Y_0''$,
and $\Lambda_j''\lhd G_Y$.
$\tldcx_0$ are all the elements of $\tldcx_j$ with second coordinate equal $\bone$.
$\tldclambda_j$ are all the elements of $\tldcx_j$ with first coordinate equal $\bone$.
\end{thm}

\begin{prf}
Since $\gamma_x(X_j)$ is the cosets of $Y_0$ in $G_X=G/Y_0$ that intersect
$X_j$, the only elements $g$ of $X_j$ for which $\gamma_x(g)=Y_0/Y_0=\bone$
are $g\in X_j\cap Y_0$.  Thus the only elements of $\tldcx_j$
that have the identity $\bone$ in the first coordinate are
$\gamma(X_j\cap Y_0)=\tldclambda_j$.
\end{prf}

We can now give a necessary and sufficient condition that guarantees
$\tldcx_j$ is a subdirect product of groups $G_X^j$ and $G_Y^j$.

\begin{thm}
\label{thm31}
For $0\le j\le\ell$,
$\tldcx_j$ is a subdirect product of groups $G_X^j$ and $G_Y^j$
\ifof\ there is an isomorphism
\be
\label{eq46}
\frac{G_X^j}{X_0'}\simeq K\simeq\frac{G_Y^j}{\Lambda_j''}
\ee
such that $(g_x,g_y)$, where $g_x\in G_X^j$ and $g_y\in G_Y^j$,
is an element of $\tldcx_j$ \ifof\ $g_x$ and $g_y$ have the
same image $k\in K$ in the homomorphisms $G_X^j\ra K$,
$G_Y^j\ra K$.
\end{thm}

\begin{prf}
The only elements of $\tldcx_j$ that have the identity $\bone$
in the second coordinate are $\tldcx_0=X_0'\times\bone$.
The only elements of $\tldcx_j$
that have the identity $\bone$ in the first coordinate are
$\tldclambda_j=\bone\times\Lambda_j''$.  Then the
theorem is just an application of the subdirect product theorem
in Hall's text \cite{MH}.
\end{prf}

From Lemma \ref{lem17}, we have $X_j^*=X_j (X_{j+1}\cap Y_0)=X_j\Lambda_{j+1}$,
for $-1\le j<\ell$.  Define $\tldcx_j^*\rmdef\gamma(X_j^*)$.
Then under the isomorphism $G\stackrel{\gamma}{\simeq} \tldcg$,
\begin{align}
\nonumber
\tldcx_j^* &=\tldcx_j \tldclambda_{j+1}  \\
\label{eq300a}
           &=\tldcx_j (\bone\times\Lambda_{j+1}'').
\end{align}
Define $G_Y^{j*}\rmdef G_Y^j\Lambda_{j+1}''$.
Since $G_Y^j\lhd G_Y^{j+1}$ and $\Lambda_{j+1}''\lhd G_Y^{j+1}$,
then $G_Y^{j*}$ is a subgroup of $G_Y^{j+1}$ and
$$
G_Y^j\lhd G_Y^{j*}\lhd G_Y^{j+1}.
$$
Then $\tldcx_j^*$ is a subdirect product of $G_X^j\times G_Y^{j*}$.

For $j=\ell-1$ we know
$$
X_{\ell-1}^*=X_{\ell-1}Y_0=X_\ell.
$$
Then under the isomorphism $G\stackrel{\gamma}{\simeq} \tldcg$,
\begin{align*}
\tldcx_{\ell-1}^* &=\tldcx_{\ell-1} (\bone\times\Lambda_\ell'') \\
                  &=\tldcx_{\ell-1} (\bone\times Y_0'').
\end{align*}
Since $\tldcx_{\ell-1}^*=\tldcx_\ell$, and
$\tldcx_\ell$ is a subdirect product of $G_X^\ell\times G_Y^\ell$,
this means
$$
G_X^{\ell-1}=G_X^\ell=G_X,
$$ and
$$
G_Y^{\ell-1}Y_0''=G_Y^{\ell-1*}=G_Y^\ell=G_Y.
$$

The isomorphism $\gamma:  G\stackrel{\gamma}{\simeq} \tldcg$
induces isomorphisms
\begin{align}
G_X    &=G/Y_0\simeq\tldcg/\tldcy_0, \\
G_Y    &=G/X_0\simeq\tldcg/\tldcx_0, \\
G_X^j  &=X_jY_0/Y_0\simeq\tldcx_j\tldcy_0/\tldcy_0, \\
G_Y^j  &=X_j/X_0\simeq\tldcx_j/\tldcx_0.
\end{align}

\begin{prop}
\label{prop30}
If a group $G$ has a \sst\ $(\xj,Y_0,\varphi)$ and $|X_0\cap Y_0|=1$, then
under the isomorphism $G\stackrel{\gamma}{\simeq} \tldcg$,
the group $\tldcg$ is a subdirect product of $G_X$ and $G_Y$.
Further group $\tldcg$ has a \sst\ $(\tldxj,\tldcy_0,\tldvphi)$,
where we have $\tldcx_j=\gamma(X_j)$, $\tldcy_0=\gamma(Y_0)$, and
the isomorphism $\tldvphi:  \tldcg/\tldcy_0\ra \tldcg/\tldcx_0$
is just the isomorphism $\varphi:  G/Y_0\ra G/X_0$.
\end{prop}

\begin{prf}
By the preceding results, we have shown there is an isomorphism
$G\stackrel{\gamma}{\simeq} \tldcg$, where
$\tldcg$ is a subdirect product of $G_X$ and $G_Y$.
From the correspondence theorem, under the isomorphism $\gamma$,
the normal chain $\xj$
gives a normal chain $\tldxj$ with $\tldcx_\ell=\tldcg$ and
each $\tldcx_j\lhd \tldcg$, and the normal subgroup $Y_0$
gives a normal subgroup $\tldcy_0$.  Under the isomorphism $\gamma$,
the isomorphism $\varphi:  G/Y_0\ra G/X_0$ induces an isomorphism
$\tldvphi:  \tldcg/\tldcy_0\ra \tldcg/\tldcx_0$ and
$\tldvphi(\tldcx_j\tldcy_0/\tldcy_0)=\tldcx_{j+1}/\tldcx_0$.
\end{prf}

We can summarize some of the results in this section as follows.

\begin{thm}
\label{thmx}
Let $G$ be a group with a \sst\ $(\xj,Y_0,\varphi)$ and $|X_0\cap Y_0|=1$.
Define $G_X^j\rmdef X_jY_0/Y_0$ and $G_Y^j\rmdef X_j/X_0$.
There is a normal chain
\begin{multline*}
\bone=G_X^{-1}\lhd X_0'=G_X^0\lhd G_X^1\lhd\cdots\lhd G_X^j\lhd\cdots  \\
\lhd G_X^{\ell-1}=G_X^\ell=G_X,
\end{multline*}
where each $G_X^j\lhd G_X$, $0\le j\le\ell$.  There are normal chains
\begin{multline*}
\bone=G_Y^0\lhd G_Y^{0*}\lhd G_Y^1\lhd G_Y^{1*}\lhd\cdots\lhd G_Y^j\lhd G_Y^{j*}\lhd \\
\cdots\lhd G_Y^{\ell-1}\lhd G_Y^{\ell-1*}=G_Y^\ell=G_Y,
\end{multline*}
and
$$
\bone=\Lambda_0''\lhd \Lambda_1''\lhd\cdots\lhd \Lambda_j''\lhd\cdots\lhd \Lambda_\ell''=Y_0'',
$$
where each $G_Y^j\lhd G_Y$, $G_Y^{j*}\lhd G_Y$, and each
$\Lambda_j''\lhd G_Y$ and $Y_0''\lhd G_Y$,
such that $G_Y^j\cap Y_0''=\Lambda_j''$
and $G_Y^{j*}=G_Y^j\Lambda_{j+1}''$.
There is an isomorphism $\varphi:  G_X\ra G_Y$ such that
$\varphi:  G_X^j\mapsto G_Y^{j+1}$ for $-1\le j<\ell$.

Under the isomorphism $G\stackrel{\gamma}{\simeq} \tldcg$,
the group $\tldcg$ is a subdirect product of $G_X$ and $G_Y$.
Further group $\tldcg$ has a \sst\ $(\tldxj,\tldcy_0,\tldvphi)$,
where we have $\tldcx_j=\gamma(X_j)$, $\tldcy_0=\gamma(Y_0)$, and
isomorphism $\tldvphi$ is closely related to $\varphi$.
We have $\tldcx_{-1}=\bone\times\bone$, $\tldcx_0=X_0'\times\bone$,
and $\tldcy_0=\bone\times Y_0''$.  For $0\le j\le\ell$,
$X_0'\times\bone$ are the only elements
of $\tldcx_j$ with $\bone$ in the second coordinate, and
$\bone\times\Lambda_j''$ are
the only elements of $\tldcx_j$ with $\bone$ in the first coordinate.
Lastly, for $0\le j\le\ell$, $\tldcx_j$ is a subdirect product
of $G_X^j$ and $G_Y^j$, and there is an isomorphism
\be
\label{thmx1}
\frac{G_X^j}{X_0'}\simeq K\simeq\frac{G_Y^j}{\Lambda_j''}
\ee
such that $(g_x,g_y)\in\tldcx_j$ \ifof\ $g_x$ and $g_y$ have the
same image $k\in K$ in the homomorphisms $G_X^j\ra K$,
$G_Y^j\ra K$.
\end{thm}

Since $G$ has a \sst\ $(\xj,Y_0,\varphi)$, we know that $X_j\subset X_{j+1}$.
Under the isomorphism $G\stackrel{\gamma}{\simeq} \tldcg$,
we have $\tldcx_j\subset\tldcx_{j+1}$ for the subdirect product group $\tldcg$.
We now give a necessary and sufficient condition for
$\tldcx_j\subset\tldcx_{j+1}$ to hold.

\begin{lem}
\label{lemma47}
Fix arbitrary integer $j$, $j\ge 0$.
Assume there are three (trivial) normal chains
\begin{align}
\label{nc1}
H_U^j & \lhd H_U^{j+1}, \\
\label{nc2}
H_V^j & \lhd H_V^{j+1}, \\
\label{nc3}
\Gamma_j'' & \lhd \Gamma_{j+1}'',
\end{align}
where $U_0'\lhd H_U^{j+1}$, $\Gamma_j''\lhd H_V^j$,
$\Gamma_{j+1}''\lhd H_V^{j+1}$, and $H_V^j\cap\Gamma_{j+1}''=\Gamma_j''$.
Since $H_V^j\lhd H_V^{j+1}$ and $\Gamma_{j+1}''\lhd H_V^{j+1}$,
there is a subgroup $H_V^{j*}=H_V^j\Gamma_{j+1}''$ of $H_V^{j+1}$ such that
$$
H_V^j\lhd H_V^{j*}\lhd H_V^{j+1}.
$$
Assume the three normal chains (\ref{nc1})-(\ref{nc3})
are related such that there are isomorphisms
$$
\beta_j:  \frac{H_U^j}{U_0'}\ra\frac{H_V^j}{\Gamma_j''}
$$
and
$$
\beta_{j+1}:  \frac{H_U^{j+1}}{U_0'}\ra\frac{H_V^{j+1}}{\Gamma_{j+1}''}.
$$
Let $\tldcu_j$ be the subdirect product of $H_U^j\times H_V^j$
implied by the isomorphism $\beta_j$, and
let $\tldcu_{j+1}$ be the subdirect product of $H_U^{j+1}\times H_V^{j+1}$
implied by the isomorphism $\beta_{j+1}$.
Let $\eta_j''$ be the isomorphism
$$
\eta_j'': H_V^j/\Gamma_j''\ra H_V^{j*}/\Gamma_{j+1}''
$$
with assignment $h_v\Gamma_j''\mapsto h_v\Gamma_{j+1}''$ for $h_v\in H_V^j$,
given by (\ref{wkhse1}) of Lemma \ref{wkhorse} using
$H_V^j\Gamma_{j+1}''=H_V^{j*}$ in the hypothesis.
Then the composition $\eta_j''\circ\beta_j$ is an isomorphism $\beta_j^*$,
$$
\beta_j^*:  H_U^j/U_0'\ra H_V^{j*}/\Gamma_{j+1}''
$$
(see Figure \ref{commut}).  We have $\tldcu_j\subset\tldcu_{j+1}$
\ifof\ the restriction of the isomorphism $\beta_{j+1}$ to $H_U^j/U_0'$
is isomorphism $\beta_j^*$.
In this case there is a group $\tldcu_j^*$ such that
$\tldcu_j\subset\tldcu_j^*\subset\tldcu_{j+1}$
where $\tldcu_j^*$ is a subdirect product of
$H_U^j\times H_V^{j*}$ implied by the isomorphism $\beta_j^*$.
\end{lem}

\begin{figure}[h]
\centering
\vspace{3ex}

\begin{picture}(100,110)
%arrows
%\put(0,70){\vector(0,-1){60}}
\put(20,100){\vector(1,0){60}}
\put(100,85){\vector(0,-1){70}}
\put(20,85){\vector(1,-1){70}}
%\put(20,0){\vector(1,0){60}}
%corner labels
%\put(0,0){\makebox(0,0){$H_U^j/U_0'$}}
\put(0,100){\makebox(0,0){$H_U^j/U_0'$}}
\put(100,100){\makebox(0,0){$H_V^j/\Gamma_j''$}}
\put(100,0){\makebox(0,0){$H_V^{j*}/\Gamma_{j+1}''$}}
%arrow labels
%\put(-5,40){\makebox(0,0)[r]{$\text{id}$}}
\put(50,105){\makebox(0,0)[b]{$\beta_j$}}
\put(50,35){\makebox(0,0)[b]{$\beta_j^*$}}
\put(105,50){\makebox(0,0)[l]{$\eta_j''$}}
\end{picture}

\caption{Commutative diagram.}
\label{commut}

\end{figure}

\begin{prf}
Since $H_V^j$ and $\Gamma_{j+1}''$ are normal
subgroups of $H_V^{j+1}$, we have $H_V^{j*}=H_V^j\Gamma_{j+1}''$ is a normal
subgroup of $H_V^{j+1}$.

Refer to Figure \ref{commut}.  Fix $cU_0'\in H_U^j/U_0'$, where $c\in H_U^j$.
Let the isomorphism $\beta_j$ make the assignment
$$
\beta_j:  cU_0'\mapsto d\Gamma_j'',
$$
where $d\in H_V^j$.  The isomorphism
$\eta_j'': H_V^j/\Gamma_j''\ra H_V^{j*}/\Gamma_{j+1}''$
gives the assignment
$$
d\Gamma_j''\mapsto d\Gamma_{j+1}''.
$$
Then the isomorphism $\beta_j^*$ makes the assignment
\be
\label{xq1}
\beta_j^*:  cU_0'\mapsto d\Gamma_{j+1}''.
\ee

First assume $\tldcu_j\subset\tldcu_{j+1}$.  We show
the restriction of $\beta_{j+1}$ to $H_U^j/U_0'$ is $\beta_j^*$.
Since $\beta_j$ makes the assignment $\beta_j:  cU_0'\mapsto d\Gamma_j''$,
the elements $cU_0'\times d\Gamma_j''$ are in $\tldcu_j$.
Since $\tldcu_j\subset\tldcu_{j+1}$,
then $cU_0'\times d\Gamma_j''\subset\tldcu_{j+1}$.
But since $\Gamma_j''\subset\Gamma_{j+1}''\subset H_V^{j+1}$ by assumption,
then $cU_0'\times d\Gamma_{j+1}''\subset\tldcu_{j+1}$.  Then the
isomorphism $\beta_{j+1}$ makes the assignment
\be
\label{xq2}
\beta_{j+1}:  cU_0'\mapsto d\Gamma_{j+1}''.
\ee
Comparing (\ref{xq1}) and (\ref{xq2}) shows that
the restriction of $\beta_{j+1}$ to $H_U^j/U_0'$ is $\beta_j^*$.

Now assume the restriction of $\beta_{j+1}$ to $H_U^j/U_0'$ is $\beta_j^*$.
We show $\tldcu_j\subset\tldcu_{j+1}$.  Let
$cU_0'\times d\Gamma_j''\subset\tldcu_j$.
Then $\beta_j$ makes the assignment $\beta_j:  cU_0'\mapsto d\Gamma_j''$,
and $\beta_j^*$ makes the assignment
$$
\beta_j^*:  cU_0'\mapsto d\Gamma_{j+1}''.
$$
Since the restriction of $\beta_{j+1}$ to $H_U^j/U_0'$ is $\beta_j^*$,
we have $\beta_{j+1}$ makes the assignment
$$
\beta_{j+1}:  cU_0'\mapsto d\Gamma_{j+1}''.
$$
Then $cU_0'\times d\Gamma_{j+1}''\subset\tldcu_{j+1}$.
Since $cU_0'\times d\Gamma_j''\subset cU_0'\times d\Gamma_{j+1}''$,
this means $\tldcu_j\subset\tldcu_{j+1}$.
\end{prf}

{\it Remark:}  Note that if $\Gamma_{j+1}''=\Gamma_j''$, Figure \ref{commut}
becomes trivial, i.e., $H_V^{j*}=H_V^j$ and $\beta_j^*=\beta_j$.

From Theorem \ref{thmx}, the conditions in Lemma \ref{lemma47} apply
to $\tldcg$, and thus $\tldcg$ has the properties given in Lemma \ref{lemma47}.
This completes the analysis of $\tldcg$.  We now give a synthesis result,
a construction of a subdirect product group which is a shift group.
We reuse the notation in Lemma \ref{lemma47}; this should not be confusing.

%%%%%%%%%%%%%%%%%%%%%%%%%%%%%%%%

\begin{thm}
\label{thmy1}
Assume there is a group $H_U$ with a normal chain
\begin{multline}
\label{nchu}
\bone=H_U^{-1}\lhd U_0'=H_U^0\lhd H_U^1\lhd\cdots\lhd H_U^j\lhd\cdots  \\
\lhd H_U^{\ell-1}=H_U^\ell=H_U,
\end{multline}
where each $H_U^j\lhd H_U$.
Assume there are groups $H_V$ and $V_0''$ and normal chains
\begin{multline}
\label{nchv}
\bone=H_V^0\lhd H_V^{0*}\lhd H_V^1\lhd H_V^{1*}\lhd\cdots\lhd H_V^j\lhd H_V^{j*}\lhd \\
\cdots\lhd H_V^{\ell-1}\lhd H_V^{\ell-1*}=H_V^\ell=H_V,
\end{multline}
\be
\label{ncgammapp}
\bone=\Gamma_0''\lhd \Gamma_1''\lhd\cdots\lhd \Gamma_j''\lhd\cdots\lhd \Gamma_\ell''=V_0'',
\ee
where each $H_V^j\lhd H_V$, and each $\Gamma_j''\lhd H_V$ and $V_0''\lhd H_V$,
such that $H_V^j\cap V_0''=\Gamma_j''$ and $H_V^{j*}=H_V^j\Gamma_{j+1}''$.
Assume there is an isomorphism
$\phi_j:  H_U^j\ra H_V^{j+1}$ for $-1\le j<\ell$, such that for $0\le j<\ell$, the
restriction of $\phi_j$ to $H_U^{j-1}$ is $\phi_{j-1}$.
Assume the three normal chains $\{H_U^j\}$, $\{H_V^j\}$, and $\{\Gamma_j''\}$
are related such that for $0\le j<\ell$ there is an isomorphism $\beta_{j+1}$,
\be
\label{thmy1a}
\beta_{j+1}:  \frac{H_U^{j+1}}{U_0'}\ra\frac{H_V^{j+1}}{\Gamma_{j+1}''},
\ee
whose restriction to $H_U^j/U_0'$ is the isomorphism
$\beta_j^*=\eta_j''\circ\beta_j$ shown in Figure \ref{commut}, where
$$
\beta_j^*:  H_U^j/U_0'\ra H_V^{j*}/\Gamma_{j+1}'',
$$
and $\eta_j''$ is the isomorphism
$$
\eta_j'': H_V^j/\Gamma_j''\ra H_V^{j*}/\Gamma_{j+1}''
$$
given by (\ref{wkhse1}) of Lemma \ref{wkhorse} using
$H_V^{j*}=H_V^j\Gamma_{j+1}''$ in the hypothesis.
Define isomorphism $\beta_0$,
$$
\beta_0:  \frac{H_U^0}{U_0'}\ra\frac{H_V^0}{\Gamma_0''},
$$
the trivial isomorphism $\beta_0:  \bone\ra\bone$.  For $0\le j<\ell$,
let $\tldcu_{j+1}$ be the subdirect product of $H_U^{j+1}\times H_V^{j+1}$ implied by
the isomorphism (\ref{thmy1a}).  In other words, $U_0'\times\bone$ are
all the elements in $\tldcu_{j+1}$ with $\bone$ in the second coordinate,
and $\bone\times\Gamma_{j+1}''$ are all the elements in $\tldcu_{j+1}$ with $\bone$
in the first coordinate, and (\ref{thmy1a}) holds.
Let $\tldcu_0$ be the subdirect product of $H_U^0\times H_V^0$ implied by
the isomorphism $\beta_0$, i.e., $\tldcu_0=U_0'\times\bone$.
Define $\tldcu_{-1}\rmdef\bone\times\bone$; define $\tldch\rmdef\tldcu_\ell$.
Then $\tldch$ is a group with a \sst\ $(\tlduj,\tldcv_0,\tldphi)$, where
$\tldcv_0\rmdef\bone\times V_0''$ and
$\tldphi:  \tldch/\tldcv_0\ra\tldch/\tldcu_0$ is an isomorphism closely
related to $\phi_{\ell-1}$.  (The precise connection is shown in the
proof below.)
\end{thm}

\begin{prf}
We need to show that $\tldch$ is a group with a \sst\
$(\tlduj,\tldcv_0,\tldphi)$.
First we show that $\tldcu_j\lhd\tldch$ for $-1\le j\le\ell$.
By assumption we know that $H_U^j\lhd H_U$ and $H_V^j\lhd H_V$ for $0\le j\le\ell$.
Now suppose $(h_u,h_v)\in\tldch$.
Since $\tldcu_j$ is a subdirect product of $H_U^j\times H_V^j$, we have
$$
(h_u,h_v)\tldcu^j(h_u,h_v)^{-1}\subset \tldcu^j.
$$
Thus $\tldcu_j\lhd\tldch$ for $0\le j\le\ell$.  Clearly $\tldcu_{-1}\lhd\tldch$.

Applying Lemma \ref{lemma47} shows that $\tldcu_j\subset\tldcu_{j+1}$
for $0\le j<\ell$.  Clearly $\tldcu_{-1}\subset\tldcu_0$.

Since $\tldcv_0=\bone\times V_0''$ are all the elements with
$\bone$ in the first coordinate, we must have $\tldcv_0\lhd\tldch$.

We now show that there is an isomorphism
$\tau:  \tldch/\tldcv_0\ra H_U$ such that the restriction
of $\tau$ to $\tldcu_j \tldcv_0/\tldcv_0$ is
$$
\tau(\tldcu_j \tldcv_0/\tldcv_0)=H_U^j
$$
for $-1\le j<\ell$.  We know that $\tldch$ is a subdirect product
of $H_U\times H_V$.  But $\bone\times V_0''$ are all the elements
in $\tldch$ with identity $\bone$ in the first coordinate.
This shows there is an isomorphism
$$
H_U\simeq\frac{\tldch}{\bone\times V_0''}=\frac{\tldch}{\tldcv_0}.
$$
Let $\tau:  \tldch/\tldcv_0\ra H_U$ be the corresponding isomorphism.
A subgroup $\grave{H}$ of $\tldch/\tldcv_0$ is just a collection of cosets
of $\tldcv_0$,
$$
\grave{H}=\{s\tldcv_0 | s\tldcv_0\in\tldch\}.
$$
Each coset $s\tldcv_0$ is of the form $h_u\times h_v V_0''$ for some
$h_u\in H_U$, $h_v\in H_V$.  Thus $\tau(\grave{H})$ is just the projection of
$\grave{H}$ onto the first coordinate $h_u$ of each coset
$s\tldcv_0\in\grave{H}$.

Now fix $j$, $0\le j<\ell$.  We know that $\tldcu_j$ is a subdirect
product of $H_U^j\times H_V^j$.  Then by construction of $\tldch$ we
know that $\tldcu_j \tldcv_0$ must be a subdirect product of $H_U^j$
and of some group $\bar{H}_V^j$ isomorphic to $\tldcu_j \tldcv_0/\tldcu_0$
such that $\bar{H}_V^j\supset H_V^j$.  Thus we must have
$\tau(\tldcu_j \tldcv_0/\tldcv_0)=H_U^j$.  Clearly
$\tau(\tldcu_{-1} \tldcv_0/\tldcv_0)=H_U^{-1}$.

We now show that there is an isomorphism
$\xi:  \tldch/\tldcu_0\ra H_V$ such that the restriction
of $\xi$ to $\tldcu_j/\tldcu_0$ is
$$
\xi(\tldcu_j/\tldcu_0)=H_V^j
$$
for $0\le j\le\ell$.  We know that $\tldch$ is a subdirect product
of $H_V\times H_V$.  But $U_0'\times\bone$ are all the elements
in $\tldcu_j$ with $\bone$ in the second coordinate.
This shows there is an isomorphism
$$
H_V\simeq\frac{\tldch}{U_0'\times\bone}=\frac{\tldch}{\tldcu_0}.
$$
Let $\xi:  \tldch/\tldcu_0\ra H_V$ be the corresponding isomorphism.
As for $\tau$, for $\acute{H}$ a collection of cosets
$\{s\tldcu_0 | s\tldcu_0\in\tldch\}$
of $\tldcu_0$, $\xi(\acute{H})$ is just the projection of
$\acute{H}$ onto the second coordinate $h_v$ of each coset
$s\tldcu_0=h_u U_0'\times h_v\in\acute{H}$.
For $0\le j\le\ell$, we know that $\tldcu_j$ is a subdirect
product of $H_U^j\times H_V^j$.  Thus we must have
$\xi(\tldcu_j/\tldcu_0)=H_V^j$ for $0\le j\le\ell$.

We now show that there is an isomorphism
$\tldphi:  \tldch/\tldcv_0\ra\tldch/\tldcu_0$ which makes $\tldch$ into a
shift group, where $\tldphi$ is closely related to $\phi_{\ell-1}$.
From the assumptions in the theorem, we know there is an isomorphism
$\phi_{\ell-1}:  H_U\ra H_V$.  Thus using $\tau$ and $\xi$ we have
\be
\label{qeqprf3}
\frac{\tldch}{\tldcv_0}\stackrel{\tau}{\simeq} H_U
\stackrel{\phi_{\ell-1}}{\simeq}H_V
\stackrel{\xi}{\simeq}\frac{\tldch}{\tldcu_0}.
\ee
This defines an isomorphism
$$
\tldphi:  \frac{\tldch}{\tldcv_0}\ra\frac{\tldch}{\tldcu_0},
$$
where $\tldphi$ is the composition $\xi^{-1}\circ\phi_{\ell-1}\circ\tau$.
We now show that
$$
\tldphi(\tldcu_j \tldcv_0/\tldcv_0)=\tldcu_{j+1}/\tldcu_0
$$
for $-1\le j<\ell$.  From the assumptions in the theorem, we have
$\phi_{\ell-1}(H_U^j)=H_V^{j+1}$ for $-1\le j<\ell$.  Then we have
\begin{align*}
\tldphi(\tldcu_j \tldcv_0/\tldcv_0)
&=(\xi^{-1}\circ\phi_{\ell-1}\circ\tau)(\tldcu_j \tldcv_0/\tldcv_0)  \\
&=\tldcu_{j+1}/\tldcu_0
\end{align*}
for $-1\le j<\ell$.  Thus $\tldphi$ is the desired isomorphism,
and $\tldch$ has a \sst\ $(\tlduj,\tldcv_0,\tldphi)$.
\end{prf}

We have just shown that Theorem \ref{thmy1} gives a shift group $\tldch$
which is a subdirect product group.  Consider a mapping $\zeta:  \tldch\ra H$,
which just regards each element $(h_u,h_v)\in\tldch$ as a single element
$h\in H$, i.e., $\zeta:  (h_u,h_v)\mapsto h$.  We require the assignment
$\zeta:  (1,1)\mapsto\bone$.  Then $H$ is a group and $\zeta$ is
an isomorphism.  Using the isomorphism $\zeta$, we can convert the
subdirect product group $\tldch$ into an abstract shift group $H$.

\begin{prop}
\label{prop35a}
The group $\tldch$ found by Theorem \ref{thmy1} is a subdirect product of
$H_U$ and $H_V$ and has a \sst\ $(\tlduj,\tldcv_0,\tldphi)$.
Under the isomorphism $\tldch\stackrel{\zeta}{\simeq} H$,
the group $H$ has a \sst\ $(\uj,V_0,\phi)$ and $|U_0\cap V_0|=1$.
\end{prop}

Note that we can make a round trip by starting with $G$, using
Theorem \ref{thmx} to obtain $\tldcg$, then using Theorem \ref{thmy1}
to obtain $\tldch=\tldcg$, and finally Proposition \ref{prop35a}
to obtain $H=G$.  Thus we can obtain any shift group $G$ by
starting with the description in Theorem \ref{thmy1}.

We now simplify Theorem \ref{thmy1} further.
From Theorem \ref{thmy1} we know that if $\tldch$ is
a shift group, there is an isomorphism $\phi_{\ell-1}:  H_U^{\ell-1}\ra H_V^\ell$
or just $\phi_{\ell-1}:  H_U\ra H_V$.  This means that $H_U$ and $H_V$
are essentially the same.  Thus the sequence of groups $\{H_V^{j*}\}$ in
$H_V$ corresponds to a dual sequence $\{H_U^{j*}\}$ in $H_U$.  We
let subgroup $H_U^{j*}$ in $H_U$ correspond to subgroup
$H_V^{j+1*}$ in $H_V$ so that $\phi_{\ell-1}(H_U^{j*})\rmdef H_V^{j+1*}$
for $-1\le j<\ell-1$.
Then we can find a refinement of the normal chain $\{H_U^j\}$ in (\ref{nchu}):
\begin{multline}
\label{eq101}
\bone=H_U^{-1}\lhd H_U^{-1*}\lhd U_0'=H_U^0\lhd H_U^{0*}\lhd H_U^1\lhd H_U^{1*}\lhd\cdots \\
\lhd H_U^j\lhd H_U^{j*}\lhd\cdots
\lhd H_U^{\ell-2}\lhd H_U^{\ell-2*}=H_U^{\ell-1}=H_U^\ell=H_U,
\end{multline}
where $\phi_{\ell-1}(H_U^{j*})=H_V^{j+1*}$ for $-1\le j<\ell-1$, and each
$H_U^{j*}\lhd H_U^{j+1}$ for $-1\le j<\ell-1$.
Note that since $H_V^{\ell-1}\lhd H_V^{\ell-1*}=H_V^\ell$,
we have $H_U^{\ell-2}\lhd H_U^{\ell-2*}=H_U^{\ell-1}$
as shown.

The normal chain $\{\Gamma_j''\}$
in $H_V$ corresponds to a dual chain $\{\Gamma_j'\}$ in $H_U$.
We let subgroup $\Gamma_j'$ in $H_U$ correspond to subgroup
$\Gamma_{j+1}''$ in $H_V$ so that $\phi_{\ell-1}(\Gamma_j')\rmdef\Gamma_{j+1}''$
for $-1\le j<\ell$.  Let $V_0''$ in $H_V$ correspond to $V_0'$ in $H_U$, so that
$\phi_{\ell-1}(V_0')=V_0''$.
Then using $\phi_{\ell-1}$ and normal chain $\{\Gamma_j''\}$ in (\ref{ncgammapp}),
we can find a normal chain
\be
\label{eq102}
\bone=\Gamma_{-1}'\lhd\Gamma_0'\lhd \Gamma_1'\lhd
\cdots\lhd \Gamma_j'\lhd\cdots\lhd \Gamma_{\ell-2}'
\lhd \Gamma_{\ell-1}'=V_0',
\ee
where $\phi_{\ell-1}(\Gamma_j')=\Gamma_{j+1}''$ and
each $\Gamma_j'\lhd H_U$ and $V_0'\lhd H_U$ for $-1\le j<\ell$,
such that $H_U^j\cap V_0'=\Gamma_j'$ for $-1\le j<\ell$,
and $H_U^{j*}=H_U^j\Gamma_{j+1}'$ for $-1\le j<\ell-1$.
Since $\Gamma_{\ell-1}''\lhd \Gamma_\ell''=V_0''$, we have
$\Gamma_{\ell-2}'\lhd \Gamma_{\ell-1}'=V_0'$ as shown.
Since $\Gamma_0''=\bone$, we have $\Gamma_{-1}'=\bone$.

Assume the two normal chains (\ref{eq101}) and (\ref{eq102})
are related such that for $0\le j<\ell$ there is an isomorphism $\alpha_{j+1}$,
\be
\label{eq103}
\alpha_{j+1}:  \frac{H_U^{j+1}}{U_0'}\ra\frac{H_U^j}{\Gamma_j'},
\ee
whose restriction to $H_U^j/U_0'$ is the isomorphism
$\alpha_j^*=\eta_{j-1}'\circ\alpha_j$, where
$$
\alpha_j^*:  H_U^j/U_0'\ra H_U^{j-1*}/\Gamma_j',
$$
and $\eta_{j-1}'$ is the isomorphism
$$
\eta_{j-1}': H_U^{j-1}/\Gamma_{j-1}'\ra H_U^{j-1*}/\Gamma_j'
$$
given by (\ref{wkhse1}) of Lemma \ref{wkhorse} using
$H_U^{j-1*}=H_U^{j-1}\Gamma_j'$ in the hypothesis (see Figure \ref{commut1}).
Define $\alpha_0$ to be the trivial isomorphism
$\alpha_0:  H_U^0/U_0'\ra H_U^{-1}/\Gamma_{-1}'$, or
$\alpha_0:  \bone\ra\bone$.

\begin{figure}[h]
\centering
\vspace{3ex}

\begin{picture}(100,110)
%arrows
%\put(0,70){\vector(0,-1){60}}
\put(20,100){\vector(1,0){52}}
\put(100,85){\vector(0,-1){70}}
\put(20,85){\vector(1,-1){70}}
%\put(20,0){\vector(1,0){60}}
%corner labels
%\put(0,0){\makebox(0,0){$H_U^j/U_0'$}}
\put(0,100){\makebox(0,0){$H_U^j/U_0'$}}
\put(100,100){\makebox(0,0){$H_U^{j-1}/\Gamma_{j-1}'$}}
\put(100,0){\makebox(0,0){$H_U^{j-1*}/\Gamma_j'$}}
%arrow labels
%\put(-5,40){\makebox(0,0)[r]{$\text{id}$}}
\put(50,105){\makebox(0,0)[b]{$\alpha_j$}}
\put(50,35){\makebox(0,0)[b]{$\alpha_j^*$}}
\put(105,50){\makebox(0,0)[l]{$\eta_{j-1}'$}}
\end{picture}

\caption{Commutative diagram.}
\label{commut1}

\end{figure}

Since $H_U$ and $H_V$ are essentially the same, this suggests
that in the construction of $\tldch$ we only need to use $H_U$.
We now show that we can recover $\tldch$ in Theorem \ref{thmy1} by
using just the two normal chains (\ref{eq101}) and (\ref{eq102}),
isomorphism $\phi_{\ell-1}$ from Theorem \ref{thmy1},
and isomorphism $\alpha_{j+1}$ in (\ref{eq103}).

\begin{thm}
\label{thmy2p}
Using the normal chain $\{H_U^j,H_U^{j*}\}$ in (\ref{eq101}),
$\{\Gamma_j'\}$ in (\ref{eq102}), isomorphism $\alpha_{j+1}$
in (\ref{eq103}), and isomorphism $\phi_{\ell-1}$ from Theorem
\ref{thmy1}, we can recover $\tldch$ in Theorem \ref{thmy1}.
\end{thm}

\begin{prf}
Clearly we can recover $\{H_U^j\}$ in (\ref{nchu}) from the refinement
in (\ref{eq101}).  Applying $\phi_{\ell-1}$
to each term in (\ref{eq101}) we can recover $\{H_V^j\}$ in (\ref{nchv}).
We know that $H_U^j\lhd H_U$ for $-1\le j\le\ell$.
Since $\phi_{\ell-1}(H_U)=H_V$, we have $H_U^j\lhd H_U$ \ifof\
$\phi_{\ell-1}(H_U^j)=H_V^{j+1}\lhd H_V$.  Then $H_V^j\lhd H_V$
for $0\le j\le\ell$.
Similarly using (\ref{eq102}) and $\phi_{\ell-1}$, we can recover $\{\Gamma_j''\}$
in (\ref{ncgammapp}).  Apply $\phi_{\ell-1}$ to $H_U^j$ and $\Gamma_j'$
on the right hand side in (\ref{eq103}); then we can recover $\beta_{j+1}$ in
(\ref{thmy1a}).  Similarly we can recover $\beta_j^*$ from
$\alpha_j^*$ and $\eta_j''$ from $\eta_{j-1}'$.
Thus we have recovered all the assumptions in Theorem \ref{thmy1},
and we can proceed to find $\tldch$ as in Theorem \ref{thmy1}.
\end{prf}

We now show that we can find a shift group
isomorphic to $\tldch$ by using just two normal chains and
isomorphism $\alpha_{j+1}$, without any overt isomorphism $\phi_{\ell-1}$.

\begin{thm}
\label{thmy3}
Using the normal chain $\{H_U^j,H_U^{j*}\}$ in (\ref{eq101}),
$\{\Gamma_j'\}$ in (\ref{eq102}),
and isomorphism $\alpha_{j+1}$ in (\ref{eq103}), we can
recover a shift group $\hatch$ isomorphic to $\tldch$.
\end{thm}

\begin{prf}
Define $\hatch_V^{j+1}\rmdef H_U^j$ for $-1\le j<\ell$,
$\hatch_V^{j+1*}\rmdef H_U^{j*}$ for $-1\le j<\ell-1$,
$\hatcgamma_{j+1}''\rmdef \Gamma_j'$ for $-1\le j<\ell$,
and $\hatcv_0''\rmdef V_0'$.  For $0\le j<\ell$,
define the isomorphism $\hatbeta_{j+1}$,
\be
\label{thmy3bb}
\hatbeta_{j+1}:  \frac{H_U^{j+1}}{U_0'}\ra\frac{\hatch_V^{j+1}}{\hatcgamma_{j+1}''},
\ee
using $\alpha_{j+1}$ and the substitutions $\hatch_V^{j+1}=H_U^j$,
$\hatcgamma_{j+1}''=\Gamma_j'$ in the right hand side of (\ref{eq103}).
In the same way, define the isomorphisms $\hatbeta_j^*$ and ${\hat{\eta}}_j''$.
Similarly define $\hatbeta_0$ using $\alpha_0$.  For $0\le j<\ell$,
let $\hatcu_{j+1}$ be the subdirect product of $H_U^{j+1}\times\hatch_V^{j+1}$
implied by the isomorphism (\ref{thmy3bb}).
Let $\hatcu_0$ be the subdirect product of $\hatch_U^0\times\hatch_V^0$ implied by
the isomorphism $\hatbeta_0$, i.e., $\hatcu_0=U_0'\times\bone$.
Define $\hatcu_{-1}=\bone\times\bone$; define $\hatch\rmdef\hatcu_\ell$.
Define the trivial isomorphism
$\hatphi_j:  H_U^j\ra \hatch_V^{j+1}$ for $-1\le j<\ell$ by the assignment
$h\mapsto h$, $h\in H_U^j$.
Then all the conditions in Theorem \ref{thmy1} are met so we see that $\hatch$ is
a group with a \sst\ $(\hatuj,\hatcv_0,\hatphi)$, where
$\hatcv_0\rmdef\bone\times \hatcv_0''$ and
$\hatphi:  \hatch/\hatcv_0\ra\hatch/\hatcu_0$ is just the isomorphism
$\hatphi_{\ell-1}$.

We have $\hatch_V^{j+1}$ is isomorphic to the group $H_V^{j+1}$ in Theorem \ref{thmy1},
and in fact
$$
\phi_{\ell-1}(H_U^j)=\phi_{\ell-1}(\hatch_V^{j+1})=H_V^{j+1},
$$
where $\phi_{\ell-1}$ is the isomorphism in Theorem \ref{thmy1}.
Similarly $\hatcgamma_{j+1}''\simeq \Gamma_{j+1}''$ since
$$
\phi_{\ell-1}(\Gamma_j')=\phi_{\ell-1}(\hatcgamma_{j+1}'')=\Gamma_{j+1}''.
$$
Thus $\hatcu_{j+1}$, implied by the isomorphism $\hatbeta_{j+1}$ in (\ref{thmy3bb}),
is isomorphic to $\tldcu_{j+1}$, implied by the isomorphism $\beta_{j+1}$ in
(\ref{thmy1a}).  Then $\hatch\simeq\tldch$.
\end{prf}

Previously we have shown that given any reduced shift group $G$,
we can use Theorem \ref{thmx} to obtain a
subdirect product group $\tldcg$ which is a shift group.  Then we
can use Theorem \ref{thmy1} to obtain $\tldch=\tldcg$,
and finally Proposition \ref{prop35a} to obtain $H=G$.
Thus we can obtain any reduced shift group $G$ by
starting with the description in Theorem \ref{thmy1}.
In Theorem \ref{thmy3}, we have shown that we can obtain a shift
group $\hatch$ such that $\hatch\simeq\tldch$.  Using
Proposition \ref{prop35a}, the subdirect product group $\hatch$
can be converted into an abstract shift group $H'$.
It is easy to show that $H'\simeq H=G$.  Thus
using the approach in Theorem \ref{thmy3}, we can find all
reduced shift groups $G$ up to isomorphism.

Having found shift group $H'$, it is clear that isomorphism
is a sufficient condition to delineate the shift structure
of any group $G$ isomorphic to $H'$.
The following proposition shows that if two groups are isomorphic
and one of them is a shift group, then the other is a shift group
and there is a 1-1 correspondence between their shift structures.
Thus Theorem \ref{thmy3} can effectively find the \sst\ of all
reduced shift groups $G$.

\begin{prop}
\label{prop48}
Let $\phi:  G\ra H$ be an isomorphism.  Then $G$ is a shift group
with a \sst\ $(\xj,Y_0,\varphi)$ \ifof\ $H$ is a shift group
with a \sst\ $(\uj,V_0,\varphi')$, where $U_j=\phi(X_j)$,
$V_0=\phi(Y_0)$, and the diagrams in Figure \ref{fig4} commute.
In Figure \ref{fig4}, $\phi_1:  G/Y_0\ra H/V_0$ is an isomorphism naturally induced
by $\phi:  G\ra H$, and $\phi_2:  G/X_0\ra H/U_0$ is an isomorphism
naturally induced by $\phi$.
\end{prop}

\begin{figure}[h]
\centering

\begin{picture}(100,210)
%arrows
%bottom box
\put(0,70){\vector(0,-1){60}}
\put(20,80){\vector(1,0){60}}
\put(100,70){\vector(0,-1){60}}
\put(20,0){\vector(1,0){60}}
%bottom box
\put(0,190){\vector(0,-1){60}}
\put(20,200){\vector(1,0){60}}
\put(100,190){\vector(0,-1){60}}
\put(20,120){\vector(1,0){60}}
%corner labels
%bottom box
\put(0,0){\makebox(0,0){$U_jV_0/V_0$}}
\put(0,80){\makebox(0,0){$X_jY_0/Y_0$}}
\put(100,80){\makebox(0,0){$X_{j+1}/X_0$}}
\put(100,0){\makebox(0,0){$U_{j+1}/U_0$}}
%top box
\put(0,120){\makebox(0,0){$H/V_0$}}
\put(0,200){\makebox(0,0){$G/Y_0$}}
\put(100,200){\makebox(0,0){$G/X_0$}}
\put(100,120){\makebox(0,0){$H/U_0$}}
%arrow labels
%bottom box
\put(-5,40){\makebox(0,0)[r]{$\phi_1$}}
\put(50,85){\makebox(0,0)[b]{$\varphi$}}
\put(105,40){\makebox(0,0)[l]{$\phi_2$}}
\put(50,5){\makebox(0,0)[b]{$\varphi'$}}
%bottom box
\put(-5,160){\makebox(0,0)[r]{$\phi_1$}}
\put(50,205){\makebox(0,0)[b]{$\varphi$}}
\put(105,160){\makebox(0,0)[l]{$\phi_2$}}
\put(50,125){\makebox(0,0)[b]{$\varphi'$}}
\end{picture}

\caption{Commutative diagrams.}
\label{fig4}

\end{figure}
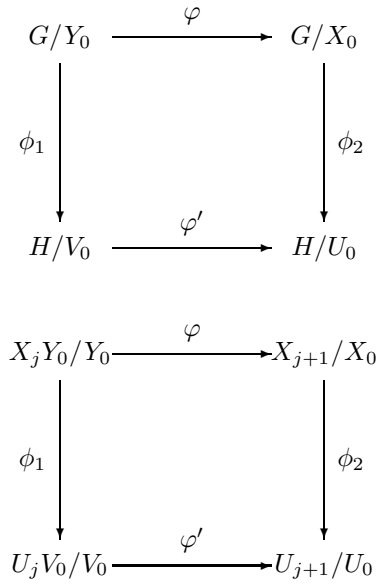

Of course the group $H_U$ in Theorems \ref{thmy2p} and \ref{thmy3}
is the state group of shift group $\tldch$ and $\hatch$, respectively.
This gives the following result.

\begin{thm}
\label{sgthm}
A group $H_U$ is the state group of a shift group that is a
subdirect product group \ifof\

(i) there is a normal chain
\begin{multline}
\label{sgcut1}
\bone=H_U^{-1}\lhd H_U^{-1*}\lhd U_0'=H_U^0\lhd H_U^{0*}\lhd H_U^1\lhd H_U^{1*}\lhd\cdots \\
\lhd H_U^j\lhd H_U^{j*}\lhd\cdots
\lhd H_U^{\ell-2}\lhd H_U^{\ell-2*}=H_U^{\ell-1}=H_U^\ell=H_U,
\end{multline}
where each $H_U^j\lhd H_U$;

(ii) there is a normal chain
$$
\bone=\Gamma_{-1}'\lhd\Gamma_0'\lhd \Gamma_1'\lhd
\cdots\lhd \Gamma_j'\lhd\cdots\lhd \Gamma_{\ell-2}'
\lhd \Gamma_{\ell-1}'=V_0',
$$
where each $\Gamma_j'\lhd H_U$ for $-1\le j<\ell$,
such that $H_U^j\cap V_0'=\Gamma_j'$ for $-1\le j<\ell$,
and $H_U^{j*}=H_U^j\Gamma_{j+1}'$ for $-1\le j<\ell-1$;

(iii) for $0\le j<\ell$, there is an isomorphism $\alpha_{j+1}$,
\be
%\label{thmy3a}
\alpha_{j+1}:  \frac{H_U^{j+1}}{U_0'}\ra\frac{H_U^j}{\Gamma_j'},
\ee
whose restriction to $H_U^j/U_0'$ is the isomorphism
$\alpha_j^*=\eta_{j-1}'\circ\alpha_j$, where
$$
\alpha_j^*:  H_U^j/U_0'\ra H_U^{j-1*}/\Gamma_j',
$$
and $\eta_{j-1}'$ is the isomorphism
$$
\eta_{j-1}': H_U^{j-1}/\Gamma_{j-1}'\ra H_U^{j-1*}/\Gamma_j'
$$
given by (\ref{wkhse1}) of Lemma \ref{wkhorse} using
$H_U^{j-1*}=H_U^{j-1}\Gamma_j'$ in the hypothesis.
Define $\alpha_0$ to be the trivial isomorphism
$\alpha_0:  H_U^0/U_0'\ra H_U^{-1}/\Gamma_{-1}'$, or
$\alpha_0:  \bone\ra\bone$.

Moreover we can find shift groups associated with $H_U$ as in Theorems
\ref{thmy2p} and \ref{thmy3}; these shift groups are isomorphic.
\end{thm}

Note that in (iii) of Theorem \ref{sgthm}, the case $j=\ell-1$ is trivial
once we have obtained $j=\ell-2$.  For $j=\ell-1$, we are required to find
an isomorphism
$$
\alpha_\ell:  H_U^\ell/U_0'\ra H_U^{\ell-1}/\Gamma_{\ell-1}',
$$
whose restriction to $H_U^{\ell-1}/U_0'$ is the isomorphism
$\alpha_{\ell-1}^*=\eta_{\ell-2}'\circ\alpha_{\ell-1}$, where
$$
\alpha_{\ell-1}^*:  H_U^{\ell-1}/U_0'\ra H_U^{\ell-2*}/\Gamma_{\ell-1}',
$$
and $\eta_{\ell-2}'$ is the isomorphism
given by (\ref{wkhse1}) of Lemma \ref{wkhorse} using
$H_U^{\ell-2*}=H_U^{\ell-2}\Gamma_{\ell-1}'$ in the hypothesis.
But the isomorphism $\eta_{\ell-2}'$ is easy to obtain from
$H_U^{\ell-2*}$.  And the case $j=\ell-2$ in (iii) gives an isomorphism
$$
\alpha_{\ell-1}:  H_U^{\ell-1}/U_0'\ra H_U^{\ell-2}/\Gamma_{\ell-2}'.
$$
Using $\alpha_{\ell-1}$ and $\eta_{\ell-2}'$ we can obtain $\alpha_{\ell-1}^*$.
Now since $H_U^{\ell-1}=H_U^\ell$ and $H_U^{\ell-2*}=H_U^{\ell-1}$, we can
trivially obtain $\alpha_\ell$ by setting $\alpha_\ell=\alpha_{\ell-1}^*$.
Thus the case $j=\ell-1$ in (iii) can be eliminated.
In addition, the group $H_U^\ell$ in (\ref{sgcut1}) is now extraneous and
can be eliminated.  This gives the following corollary.

\begin{cor}
\label{cor51}
A group $H_U$ is the state group of a shift group that is a
subdirect product group \ifof\

(i) there is a normal chain
\begin{multline}
%\label{sgcut1}
\bone=H_U^{-1}\lhd H_U^{-1*}\lhd U_0'=H_U^0\lhd H_U^{0*}\lhd H_U^1\lhd H_U^{1*}\lhd\cdots \\
\lhd H_U^j\lhd H_U^{j*}\lhd\cdots
\lhd H_U^{\ell-2}\lhd H_U^{\ell-2*}=H_U^{\ell-1}=H_U,
\end{multline}
where each $H_U^j\lhd H_U$;

(ii) there is a normal chain
$$
\bone=\Gamma_{-1}'\lhd\Gamma_0'\lhd \Gamma_1'\lhd
\cdots\lhd \Gamma_j'\lhd\cdots\lhd \Gamma_{\ell-2}'
\lhd \Gamma_{\ell-1}'=V_0',
$$
where each $\Gamma_j'\lhd H_U$ for $-1\le j<\ell$,
such that $H_U^j\cap V_0'=\Gamma_j'$ for $-1\le j<\ell$,
and $H_U^{j*}=H_U^j\Gamma_{j+1}'$ for $-1\le j<\ell-1$;

(iii) for $0\le j<\ell-1$, there is an isomorphism $\alpha_{j+1}$,
\be
%\label{thmy3a}
\alpha_{j+1}:  \frac{H_U^{j+1}}{U_0'}\ra\frac{H_U^j}{\Gamma_j'},
\ee
whose restriction to $H_U^j/U_0'$ is the isomorphism
$\alpha_j^*=\eta_{j-1}'\circ\alpha_j$, where
$$
\alpha_j^*:  H_U^j/U_0'\ra H_U^{j-1*}/\Gamma_j',
$$
and $\eta_{j-1}'$ is the isomorphism
$$
\eta_{j-1}': H_U^{j-1}/\Gamma_{j-1}'\ra H_U^{j-1*}/\Gamma_j'
$$
given by (\ref{wkhse1}) of Lemma \ref{wkhorse} using
$H_U^{j-1*}=H_U^{j-1}\Gamma_j'$ in the hypothesis.
Define $\alpha_0$ to be the trivial isomorphism
$\alpha_0:  H_U^0/U_0'\ra H_U^{-1}/\Gamma_{-1}'$, or
$\alpha_0:  \bone\ra\bone$.
\end{cor}

Thus we can find all reduced shift groups $G$ up to isomorphism
by first finding all state groups $H_U$ with the properties in
Corollary \ref{cor51}, and then finding associated shift groups as in
Theorem \ref{thmy3}.

Note that even though $|X_0\cap Y_0|=1$ for $G$, we do not necessarily have
$|U_0'\cap V_0'|=1$ for $H_U$.  From (ii) of Corollary \ref{cor51},
we have $H_U^0\cap V_0'=\Gamma_0'$, which implies $U_0'\cap V_0'=\Gamma_0'$.
Note that the state group
has one less degree of freedom than the shift
group; i.e., we have $H_U^{\ell-1}=H_U$.  We can think of the state
group as being ``$\ell-1$-controllable'' \cite{FT}.

Corollary \ref{cor51} suggests a method to construct any state group $H_U$.
We start with a group $U_0'$ and then construct
a chain of groups $H_U^j$ that converges to $H_U^{\ell-2}$;
then we find $H_U^{\ell-2*}=H_U$.
Roughly, we can do this as follows (in the rough sketch here, we
neglect any discussion of normality requirements).  Let $H_U^0=U_0'$ and define
$\Gamma_{-1}'\rmdef\bone$.  Then $\alpha_0:  H_U^0/U_0'\ra H_U^{-1}/\Gamma_{-1}'$,
which is just the isomorphism $\alpha_0:  \bone\ra\bone$.
We have $H_U^{-1*}=H_U^{-1}\Gamma_0'=\Gamma_0'$.  Thus we have
obtained $H_U^0$, $\Gamma_0'$, and $\alpha_0$.

In general assume we have found $H_U^j$, $\Gamma_j'$, and
an isomorphism $\alpha_j$.  We now show how to find $H_U^{j+1}$,
$\Gamma_{j+1}'$, and an isomorphism
$\alpha_{j+1}$ that satisfies
the restrictions in (iii) of Corollary \ref{cor51}.
Please refer to Figure \ref{ladder} where
isomorphism $\alpha_j$ is shown in the bottom line.
Note that subgroup $H_U^{j-1*}$ of $H_U^j$ satisfies
$H_U^{j-1*}=H_U^{j-1}\Gamma_j'$.  Then by (\ref{wkhse1}) of Lemma \ref{wkhorse}
there is an isomorphism $\eta_{j-1}'$,
$$
\eta_{j-1}': H_U^{j-1}/\Gamma_{j-1}'\ra H_U^{j-1*}/\Gamma_j'.
$$
This gives an isomorphism $\alpha_j^*$,
$$
\alpha_j^*:  H_U^j/U_0'\ra H_U^{j-1*}/\Gamma_j',
$$
which is the next line of Figure \ref{ladder}.
Now construct a group $H_U^{j*}=H_U^j\Gamma_{j+1}'$, where
$\Gamma_{j+1}'\supset\Gamma_j'$, such that
$H_U^{j*}$ is an extension of $U_0'$ by $\dotch/\Gamma_j'$, where
$$
\frac{H_U^{j-1*}}{\Gamma_j'}\subset\frac{\dotch}{\Gamma_j'}\subset\frac{H_U^j}{\Gamma_j'}.
$$
In other words there is an isomorphism
$$
\alpha_j^{**}:  \frac{H_U^{j*}}{U_0'}\ra\frac{\dotch}{\Gamma_j'},
$$
which is the next line of Figure \ref{ladder}.
We require that the restriction of $\alpha_j^{**}$ to $H_U^j/U_0'$
is the isomorphism $\alpha_j^*$.  Now find a group
$H_U^{j+1}$ such that $H_U^{j+1}\supset H_U^{j*}$ and $H_U^{j+1}$
is an extension of $U_0'$ by $H_U^j/\Gamma_j'$; in other words
there is an isomorphism
$$
\alpha_{j+1}:  \frac{H_U^{j+1}}{U_0'}\ra\frac{H_U^j}{\Gamma_j'},
$$
which is the top line in Figure \ref{ladder}.
We require that the restriction of $\alpha_{j+1}$ to $H_U^{j*}/U_0'$
is the isomorphism $\alpha_j^{**}$.  In general
this restriction is easy to meet since $H_U^{j+1}\supset H_U^{j*}$.

Thus we have obtained $H_U^{j+1}$, $\Gamma_{j+1}'$, and
an isomorphism $\alpha_{j+1}$ that meets the restrictions in (iii)
of Corollary \ref{cor51}.  Continuing in this way gives $H_U^{\ell-2}$,
$\Gamma_{\ell-2}'$, and isomorphism
$$
\alpha_{\ell-2}:  H_U^{\ell-2}/U_0'\ra H_U^{\ell-3}/\Gamma_{\ell-3}'.
$$
In the last step, the top two lines of Figure \ref{ladder} are the same, and
the algorithm becomes degenerate.  We have $H_U^{\ell-2*}=H_U^{\ell-1}$,
$\dotch=H_U^{\ell-2}$, and $\alpha_{\ell-2}^{**}=\alpha_{\ell-1}$.
First find
$$
\alpha_{\ell-2}^*:  H_U^{\ell-2}/U_0'\ra H_U^{\ell-3*}/\Gamma_{\ell-2}'.
$$
Next construct a group $H_U^{\ell-2*}=H_U^{\ell-2}\Gamma_{\ell-1}'$
such that there is an isomorphism
$$
\alpha_{\ell-2}^{**}:  \frac{H_U^{\ell-2*}}{U_0'}\ra
\frac{H_U^{\ell-2}}{\Gamma_{\ell-2}'}.
$$
We require that the restriction of $\alpha_{\ell-2}^{**}$ to
$H_U^{\ell-2}/U_0'$ is $\alpha_{\ell-2}^*$.  Again, since
the last step is degenerate,
$\alpha_{\ell-2}^{**}$ is $\alpha_{\ell-1}$ and $H_U^{\ell-2*}$ is
$H_U^{\ell-1}$, which is just state group $H_U$.

\begin{figure}[htbp]
\centering
%\vspace{3ex}

$$
\begin{array}{rlll}
\alpha_{j+1}:  & H_U^{j+1}/U_0' & \ra & H_U^j/\Gamma_j'          \\
\alpha_j^{**}: & H_U^{j*}/U_0'  & \ra & \dotch/\Gamma_j'        \\
\alpha_j^*:    & H_U^j/U_0'     & \ra & H_U^{j-1*}/\Gamma_j'     \\
\alpha_j:      & H_U^j/U_0'     & \ra & H_U^{j-1}/\Gamma_{j-1}'
\end{array}
$$

\caption{Isomorphisms and groups used in construction of state group $H_U$.}
\label{ladder}

\end{figure}

For shift group $G$, we saw that $X_0$ and the normal chain $\{X_j\cap Y_0\}$
were related.  This suggests that for a state group, $U_0'$ and $\{\Gamma_j'\}$
are related.  We now prove this result.  This approach shows finer details of
the group $H_U$ and gives a more elaborate version of Figure \ref{ladder},
allowing us to improve Corollary \ref{cor51} and the algorithm.

\begin{lem}
\label{sglem8}
Let $H_U$ be the state group of a shift group.  Fix $j$, $-1\le j<\ell-2$.
If there is a normal chain
\be
\label{sgcapa}
H_U^{j+1}=Q_{j+1}^0\lhd Q_{j+1}^1\lhd Q_{j+1}^2\lhd\cdots\lhd Q_{j+1}^{p-1}\lhd Q_{j+1}^p=H_U^{j+2},
\ee
then there is a normal chain
\begin{multline}
\label{sgcapb}
H_U^j\lhd Q_j^a\lhd\cdots\lhd Q_j^b\lhd Q_j^0\lhd Q_j^1\lhd Q_j^2\lhd\cdots \\
\lhd Q_j^{p-1}\lhd Q_j^p=H_U^{j+1},
\end{multline}
where $Q_j^0=H_U^{j*}$ and the normal chain
\be
\label{sgspecnc}
H_U^j\lhd Q_j^a\lhd\cdots\lhd Q_j^b\lhd Q_j^0
\ee
is an arbitrary refinement of the trivial normal chain $H_U^j\lhd Q_j^0$.
We have $H_U^j=Q_j^0$ \ifof\ $H_U^j=H_U^{j*}$; in this case any refinement in
(\ref{sgspecnc}) is trivial.  Although there is no restriction on the choice of the
normal chain in (\ref{sgspecnc}), there are dependent relations among the
$Q_j^n$ and $Q_{j+1}^n$, $0\le n\le p$.  We have
\be
\label{sgeqx4}
\frac{Q_j^n}{Q_j^m}\simeq\frac{Q_{j+1}^n}{Q_{j+1}^m}
\ee
for $m,n$ satisfying $0\le m\le n\le p$.
Moreover $Q_j^n\lhd H_U$ if $Q_{j+1}^n\lhd H_U$, for $n$ satisfying $0\le n\le p$.
In addition, $Q_j^n$ and $Q_{j+1}^n$ are related by the isomorphism
$\alpha_{j+2}$,
\be
\label{sgeqx5}
\alpha_{j+2}(Q_{j+1}^n/U_0')=Q_j^n/\Gamma_{j+1}',
\ee
for $n$ satisfying $0\le n\le p$.

Conversely, if there is a normal chain as in (\ref{sgcapb}) with
$Q_j^0=H_U^{j*}$, then there is a normal chain as in (\ref{sgcapa}),
and $Q_{j+1}^n\lhd H_U$ if $Q_j^n\lhd H_U$, for $n$ satisfying $0\le n\le p$,
and properties (\ref{sgeqx4})-(\ref{sgeqx5}) hold.
\end{lem}

\begin{prf}
Fix $j$, $-1\le j<\ell-2$.
We first show that if (\ref{sgcapa}) holds, then (\ref{sgcapb}) holds.
As in (\ref{sgcapa}), let
$$
Q_{j+1}^0\lhd Q_{j+1}^1\lhd Q_{j+1}^2\lhd\cdots\lhd Q_{j+1}^p
$$
be a normal chain with each $Q_{j+1}^n\lhd H_U$.
We know $U_0'\lhd H_U$ and $U_0'\subset Q_{j+1}^0$.
Then from the correspondence theorem, there is a normal chain
$$
\frac{Q_{j+1}^0}{U_0'}\lhd \frac{Q_{j+1}^1}{U_0'}\lhd \frac{Q_{j+1}^2}{U_0'}\lhd
\cdots\lhd \frac{Q_{j+1}^p}{U_0'}
$$
where
\be
\label{sgeqx1}
\frac{Q_{j+1}^n/U_0'}{Q_{j+1}^m/U_0'}\simeq\frac{Q_{j+1}^n}{Q_{j+1}^m},
\ee
for $m\ge 0,n\ge 0$ satisfying $0\le m\le n\le p$, and each $Q_{j+1}^n/U_0'\lhd H_U/U_0'$.

Since for a state group there is an isomorphism
$\alpha_{j+2}:  H_U^{j+2}/U_0'\ra H_U^{j+1}/\Gamma_{j+1}'$, for each
$n$, $0\le n\le p$, there is a subgroup $\dotcq_j^n/\Gamma_{j+1}'$ such that
$\alpha_{j+2}(Q_{j+1}^n/U_0')=\dotcq_j^n/\Gamma_{j+1}'$.
Thus the isomorphism $\alpha_{j+2}$ gives
a normal chain
\be
\label{sgeqxx}
\frac{\dotcq_j^0}{\Gamma_{j+1}'}\lhd \frac{\dotcq_j^1}{\Gamma_{j+1}'}\lhd
\frac{\dotcq_j^2}{\Gamma_{j+1}'}\lhd\cdots\lhd \frac{\dotcq_j^p}{\Gamma_{j+1}'},
\ee
where each $\dotcq_j^n/\Gamma_{j+1}'\lhd H_U^{j+1}/\Gamma_{j+1}'$, and
\be
\label{sgeqx2}
\frac{\dotcq_j^n/\Gamma_{j+1}'}{\dotcq_j^m/\Gamma_{j+1}'}\simeq\frac{Q_{j+1}^n/U_0'}{Q_{j+1}^m/U_0'}.
\ee
Since $H_U$ is a state group, we have $\dotcq_j^0/\Gamma_{j+1}'=H_U^{j*}/\Gamma_{j+1}'$ and
$\dotcq_j^p/\Gamma_{j+1}'=H_U^{j+1}/\Gamma_{j+1}'$.

Consider the natural map $\nu_{j+1}:  H_U^{j+1}\ra H_U^{j+1}/\Gamma_{j+1}'$ defined
by the assignment $h\mapsto h\Gamma_{j+1}'$.
Define $Q_j^n\rmdef (\nu_{j+1})^{-1}(\dotcq_j^n/\Gamma_{j+1}')$.
Then $Q_j^0=H_U^{j*}$ and $Q_j^p=H_U^{j+1}$.  Then using (\ref{sgeqxx}) and the
correspondence theorem, we have a normal chain
\be
\label{sgeqx2a}
Q_j^0\lhd Q_j^1\lhd Q_j^2\lhd\cdots\lhd Q_j^p,
\ee
where
\be
\label{sgeqx3}
\frac{Q_j^n}{Q_j^m}\simeq\frac{\dotcq_j^n/\Gamma_{j+1}'}{\dotcq_j^m/\Gamma_{j+1}'}.
\ee
Since $Q_j^0=H_U^{j*}$, we have $H_U^j\subset Q_j^0$, and combining this
with (\ref{sgeqx2a}) gives (\ref{sgcapb}).
From the correspondence theorem, we have each $Q_j^n\lhd H_U$.  Collecting (\ref{sgeqx1}), (\ref{sgeqx2}),
and (\ref{sgeqx3}) gives (\ref{sgeqx4}).  Finally we have that
(\ref{sgeqx5}) holds by construction.

Now assume (\ref{sgcapb}) holds.  We can show that (\ref{sgcapa}) holds
by essentially reversing the above steps.
\end{prf}

\newcommand{\huj}{{\{H_U^j\}}}

We see there are two cases to consider in Lemma \ref{sglem8} depending
on whether $H_U^{j*}=H_U^j$ or $H_U^{j*}>>H_U^j$.
Formally, we introduce a parameter $\epsilon_j$ for $-1\le j<\ell-1$.
We set $\epsilon_j=1$ if $H_U^{j*}>>H_U^j$, and $\epsilon_j=0$ if $H_U^{j*}=H_U^j$.

Note that parameter $\epsilon_j$ is not the same as parameter $\varepsilon_j$.
We have $H_U^{j*}>>H_U^j$ \ifof\ $H_V^{j+1*}>>H_V^{j+1}$.
Therefore $H_U^{j*}>>H_U^j$ \ifof\ $\tldcu_{j+1}^*>>\tldcu_{j+1}$ in $\tldch$.
Under the isomorphism $\tldch\stackrel{\zeta}{\simeq} H$,
we have $\tldcu_{j+1}^*>>\tldcu_{j+1}$ \ifof\
$U_{j+1}^*>>U_{j+1}$ in $H$.  Therefore $\epsilon_j$ corresponds
to $\varepsilon_{j+1}$.  Note that $\epsilon_{\ell-2}=\varepsilon_{\ell-1}=1$ always.
We have $\epsilon_{-1}=\varepsilon_0$.  We have $\epsilon_{-1}=0$ \ifof\
$\Gamma_0'=\bone$.  We always have $\varepsilon_{-1}=0$ since
$U_{-1}^*=U_0\cap V_0=\bone$ for a reduced shift group.

In the next theorem, we use Lemma \ref{sglem8} to find a refinement of
(\ref{sgcut1}).  It is convenient to write the refinement using slightly
different notation than in Lemma \ref{sglem8}.  Thus in place of (\ref{sgcapa}),
we write the portion of the refinement between $H_U^{j+1}$ and $H_U^{j+2}$ as
\begin{multline}
\label{sgreft2}
H_U^{j+1}=H_U^{j+1,(k_{j+1})}\lhd H_U^{j+1,(k_{j+1}+1)}
\lhd H_U^{j+1,(k_{j+1}+2)}\lhd\cdots  \\
\lhd H_U^{j+1,(\ell'-1)}\lhd H_U^{j+1,(\ell')}=H_U^{j+2},
\end{multline}
where $k_{j+1}$ and $\ell'$ are positive integers.
Using (\ref{sgreft2}) in Lemma \ref{sglem8}, we obtain the portion of the
refinement between $H_U^j$ and $H_U^{j+1}$ as
\begin{multline}
\label{sgreft1}
H_U^j\lhd H_U^{j,(k_{j+1})}\lhd H_U^{j,(k_{j+1}+1)}\lhd H_U^{j,(k_{j+1}+2)}\lhd\cdots  \\
\lhd H_U^{j,(\ell'-1)}\lhd H_U^{j,(\ell')}=H_U^{j+1},
\end{multline}
where $H_U^{j,(k_{j+1})}=H_U^{j*}$.
We only use Lemma \ref{sglem8} for a trivial refinement in (\ref{sgspecnc}),
that is, when $H_U^j=Q_j^a=\cdots=Q_j^b$.  In (\ref{sgreft1}),
we have $H_U^{j,(k_{j+1})}=H_U^{j*}$ if $\epsilon_j=1$, and
$H_U^{j,(k_{j+1})}=H_U^{j*}=H_U^j$ if $\epsilon_j=0$.

In general for each $j$, $-1\le j\le\ell-2$, we define a refinement
in which the superscript
$m$ of $H_U^{j,(m)}$ runs from integer $k_j$ to integer $\ell'$.
For $0\le j\le\ell-1$, we define
$H_U^{j-1,(\ell')}\rmdef H_U^j\rmdef H_U^{j,(k_j)}$;
then $H_U^{\ell-2,(\ell')}=H_U^{\ell-1}=H_U^{\ell-1,(k_{\ell-1})}$.
We also define $H_U^{-1}\rmdef H_U^{-1,(k_{-1})}$.
In this notation,
the portion of the refinement between $H_U^j$ and $H_U^{j+1}$ is
\begin{multline}
\label{sgreft3}
H_U^j=H_U^{j,(k_j)}\lhd H_U^{j,(k_j+1)}\lhd H_U^{j,(k_j+2)}\lhd\cdots  \\
\lhd H_U^{j,(\ell'-1)}\lhd H_U^{j,(\ell')}=H_U^{j+1}.
\end{multline}
Comparing (\ref{sgreft1}) and (\ref{sgreft3}) shows that we must have
$H_U^j=H_U^{j,(k_j)}=H_U^{j,(k_{j+1})}=H_U^{j*}$ if $\epsilon_j=0$ and
$H_U^{j,(k_j+1)}=H_U^{j,(k_{j+1})}=H_U^{j*}$ if $\epsilon_j=1$.
This means $k_j+\epsilon_j=k_{j+1}$.  If we use the above
procedure and apply Lemma \ref{sglem8} recursively starting with
the normal chain
$$
H_U^{\ell-2}=H_U^{\ell-2,(k_{\ell-2})}\lhd H_U^{\ell-2,(\ell')}=H_U^{\ell-1}=H_U,
$$
we obtain
\be
\label{sgthm14a1plus}
k_j=\ell'-\sum_{j\le i<\ell-1}\epsilon_i
\ee
for $-1\le j<\ell-1$.  Define
$$
\ell'\rmdef \sum_{-1\le i<\ell-1}\epsilon_i.
$$
Then from (\ref{sgthm14a1plus}) we see $k_{-1}=0$.  If $j=\ell-1$,
we define $k_j=k_{\ell-1}\rmdef\ell'$ trivially.
Thus as $j$ runs from $-1$ to $\ell-1$, $k_j$ takes all values in
the range $[0,\ell']$.
Since
$$
\sum_{-1\le i<\ell-1}\epsilon_i=\sum_{-1\le i<\ell}\varepsilon_i,
$$
we see the above definition of $\ell'$ is consistent with the previous definition.

\begin{thm}
\label{sgthm14a}
Let a shift group have a state group $H_U$.
There is a refinement of $\huj$, and of the normal chain in (\ref{sgcut1}),
given by
\begin{multline}
\label{sgeqbb}
H_U^{-1}=H_U^{-1,(k_{-1})}\lhd\cdots\lhd H_U^{-1,(\ell')}=H_U^0=H_U^{0,(k_0)}\lhd\cdots  \\
\lhd H_U^{j-1,(\ell')}=H_U^j=H_U^{j,(k_j)}\lhd H_U^{j,(k_j+1)}\lhd H_U^{j,(k_j+2)}\lhd\cdots  \\
\lhd H_U^{j,(\ell'-1)}\lhd H_U^{j,(\ell')}=H_U^{j+1}=H_U^{j+1,(k_{j+1})}\lhd\cdots  \\
\lhd H_U^{\ell-2,(k_{\ell-2})}\lhd H_U^{\ell-2,(k_{\ell-2}+1)}=H_U^{\ell-2,(\ell')}=H_U^{\ell-1}=  \\
H_U^{\ell-1,(k_{\ell}-1)}=H_U,
\end{multline}
where each $H_U^{j,(k_j+n)}\lhd H_U$ and $H_U^{j,(k_j+1)}=H_U^{j*}$ if $\epsilon_j=1$.
Moreover
\be
\label{sgthm14a1}
\frac{H_U^{-1,(k_j+n)}}{H_U^{-1,(k_j+m)}}\simeq\frac{H_U^{j,(k_j+n)}}{H_U^{j,(k_j+m)}}
\ee
for $-1\le j<\ell-1$ and $m,n$ satisfying $k_j\le k_j+m\le k_j+n\le\ell'$.
In addition, the isomorphism $\alpha_{j+2}$ satisfies
\be
\label{sgthm14a2}
\alpha_{j+2}(H_U^{j+1,(k_{j+1}+n)}/U_0')=H_U^{j,(k_j+\epsilon_j+n)}/\Gamma_{j+1}'
\ee
for $-1\le j<\ell-2$ and $n$ satisfying $k_{j+1}\le k_{j+1}+n\le\ell'$.
\end{thm}

\begin{prf}
Starting from the normal chain
$H_U^{\ell-2}=H_U^{\ell-2,(k_{\ell-2})}\lhd H_U^{\ell-2,(\ell')}=H_U^{\ell-1}$,
where $H_U^{\ell-2}\lhd H_U$ and $H_U^{\ell-1}\lhd H_U$, we can use Lemma \ref{sglem8}
to go `backwards' and for each $j$, $-1\le j<\ell-2$, obtain a
normal chain from $H_U^j$ to $H_U^{j+1}$ as in (\ref{sgeqbb}),
where  each $H_U^{j,(k_j+n)}\lhd H_U$ for $n$ satisfying
$k_j\le k_j+n\le\ell'$, and $H_U^{j,(k_j+1)}=H_U^{j*}$ if $\epsilon_j=1$.

Since $k_{j+1}=k_j+\epsilon_j$, we can restate (\ref{sgeqx5}) of Lemma \ref{sglem8} as in
(\ref{sgthm14a2}), for $n$ satisfying $k_{j+1}\le k_{j+1}+n\le\ell'$.

It only remains to show (\ref{sgthm14a1}).  We can do this by induction.
We assume (\ref{sgthm14a1}) holds for $q+1$, that is, we assume
\be
\label{sgzeq40}
\frac{H_U^{q+1,(k_j+n)}}{H_U^{q+1,(k_j+m)}}\simeq\frac{H_U^{j,(k_j+n)}}{H_U^{j,(k_j+m)}}
\ee
for $q+1\le j<\ell-1$ and $m,n$ satisfying $k_j\le k_j+m\le k_j+n\le\ell'$.
Note that the left hand side of (\ref{sgzeq40}) is well defined since
$k_{q+1}\le k_j$ for $q+1\le j$.
Then we show (\ref{sgthm14a1}) holds for $q$, that is, we show
\be
\label{sgzeq41}
\frac{H_U^{q,(k_j+n)}}{H_U^{q,(k_j+m)}}\simeq\frac{H_U^{j,(k_j+n)}}{H_U^{j,(k_j+m)}}
\ee
for $q\le j<\ell-1$ and $m,n$ satisfying $k_j\le k_j+m\le k_j+n\le\ell'$.

Assume that $j$ satisfies $q+1\le j<\ell-1$ and $m,n$ satisfy
$k_j\le k_j+m\le k_j+n\le\ell'$.  Assume that (\ref{sgzeq40}) holds.
We can write the portion of the normal chain in (\ref{sgeqbb})
between $H_U^q$ and $H_U^{q+1}$ as
\begin{multline}
\label{sgzeq42minus}
H_U^q=
H_U^{q,(k_q)}\lhd H_U^{q,(k_q+1)}\lhd H_U^{q,(k_q+2)}\lhd\cdots  \\
\lhd H_U^{q,(\ell'-1)}\lhd H_U^{q,(\ell')}=H_U^{q+1},
\end{multline}
and between $H_U^{q+1}$ and $H_U^{q+2}$ as
\begin{multline}
\label{sgzeq42}
H_U^{q+1}=
H_U^{q+1,(k_{q+1})}\lhd H_U^{q+1,(k_{q+1}+1)}\lhd H_U^{q+1,(k_{q+1}+2)}\lhd\cdots  \\
\lhd H_U^{q+1,(\ell'-1)}\lhd H_U^{q+1,(\ell')}=H_U^{q+2}.
\end{multline}
Then using Lemma \ref{sglem8} with (\ref{sgzeq42}) in place of (\ref{sgcapa}) and
(\ref{sgzeq42minus}) in place of (\ref{sgcapb}), we have from (\ref{sgeqx4})
\be
\label{sgzeq43}
\frac{H_U^{q,(k_j+n)}}{H_U^{q,(k_j+m)}}\simeq\frac{H_U^{q+1,(k_j+n)}}{H_U^{q+1,(k_j+m)}}.
\ee
Note that all terms in (\ref{sgzeq43}) are well defined since
$k_q\le k_{q+1}\le k_j$ for $q+1\le j$.
Combining (\ref{sgzeq43}) with (\ref{sgzeq40}) gives
\be
\label{sgzeq44}
\frac{H_U^{q,(k_j+n)}}{H_U^{q,(k_j+m)}}\simeq\frac{H_U^{j,(k_j+n)}}{H_U^{j,(k_j+m)}}.
\ee
We know that (\ref{sgzeq44}) holds
for $q+1\le j<\ell-1$ and $m,n$ satisfying $k_j\le k_j+m\le k_j+n\le\ell'$.
But (\ref{sgzeq44}) also holds trivially for $j=q$.  Then (\ref{sgzeq44}) holds
for $q\le j<\ell-1$ and $m,n$ satisfying $k_j\le k_j+m\le k_j+n\le\ell'$,
giving (\ref{sgzeq41}).

We start the induction by proving (\ref{sgzeq41}) for $q=\ell-3$.
But from Lemma \ref{sglem8}, we know there are
normal chains $H_U^{\ell-2}\lhd H_U^{\ell-1}$ and
$H_U^{\ell-3}\lhd H_U^{\ell-3*}\lhd H_U^{\ell-2}$ with
$$
\frac{H_U^{\ell-2}}{H_U^{\ell-3*}}\simeq\frac{H_U^{\ell-1}}{H_U^{\ell-2}}.
$$
Rewriting this as
$$
\frac{H_U^{\ell-3,(\ell')}}{H_U^{\ell-3,(k_{\ell-2})}}\simeq
\frac{H_U^{\ell-2,(\ell')}}{H_U^{\ell-2,(k_{\ell-2})}}
$$
gives (\ref{sgzeq41}) for $q=\ell-3$.
\end{prf}

We can illustrate Theorem \ref{sgthm14a} as previously done for
Theorem \ref{thm14a} in Figure \ref{fig1a}.

We are particularly interested in the portion of the normal chain from
$H_U^{-1}$ to $H_U^0$:
\begin{multline}
\label{sgnrml1}
H_U^{-1}=H_U^{-1,(k_{-1})}\lhd H_U^{-1,(k_{-1}+1)}\lhd\cdots\lhd H_U^{-1,(k_{-1}+n)}\lhd\cdots \\
\lhd H_U^{-1,(\ell'-1)}\lhd H_U^{-1,(\ell')}=H_U^0.
\end{multline}
In (\ref{sgnrml1}),
the superscript $m$ of $H_U^{-1,(m)}$ takes all values in the interval
$[k_{-1},\ell']$ or $[0,\ell']$.  Using (\ref{sgthm14a1plus}), for $j$
satisfying $-1\le j\le\ell-1$, we know $k_j$ takes all values in the interval
$[0,\ell']$.  Then for $-1\le j\le\ell-1$, the term $H_U^{-1,(k_j)}$
appears in (\ref{sgnrml1}), and we can make the definition
$$
\Delta_j'\rmdef H_U^{-1,(k_j)}.
$$
Then
\be
\label{sgr0}
H_U^{-1}=\Delta_{-1}'\lhd \Delta_0'\lhd\cdots\lhd\Delta_j'\lhd
\cdots\lhd\Delta_{\ell-1}'=H_U^0
\ee
is a refinement of (\ref{sgnrml1}) which at most just repeats terms in
(\ref{sgnrml1}).  Since each $H_U^{-1,(k_{-1}+n)}\lhd H_U$, we know that each
$\Delta_j'\lhd H_U$.

Given a state group $H_U$,
the normal chain in (\ref{sgeqbb}) is uniquely determined, and so the normal
chains (\ref{sgnrml1}) and (\ref{sgr0}) are uniquely determined.  We
say the normal chain in (\ref{sgr0}) is a {\it signature chain} of state group $H_U$.
We now give some properties of the signature chain.

\begin{thm}
\label{sgthm3}
Let a shift group have a state group $H_U$.  Fix $j$, $-1\le j<\ell-1$.
The signature chain of the state group has the property that
\be
\label{sgr1}
\frac{H_U^{j+1}}{H_U^j}\simeq\frac{H_U^0}{\Delta_j'},
\ee
\be
\label{sgr2}
\frac{H_U^{j+1}}{H_U^{j*}}\simeq\frac{H_U^0}{\Delta_{j+1}'},
\ee
and
\be
\label{sgr3}
\frac{H_U^{j*}}{H_U^j}\simeq\frac{\Delta_{j+1}'}{\Delta_j'}.
\ee
We have
\be
\label{sgr4b}
\Delta_0'=H_U^{-1*}=\Gamma_0',
\ee
\be
\label{sgr4}
\frac{\Delta_{j+1}'}{\Delta_j'}\simeq\frac{\Gamma_{j+1}'}{\Gamma_j'},
\ee
and
\be
\label{sgr5}
|\Delta_{j+1}'|=|\Gamma_{j+1}'|.
\ee
\end{thm}

\begin{prf}
Results (\ref{sgr1})-(\ref{sgr3}) follow from (\ref{sgthm14a1}) of Theorem
\ref{sgthm14a} using the definition of $\Delta_j'$.

We now show (\ref{sgr4b}).  We have $H_U^{-1,(k_{-1}+1)}=H_U^{-1*}$
if $\epsilon_{-1}=1$, and
$H_U^{-1,(k_{-1})}=H_U^{-1*}=H_U^{-1}$ if $\epsilon_{-1}=0$.
Also $k_0$ and $k_{-1}$ are related by $k_0=k_{-1}+\epsilon_{-1}$.
Thus $H_U^{-1,(k_0)}=H_U^{-1*}$ if $\epsilon_{-1}=1$ or $\epsilon_{-1}=0$.
But $\Delta_0'=H_U^{-1,(k_0)}$ by definition,
and $H_U^{-1*}=H_U^{-1}\Gamma_0'=\Gamma_0'$ using (ii) of Theorem \ref{sgthm}.
Then (\ref{sgr4b}) follows.

Now use Lemma \ref{wkhorse} with $Q'=\Gamma_j'$; $Q=H_U^j$, $R'=\Gamma_{j+1}'$,
and $R=H_U^{j*}$.  The conditions in Lemma \ref{wkhorse} are satisfied
because $H_U$ is a state group.  Then (\ref{wkhse2}) of Lemma \ref{wkhorse}
gives
\be
\label{sgr6}
\frac{H_U^{j*}}{H_U^j}\simeq\frac{\Gamma_{j+1}'}{\Gamma_j'}.
\ee
Combining (\ref{sgr3}) and (\ref{sgr6}) gives (\ref{sgr4}).  Now use induction with
(\ref{sgr4b}) and (\ref{sgr4}) to obtain (\ref{sgr5}).
\end{prf}

{\it Remark:}  Note from (\ref{sgr4}) that if
$\Delta_{-1}'=\cdots=\Delta_j'=\bone$ and $\Delta_{j+1}'\ne\bone$,
then $\Gamma_{-1}'=\cdots=\Gamma_j'=\bone$ and
$\Delta_{j+1}'\simeq\Gamma_{j+1}'$.
Since $H_U^{\ell-2}<< H_U^{\ell-2*}$,
we always have $|\Delta_{\ell-2}'|<|\Delta_{\ell-1}'|$.

We have the following easy corollary of Theorem \ref{sgthm3}.

\begin{cor}
\label{sgcor11}
If $H_U$ is a state group, the factor groups $H_U^{j+1}/H_U^j$ in the
normal chain $\huj$ are abelian if $U_0'=H_U^0$ is abelian.  In this case then,
$\huj$ is a solvable series and $H_U$ is solvable.
\end{cor}

%%%%%%%%%%%%%%%%%%insert here

We can now include the results of Theorem \ref{sgthm14a} and Theorem \ref{sgthm3}
in Corollary \ref{cor51}.

\begin{thm}
\label{thm60}
A group $H_U$ is the state group of a shift group that is a
subdirect product group \ifof\

(i) there is a normal chain
\begin{multline}
%\label{sgeqbb}
H_U^{-1}=H_U^{-1,(k_{-1})}\lhd\cdots\lhd H_U^{-1,(\ell')}=H_U^0=H_U^{0,(k_0)}\lhd\cdots  \\
\lhd H_U^{j-1,(\ell')}=H_U^j=H_U^{j,(k_j)}\lhd H_U^{j,(k_j+1)}\lhd H_U^{j,(k_j+2)}\lhd\cdots  \\
\lhd H_U^{j,(\ell'-1)}\lhd H_U^{j,(\ell')}=H_U^{j+1}=H_U^{j+1,(k_{j+1})}\lhd\cdots  \\
\lhd H_U^{\ell-2,(k_{\ell-2})}\lhd H_U^{\ell-2,(k_{\ell-2}+1)}=H_U^{\ell-2,(\ell')}=  \\
H_U^{\ell-2*}=H_U^{\ell-1}=H_U,
\end{multline}
where each $H_U^{j,(k_j+n)}\lhd H_U$ and $H_U^{j,(k_j+1)}=H_U^{j*}$ if $\epsilon_j=1$;

(ii) there is a refinement of the portion of the normal chain from
$\bone=H_U^{-1}$ to $U_0'=H_U^0$, given by
$$
\bone=\Delta_{-1}'\lhd\Delta_0'\lhd \Delta_1'\lhd \cdots \Delta_j'\lhd\cdots\lhd\Delta_{\ell-1}'=U_0'=H_U^0,
$$
where each $\Delta_j'\lhd H_U$ and $\Delta_j'\rmdef H_U^{-1,(k_j)}$ for $-1\le j<\ell$;

(iii) there is a normal chain
$$
\bone=\Gamma_{-1}'\lhd\Gamma_0'\lhd \Gamma_1'\lhd
\cdots\lhd \Gamma_j'\lhd\cdots\lhd \Gamma_{\ell-2}'
\lhd \Gamma_{\ell-1}'=V_0',
$$
where each $\Gamma_j'\lhd H_U$ for $-1\le j<\ell$,
such that $H_U^j\cap V_0'=\Gamma_j'$ for $-1\le j<\ell$,
and $H_U^{j*}=H_U^j\Gamma_{j+1}'$ for $-1\le j<\ell-1$, and
$$
\Gamma_{j+1}'/\Gamma_j'\simeq\Delta_{j+1}'/\Delta_j'
$$
for $-1\le j<\ell-1$;

(iv) for $0\le j<\ell-1$, there is an isomorphism $\alpha_{j+1}$,
\be
%\label{thmy3a}
\alpha_{j+1}:  \frac{H_U^{j+1}}{U_0'}\ra\frac{H_U^j}{\Gamma_j'},
\ee
whose restriction to $H_U^j/U_0'$ is the isomorphism
$\alpha_j^*=\eta_{j-1}'\circ\alpha_j$, where
$$
\alpha_j^*:  H_U^j/U_0'\ra H_U^{j-1*}/\Gamma_j',
$$
and $\eta_{j-1}'$ is the isomorphism
$$
\eta_{j-1}': H_U^{j-1}/\Gamma_{j-1}'\ra H_U^{j-1*}/\Gamma_j'
$$
given by (\ref{wkhse1}) of Lemma \ref{wkhorse} using
$H_U^{j-1*}=H_U^{j-1}\Gamma_j'$ in the hypothesis;
define $\alpha_0$ to be the trivial isomorphism
$\alpha_0:  H_U^0/U_0'\ra H_U^{-1}/\Gamma_{-1}'$, or
$\alpha_0:  \bone\ra\bone$;

(v) for $0\le j<\ell-1$, the isomorphism $\alpha_{j+1}$ satisfies
\be
\label{eq142}
\alpha_{j+1}(H_U^{j,(k_j+n)}/U_0')=H_U^{j-1,(k_{j-1}+\epsilon_{j-1}+n)}/\Gamma_j'
\ee
for $n$ satisfying $k_j\le k_j+n\le\ell'$.
\end{thm}

We now restate Theorem \ref{thm60} by combining (iv) and (v).

%\newpage

\begin{cor}
\label{cor58}
A group $H_U$ is the state group of a shift group that is a
subdirect product group \ifof\ (i), (ii), and (iii) of Theorem \ref{thm60}
hold, and

(iv) for $0\le j<\ell-1$ and $n$ satisfying $k_j\le k_j+n\le\ell'$,
there is an isomorphism $\alpha_j^{(k_j+n)}$, given by
\be
\label{eq143}
\alpha_j^{(k_j+n)}:  \frac{H_U^{j,(k_j+n)}}{U_0'}\ra\frac{H_U^{j-1,(k_{j-1}+\epsilon_{j-1}+n)}}{\Gamma_j'},
\ee
such that for $n=1,\ldots,\ell'-k_j$, the restriction of $\alpha_j^{(k_j+n)}$ to
$H_U^{j,(k_j+n-1)}/U_0'$ is $\alpha_j^{(k_j+n-1)}$.  The isomorphism
$$
\alpha_j^{(k_j)}:  H_U^j/U_0'\ra H_U^{j-1*}/\Gamma_j'
$$
is the isomorphism $\alpha_j^{(k_j)}=\eta_{j-1}'\circ\alpha_{j-1}^{(\ell')}$,
where $\eta_{j-1}'$ is the isomorphism
$$
\eta_{j-1}': H_U^{j-1}/\Gamma_{j-1}'\ra H_U^{j-1*}/\Gamma_j'
$$
given by (\ref{wkhse1}) of Lemma \ref{wkhorse} using
$H_U^{j-1*}=H_U^{j-1}\Gamma_j'$ in the hypothesis.  For $j=0$, note that
$\alpha_0^{(k_0)}:  H_U^0/U_0'\ra H_U^{-1*}/\Gamma_0'$ is the trivial
isomorphism $\alpha_0^{(k_0)}:  \bone\ra\bone$, and we define $\alpha_0^{(k_0)}$
this way.
\end{cor}

\begin{prf}
For $0\le j<\ell-1$ and $n$ satisfying $k_j\le k_j+n\le\ell'$,
we define $\alpha_j^{(k_j+n)}$ to be an isomorphism with domain and
range as in (\ref{eq143}) such that
$$
\alpha_j^{(k_j+n)}(H_U^{j,(k_j+n)}/U_0')=\alpha_{j+1}(H_U^{j,(k_j+n)}/U_0').
$$
Now note that $\alpha_j^{(k_j)}$ is just $\alpha_j^*$ and
$\alpha_{j-1}^{(\ell')}$ is just $\alpha_j$.
\end{prf}

\section{Algorithms}
\label{sec5}

Arpasi and Palazzo \cite{ARP} have previously given an algorithm to
construct a strongly controllable group code starting with a given
group $G$ (if it is possible).  Sarvis and Trott \cite{ST} and
Sindhushayana, Marcus, and Trott \cite{SMT} have given algorithms to
construct all \htc s and all \hs s, respectively.  In this section,
we give an algorithm to construct the state group of a shift group.
Using the state group, it is easy to construct the \sctlsg\ and group
code.  We start with the group $U_0'$ and work up to state group
$H_U$.  This approach may have an advantage in
constructing a \lgc\ since we can specify a group $U_0'$ with
the desired properties at the start.  In the approach here,
all intermediate calculations take place inside the final group $H_U$, whereas
the approach of \cite{ST,SMT} uses a sequence of derivative
codes or derived shifts which are indirectly related to the final group.

We give an algorithm to find all state groups $H_U$ having a given $U_0'$
and a given signature chain
$$
\bone=\Delta_{-1}'\lhd\Delta_0'\lhd \Delta_1'\lhd
\cdots \Delta_j'\lhd\cdots\lhd\Delta_{\ell-1}'=U_0'.
$$
Then it is easy to find the reduced shift group associated with $H_U$.
The algorithm is loosely based on Algorithm 1 in version 1 of this paper.
We can find a \lsg\ and \lgc\ by modifying Algorithms 2 and 3 in
version 1 of this paper.

The algorithm is just a literal implementation of Corollary \ref{cor58}.
The algorithm has three parts, I, II, and III, which cover the index
step range $j=-1,\ldots,\ell-2$.  Part I is an initialization; this is
index step $j=-1$.  Part II is the main portion of the algorithm; it covers
index steps $j=0,\ldots,\ell-2$.  Part III just states the final result.

\begin{alg}{\bf to find state group:}
\vspace{1ex}

I.  Pick a group $U_0'$ and a normal chain
\be
\label{eq76a}
\bone=\Delta_{-1}'\lhd\Delta_0'\lhd \Delta_1'\lhd
\cdots \Delta_j'\lhd\cdots\lhd\Delta_{\ell-1}'=U_0'
\ee
where each $\Delta_j'\lhd U_0'$.  Construct the parameters
$\epsilon_j$ for $-1\le j<\ell-1$.
Thus using (\ref{eq76a}), we set
$\epsilon_j=1$ if $|\Delta_{j+1}'|/|\Delta_j'|>1$ and $\epsilon_j=0$
if $|\Delta_{j+1}'|/|\Delta_j'|=1$.
There is a subsequence of (\ref{eq76a}),
\be
\label{eq76b}
\bone=\Delta_{-1}'\lhd\cdots\lhd\Delta_m'\lhd \Delta_{m'}'\lhd
\cdots\lhd\Delta_{\ell-1}'=U_0',
\ee
consisting of terms $\Delta_{m+1}'$ for which $\epsilon_m=1$, or
$|\Delta_{m+1}'|/|\Delta_m'|>1$, and an initial term $\bone=\Delta_{-1}'$.
Define parameter $\ell'$,
$$
\ell'\rmdef |\{j | \epsilon_j=1,-1\le j<\ell-1\}|.
$$
There are $\ell'+1$ terms in (\ref{eq76b}).  We reindex the subscripts in
(\ref{eq76b}) with integers $0,1,\ldots,\ell'$ so that order is preserved,
and define this to be the sequence
\begin{multline*}
\bone=H_U^{-1,(0)}\lhd H_U^{-1,(1)}\lhd\cdots
\lhd H_U^{-1,(j)}\lhd H_U^{-1,(j+1)}\lhd\cdots  \\
\lhd H_U^{-1,(\ell')}=U_0'.
\end{multline*}
In other words, $H_U^{-1,(j)}=\Delta_m'$ \ifof\ $H_U^{-1,(j+1)}=\Delta_{m'}'$.
Note that $H_U^{-1,(0)}\rmdef\Delta_{-1}'=\bone$ and
$H_U^{-1,(\ell')}\rmdef\Delta'_{\ell-1}=U_0'=H_U^0$.
In general, for $-1\le j<\ell-1$ define
$$
k_j\rmdef\ell'-\sum_{j\le i<\ell-1}\epsilon_i.
$$
Then $k_{-1}=0$.  With $k_{-1}=0$, note that we have defined
$H_U^{-1,(k_{-1}+n)}$ for $n=0,\ldots,\ell'$.

Define $\Gamma_{-1}'=\bone$.  Note that $\Gamma'_0=\Delta_0'$.
\vspace{1ex}

II.  For $j=0,\ldots, \ell-2$:

{\bf DO}

1.  We are given $H_U^j$ and $\Gamma_j'$.
We have found $H_U^j$ as the sequence of subgroups
\begin{multline*}
H_U^{j-1}=H_U^{j-1,(k_{j-1})},H_U^{j-1,(k_{j-1}+1)},\ldots, \\
H_U^{j-1,(k_{j-1}+n)},\ldots,H_U^{j-1,(\ell')}=H_U^j.
\end{multline*}

2.  We now find $H_U^{j+1}$.  We can do this in increments,
finding $H_U^{j,(k_j+n)}$ and isomorphism $\alpha_j^{(k_j+n)}$
for $k_j\le k_j+n\le\ell'$.
We already know $H_U^{j,(k_j)}=H_U^j=H_U^{j-1,(\ell')}$.
Define the isomorphism
$\alpha_j^{(k_j)}=\eta_{j-1}'\circ\alpha_{j-1}^{(\ell')}$,
where $\eta_{j-1}'$ is the isomorphism
$$
\eta_{j-1}': H_U^{j-1}/\Gamma_{j-1}'\ra H_U^{j-1*}/\Gamma_j'
$$
given by (\ref{wkhse1}) of Lemma \ref{wkhorse} using
$H_U^{j-1*}=H_U^{j-1}\Gamma_j'$ in the hypothesis.  Then
$$
\alpha_j^{(k_j)}:  H_U^j/U_0'\ra H_U^{j-1*}/\Gamma_j'.
$$
(For $j=0$, define $\alpha_0^{(k_0)}$ to be the trivial
isomorphism $\alpha_0^{(k_0)}:  \bone\ra\bone$.)

We now consider some specific details of each increment $n$.
First consider $k_j+n=k_j+\epsilon_j$.  If $\epsilon_j=0$, there
is nothing to do except define $\Gamma_{j+1}'\rmdef\Gamma_j'$.

If $\epsilon_j=1$, we find $H_U^{j,(k_j+\epsilon_j)}$ such that

(i) $H_U^{j,(k_j+\epsilon_j)}\supset H_U^{j,(k_j)}$.

(ii) $H_U^{j,(k_j+\epsilon_j)}$ is an extension of $U_0'$ such that
there is an isomorphism $\alpha_j^{(k_j+\epsilon_j)}$,
$$
\alpha_j^{(k_j+\epsilon_j)}:  \frac{H_U^{j,(k_j+\epsilon_j)}}{U_0'}\ra
\frac{H_U^{j-1,(k_{j-1}+\epsilon_{j-1}+\epsilon_j)}}{\Gamma_j'},
$$
whose restriction to $H_U^{j,(k_j)}/U_0'$ is $\alpha_j^{(k_j)}$.

(iii) $H_U^{j,(k_j+\epsilon_j)}=H_U^{j,(k_j)}(\Gamma_{j+1}')$,
where subgroup $\Gamma_{j+1}'\subset H_U^{j,(k_j+\epsilon_j)}$ satisfies
\begin{align*}
%\nonumber
\Gamma_{j+1}'\cap H_U^{j,(k_j)}&=\Gamma_j', \\
%\nonumber
\Gamma_{j+1}'/\Gamma_j' &\simeq\Delta_{j+1}'/\Delta_j', \\
%\nonumber
\Gamma_{j+1}' &\lhd H_U^{j,(k_j+\epsilon_j)}.
\end{align*}

We also require that
$$
H_U^0,H_U^1,\ldots, H_U^j\lhd H_U^{j,(k_j+\epsilon_j)}.
$$
%%%%%%%%%%%%%%%%%%%%%%%%%%%%%%%%

For the remaining increments, for $n$ satisfying
$k_j+\epsilon_j<k_j+n\le\ell'$, we just need to find
$H_U^{j,(k_j+n)}$ such that

(i) $H_U^{j,(k_j+n)}\supset H_U^{j,(k_j+n-1)}$.

(ii) $H_U^{j,(k_j+n)}$ is an extension of $U_0'$ such that
there is an isomorphism $\alpha_j^{(k_j+n)}$,
$$
%\label{eq142}
\alpha_j^{(k_j+n)}:  \frac{H_U^{j,(k_j+n)}}{U_0'}\ra\frac{H_U^{j-1,(k_{j-1}+\epsilon_{j-1}+n)}}{\Gamma_j'},
$$
whose restriction to $H_U^{j,(k_j+n-1)}/U_0'$ is $\alpha_j^{(k_j+n-1)}$.

We also require that
$$
H_U^0,H_U^1,\ldots, H_U^j\lhd H_U^{j,(k_j+n)}
$$
and $\Gamma_{j+1}'\lhd H_U^{j,(k_j+n)}$.

{\bf ENDDO}
\vspace{1ex}

III.  For $j=\ell-2$, part II is abbreviated since
$k_{\ell-2}+\epsilon_{\ell-2}=k_{\ell-2}+1=\ell'$.
Then $H_U^{\ell-2,(k_{\ell-2}+\epsilon_{\ell-2})}$ is the state
group $H_U$ of a shift group that is a subdirect product group.
\end{alg}

We can implement increment $k_j+n=k_j+\epsilon_j$ as follows.
Since $H_U^{j,(k_j+\epsilon_j)}=H_U^{j,(k_j)}(\Gamma_{j+1}')$, from
(\ref{wkhse3}) of Lemma \ref{wkhorse} we have
$$
\frac{H_U^{j,(k_j+\epsilon_j)}}{\Gamma_j'}\simeq
\frac{H_U^{j,(k_j)}}{\Gamma_j'}\times\frac{\Gamma_{j+1}'}{\Gamma_j'}
\rmdef H^\times.
$$
Thus we first find a group $\Gamma_{j+1}'/\Gamma_j'$ isomorphic to
$\Delta_{j+1}'/\Delta_j'$.  Then form the direct product group $H^\times$.
Now find $H_U^{j,(k_j+\epsilon_j)}\supset H_U^{j,(k_j)}$ an extension
of $\Gamma_j'$ by $H^\times$ such that $H_U^{j,(k_j+\epsilon_j)}$ contains
a normal subgroup $\Gamma_{j+1}'$ which is an extension of $\Gamma_j'$
by $\Gamma_{j+1}'/\Gamma_j'$.  Now check whether (ii) is satisfied.
Note that the direct product group $H^\times$ gives some insight into the
structure of the state group and explains why $D_8$ can be the state
group of the V.32 code \cite{MT}.

The algorithm can be improved by using a composition chain of $H_U$,
as obtained for $G$ in Theorem \ref{thm15}; this approach somewhat
resembles the cyclic extension method \cite{GB}.


\begin{thebibliography}{99}


%\parskip 2mm
\bibitem{KIT}
B. Kitchens, ``Expansive dynamics on zero-dimensional groups,'' {\it
Ergodic Theory and Dynamical Systems} {\bf 7}, pp. 249-261, 1987.

\bibitem{FT}
G. D. Forney, Jr. and M. D. Trott, ``The dynamics of group
codes:  state spaces, trellis diagrams, and canonical encoders,''
{\it IEEE Trans. Inform. Theory}, vol. 39, pp.
1491-1513, Sept. 1993.

\bibitem{LM}
H.-A. Loeliger and T. Mittelholzer, ``Convolutional codes over
groups,'' {\it IEEE Trans. Inform. Theory, Part I}, vol. 42, pp.
1660-1686, Nov. 1996.

\bibitem{UNG}
G. Ungerboeck, ``Channel coding with multilevel/phase
signals,'' {\it IEEE Trans. Inform. Theory}, vol. IT-28,
pp. 55-67, January 1982.

\bibitem{FY2}
G. D. Forney, Jr., ``Geometrically uniform codes,''
{\it IEEE Trans. Inform. Theory}, vol. 37, pp.
1241-1260, Sept. 1991.

\bibitem{MT}
M. D. Trott, ``The algebraic structure of trellis codes,'' Ph.D.
thesis, Stanford Univ., Aug. 1992.

\bibitem{RSH}
E. J. Rossin, N. T. Sindhushayana, and C. D. Heegard,
``Trellis group codes for the Gaussian channel,''
{\it IEEE Trans. Inform. Theory}, vol. 41, pp.
1217-1245, Sept. 1995.

\bibitem{TS}
M. D. Trott and J. P. Sarvis, ``Homogeneous trellis codes,'' in {\it
32nd Annual Allerton Conference on Communication, Control, and
Computing}, Monticello, IL, September 28-30, 1994, pp. 210-219.

\bibitem{JPS}
J. P. Sarvis, ``Symmetries of trellis codes,''
M.E. thesis, MIT, June 1995.

\bibitem{ST}
J. P. Sarvis and M. D. Trott, ``Useful groups for trellis codes,''
in {\it Proc. IEEE Int. Symp. Inform. Theory},
Whistler, BC, Canada, Sept. 17-22, 1995, p. 308.

\bibitem{SMT}
N. T. Sindhushayana, B. Marcus, and M. Trott, ``Homogeneous
shifts,'' {\it IMA J. Math. Contr. Inform.}, vol. 14, pp. 255-287, 1997.

\bibitem{ARP}
J. P. Arpasi and R. Palazzo, Jr., ``An algorithm to construct strongly
controllable group codes,''
{\it 1998 IEEE International Symposium on Information Theory},
Boston, MA, August 1998, p. 154.

\bibitem{MAC}
K. M. Mackenthun, Jr., ``On groups with a shift structure:  the Schreier matrix
and an algorithm,'' in {\it 41st Annual Conf. on Information Sciences and Systems},
Baltimore, MD, March 14-16, 2007.

\bibitem{MAC1}
K. M. Mackenthun, Jr., ``A simple approach to groups with a shift structure
and related group shifts and group codes,'' submitted to {\it
45th Annual Allerton Conference on Communication, Control, and
Computing}, June 28, 2007.

\bibitem{MH}
M. Hall, Jr., {\it The Theory of Groups}, Chelsea, New York, 1959.

\bibitem{ROT}
J. J. Rotman, {\it An Introduction to the Theory of Groups}
($4^{\rm th}$ edition), Springer, New York, 1995.

\bibitem{LMr}
D. Lind and B. Marcus, {\it An Introduction to Symbolic Dynamics and
Coding}, Cambridge Univ. Press, New York, 1995.

\bibitem{DJ}
D. Jungnickel, ``Latin squares, their geometries and their groups.  A
survey,'' in {\it Coding Theory and Design Theory, Part II} (D.
Ray-Chaudhuri, ed.), vol. 21 of {\it IMA Volumes in Mathematics and its
Applications}, pp. 166-225, Springer, 1992.

\bibitem{MN1}
H. B. Mann, ``The construction of orthogonal Latin squares,'' {\it Ann. Math.
Stat.}, vol. 13, 1942, pp. 418-423.

\bibitem{MN2}
H. B. Mann, ``On the construction of sets of mutually orthogonal Latin squares,''
{\it Ann. Math. Stat.}, vol. 14, 1943, pp. 401-414.

\bibitem{APS}
A. P. Sprague, ``Translation nets,'' {\it Mitt. Math. Sem. Giessen}, vol. 157,
1982, pp. 46-68.

\bibitem{BJ}
R. A. Bailey and D. Jungnickel, ``Translation nets and fixed-point-free
group automorphisms,'' {\it J. Comb. Th. (A)}, vol. 55, no. 1,
Sept. 1990, pp. 1-13.

\bibitem{BS}
A. Barlotti and K. Strambach, ``The geometry of binary systems,''
{\it Advances Math.}, vol. 49, 1983, pp. 1-105.

\bibitem{GB}
G. Butler, {\it Fundamental Algorithms for Permutation Groups},
Springer-Verlag, New York, 1991.


\end{thebibliography}
\end{document}